# DECIPHERING NEURAL REPARAMETERIZED FULL-WAVEFORM INVERSION WITH NEURAL SENSITIVITY KERNEL AND WAVE TANGENT KERNEL

RUIHUA CHEN*, BANGYU WU*, YISI LUO‡, XILE ZHAO†, AND DEYU MENG‡

**Abstract.** Full-waveform inversion (FWI) estimates unknown parameters in the wave equation from limited boundary measurements. Recent advances in neural reparameterized FWI (NeurFWI) demonstrate that representing the parameters using a neural network can reduce the reliance on the high-quality initial model and wavefield data, at the cost of slow high-resolution convergence. However, its underlying theoretical mechanism remains unclear. In this study, we establish the neural sensitivity kernel (NSK) and the wave tangent kernel (WTK) to analyze their convergence behavior from both model and data domains. These theoretical frameworks show that the neural tangent kernel (NTK) induced by neural representation adaptively modulates the original sensitivity and wave tangent kernels. This modulation leads to several key outcomes, i.e., the spectral filtering effect, the gradient wavenumber modulation, and the wave frequency bias, connecting the convergence behavior of NeurFWI with the eigen-structures of NSK and WTK. Building on these insights, we propose several enhanced NeurFWI methods with tailored eigen-structures in NSK and WTK to improve inversion performances and efficiency. We numerically validate these theoretical claims and the proposed methods in seismic exploration, and firstly extend their application to medical imaging.

**Key words.** Full-waveform inversion, seismic inversion, wave equation, neural reparameterization, neural tangent kernel, sensitivity kernel, deep learning, nonlinear inverse problem.

**AMS subject classifications.** 86A22, 35R30, 68T07, 65M32, 35Q86

**1. Introduction.** Many applications in computational science involve the parameter estimation in partial differential equations (PDEs) [21, 23, 50]. One prominent and computationally challenging nonlinear problem is full-waveform inversion (FWI) [8, 46], as shown in Fig. 1, which leverages the entire waveform content for high-resolution imaging in geophysical and industrial applications [61], including seismic exploration [49], medical imaging [20], and ground penetrating radar (GPR) [41]. The unknown parameters in the wave equation are determined by iteratively minimizing the following misfit between simulated and recorded wavefield data, i.e.,

$$m^{\dagger} = \arg\min_{m\in\mathcal{M}} \{\mathcal{J}(\mathcal{P}[u(m)], d_{\text{obs}}) + \mathcal{R}(m)\}, \tag{1.1}$$

where $m$ denotes the parameters, $\mathcal{M}$ represents the admissible set, $u$ is the simulated data, $\mathcal{P}$ is the sampling operator, $d_{\text{obs}}$ is the observed data, and the objective function includes data fidelity term $\mathcal{J}$ and regularization term $\mathcal{R}$. The data $u(m)$ is obtained by solving the wave equation, which is nonlinear w.r.t the parameters. In this study, we consider the following acoustic wave equation without loss of generality [16]:

$$\begin{cases} m\partial_{tt}u - \Delta u + \mu\partial_t u = s, & \text{in } T\times\Omega, \\ \mathbf{n}\cdot\nabla u|_{\Gamma_N} = 0, u|_{\Gamma_D} = 0, & \text{in } T\times\partial\Omega, \\ u(0,\cdot) = u_0, \partial_t u(0,\cdot) = u_1, & \text{in } \Omega, \end{cases} \tag{1.2}$$

where $T$ is the time domain and $\Omega$ is the spatial domain with boundary $\partial\Omega$ (i.e., Neumann boundary $\Gamma_N$ and Dirichlet boundary $\Gamma_D$), $\Delta$ be the Laplacian operator, $\nabla$ be the differential operator, $\mathbf{n}$ is the unit outward normal, $\mu$ is the damping term, $s$ is the source function, $u_0$ and $u_1$ represent the initial wavefield and accelerated speed, respectively. We define the model domain as the space of parameters, and the data domain as the space of wavefield recordings. The wave equation maps the parameters $m$ in the model domain to the wavefield $u(m)$ in the data domain. To update the parameters based on $L_2$-norm misfit, the sensitivity kernel maps the misfit in the data domain back to the gradients $\partial \mathcal{J}(m)/\partial m$ in the model domain [50], i.e.,

$$\frac{\partial \mathcal{J}(m)}{\partial m}(\mathbf{y}) = \int_T \int_\Omega \frac{\partial u(\mathbf{x},t)}{\partial m(\mathbf{y})} \cdot \Delta D_m(\mathbf{x},t)\, d\mathbf{x}dt,\ \mathbf{y} \in \Omega, \tag{1.3}$$

where $\Delta D_m = \mathcal{P}^\dagger\left(\mathcal{P}[u(m)] - d_{\text{obs}}\right)$ denotes the data misfit and $\partial u/\partial m$ represents the sensitivity kernel. Although the adjoint state method [50] can efficiently calculate the gradient, the sensitivity analysis remains essential to understand FWI [9, 58, 66].

*School of Mathematics and Statistics, Xi'an Jiaotong University, Xi'an, Shaanxi 710049, China (Ruihua.Chen@stu.xjtu.edu.cn, bangyuwu@xjtu.edu.cn).

†School of Mathematical Sciences, University of Electronic Science and Technology of China, Chengdu 610000, China (xlzhao122003@163.com).

‡Corresponding author. School of Mathematics and Statistics, Xi'an Jiaotong University, Xi'an, Shaanxi, China (yisiluo1221@foxmail.com, dymeng@mail.xjtu.edu.cn)

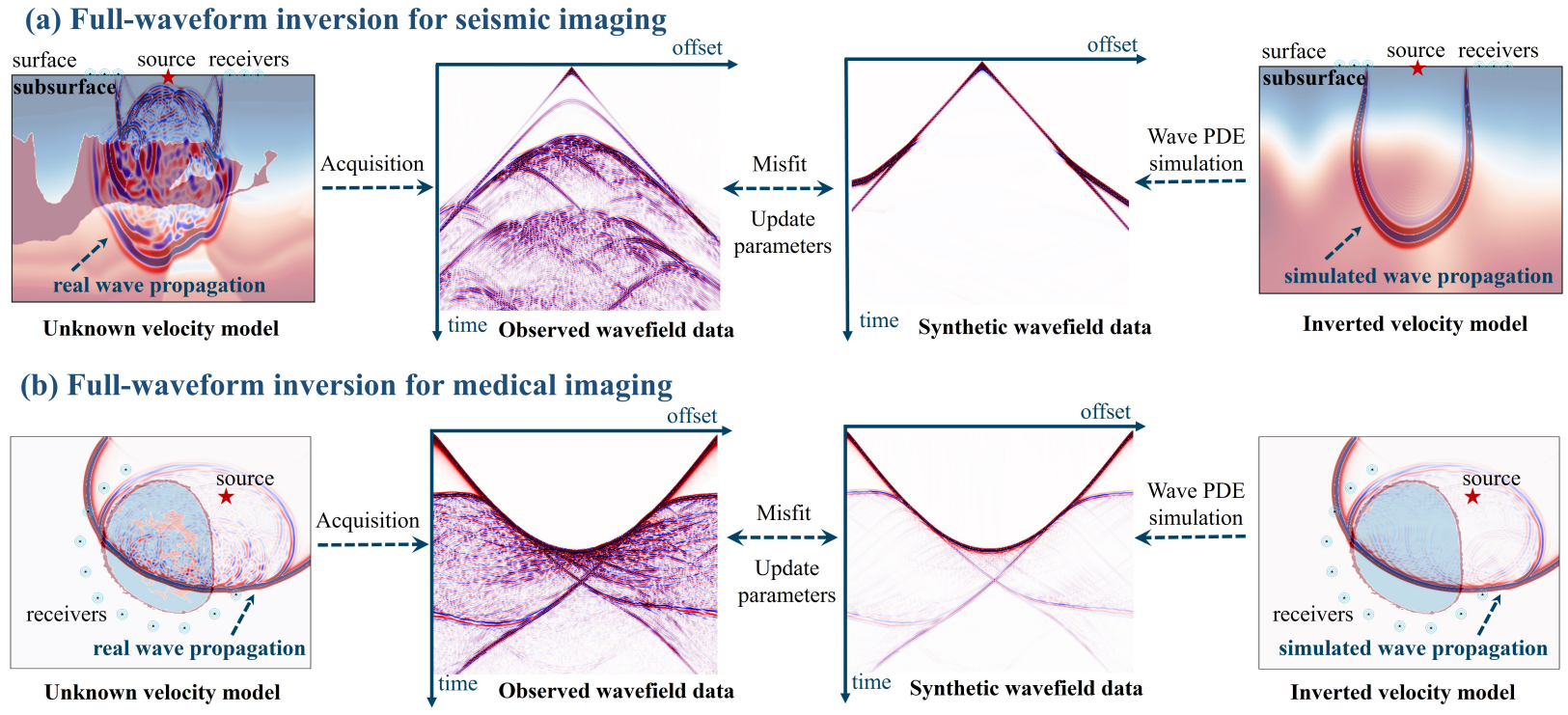

FIG. 1. *The process of FWI. (a) Seismic imaging. (b) Medical imaging. The acquisition of wavefield data involves activating a source and recording the data with receivers. FWI inverts the unknown parameters of the model, starting from an initial model based on wavefield measurements.*

**1.1. Challenges of FWI.** FWI is a highly ill-posed PDE constrained optimization problem, particularly when starting from inaccurate initial models [50]. The key challenges lie in the wavenumber distribution of sensitivity kernel [52, 45, 66, 69] and cycle-skipping [50, 61, 54, 19]. Next, we detail these two challenges below.

**1.1.1. Model Domain Challenge.** In realistic scenarios, the high-wavenumber mode dominates FWI gradients [52, 45, 66] due to the inherent suppression of low-frequency sensitivity kernel from the derivative operator [69]. Theoretically, for a homogeneous velocity background medium with an incident monochromatic plane wave along $\hat{\mathbf{s}}$ and a scattered plane wave along $\hat{\mathbf{r}}$ under the far-field assumption, the gradient of FWI in the equation (1.3) can be expressed as follows[1]:

$$\frac{\delta \mathcal{J}(m)}{\delta m}(\mathbf{y}) = \sum_\omega \sum_s \sum_r \omega^2 \underbrace{\Re\{\exp(-i\frac{\omega}{c_0}(\hat{\mathbf{s}}+\hat{\mathbf{r}})\cdot\mathbf{y})\}}_{\text{Wavenumber oscillatory kernel}} \underbrace{\Delta D_m(\mathbf{x}_s,\mathbf{x}_r,\omega)}_{\text{Data residual}}, \tag{1.4}$$

where $\Re\{\cdot\}$ denotes the real part, $\omega, s, r$ denote the frequency, source, and receiver index, respectively. The factor $\omega^2$ amplifies higher-frequency sensitivity contributions, and $\exp(-i\frac{\omega}{c_0}(\hat{\mathbf{s}}+\hat{\mathbf{r}})\cdot\mathbf{y})$ introduces wavenumber oscillations, causing higher frequencies to produce high-wavenumber gradients. Hence, the gradients are inherently biased toward high-wavenumber updates, hindering the background reconstruction. To exhibit this issue, we compute frequency-domain sensitivity kernels for a two-layer model[2], as shown in Figs. 2 (a) and (b). The low-frequency kernel tends to favor low-wavenumber updates, while the high-frequency kernel tends to govern high-wavenumber updates. The high-frequency component dominates in amplitude (Fig. 2 (c)), leading to excessive high-wavenumber updates. This imbalance steers the inversion toward a local minimum [50, 52, 45] that captures structural details but lacks kinematic accuracy.

[1]The detailed derivation can be found in the supplementary materials SM1.1.

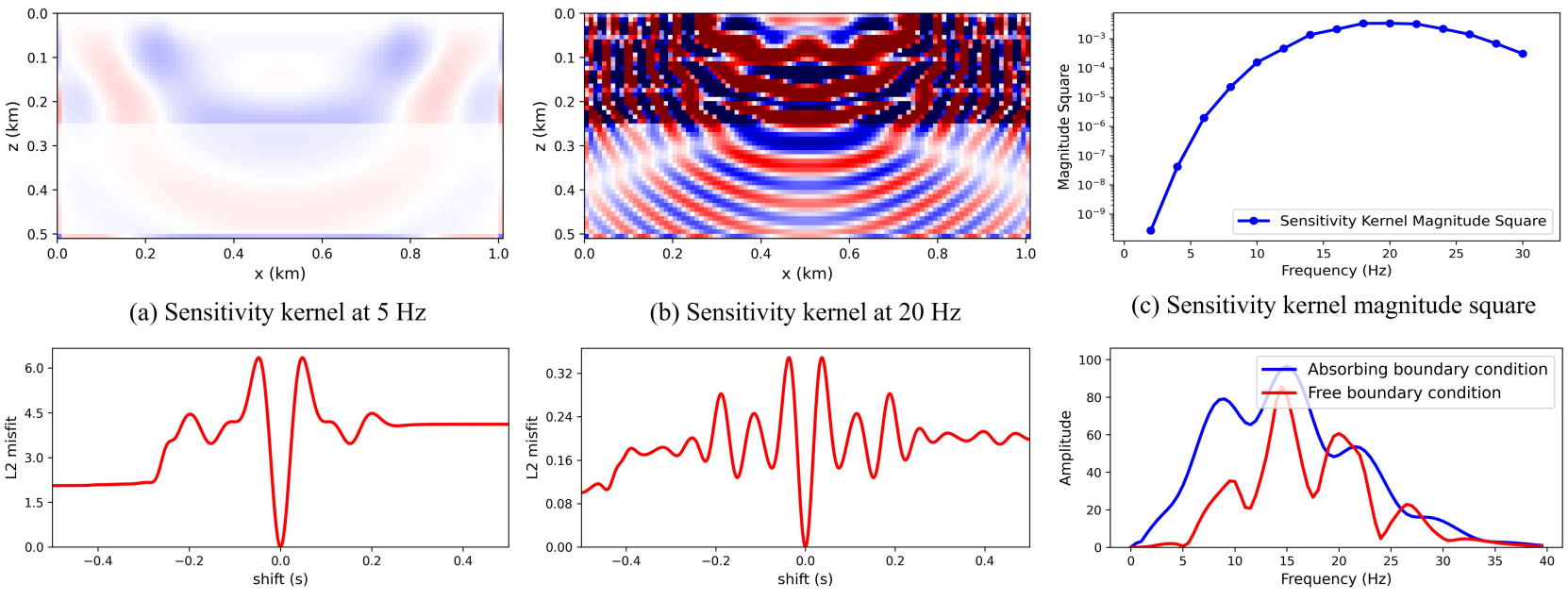

(a) Sensitivity kernel at 5 Hz (b) Sensitivity kernel at 20 Hz (c) Sensitivity kernel magnitude square

(d) $L^2$ misfit with absorbing boundary (e) $L^2$ misfit with free boundary (f) Spectral distribution of wavefield data

FIG. 2. *Traditional FWI may suffer from unbalanced wavenumber distributions in the model domain (a)-(c) and cycle-skipping in the data domain (d)-(f) (a two-layer model example).*

**1.1.2. Data Domain Challenge.** From the data-fitting perspective, FWI faces the notorious cycle-skipping (i.e., half-cycle waveform misfit causing inversion failure) due to the wavefield oscillatory [61]. To illustrate this challenge, we compute the $L^2$ misfit as a function of the time shift $s$ under absorbing boundary conditions, as shown in Fig. 2 (d). The result shows that there are multiple local minima around the true time shift, which may trap the optimization in local minima. The issue becomes particularly pronounced in complex regions [7] with free-surface boundary conditions. In such settings, the reflections from the free surface and subsurface interfaces undergo multiple reflections, scattering, and coupling [39]. These surface-related ghosts [53] introduce additional local minima into the inversion and generate spectral notches that attenuate essential low-frequency components, as shown in Figs. 2 (e) and (f).

**1.2. Related Works and Motivations.** To mitigate the above challenges, conventional researches focus on the sensitivity kernel decomposition or gradient preconditioning [9, 58, 66], multi-frequency continuation strategies [8, 18], more convex misfit functions [54, 63], hybrid travel time tomography methods [2, 48, 42], and explicit regularization approaches [1, 60, 2, 17, 3]. The effectiveness and mechanism of most methods can be validated by analyzing the sensitivity kernel and gradient [50, 54].

Deep learning-based FWI methods have also been proposed to address these challenges, including data-driven approaches [55, 68], learnable loss functions [62, 36], and adaptive regularization [43, 51, 47]. Recently, neural reparameterized FWI (NeurFWI) demonstrated that representing parameters via neural networks rather than grid-based discretization (Fig. 3 (a)) mitigates dependence on high-quality initial models without explicit regularization [57, 71, 40, 31]. NeurFWI architectures primarily include convolutional neural networks (CNNs) [14, 26, 29, 30, 67, 71] and implicit neural representations (INRs) [40, 64, 15, 27, 12, 11]. We focus on INR-based NeurFWI[3] because it effectively alleviates cycle-skipping even with constant initial models [40]. However, existing INR implementations using tanh [71], sinusoidal [40], or Gabor [64] activations may suffer from slow high-wavenumber convergence, rendering them computationally prohibitive for large-scale datasets. As shown in Fig. 3 (b)-(h), comparisons between implicit FWI (IFWI) [40] and automatic differentiable FWI (ADFWI) [30] on the 2D Overthrust model[4] confirm that while NeurFWI reduces initial model dependency, it struggles with slow high-wavenumber convergence.

[2]Detailed settings can be found in the supplementary materials SM3.2.

[3]We focus on INRs due to the natural coordinate-to-value mapping, which facilitates theoreti-

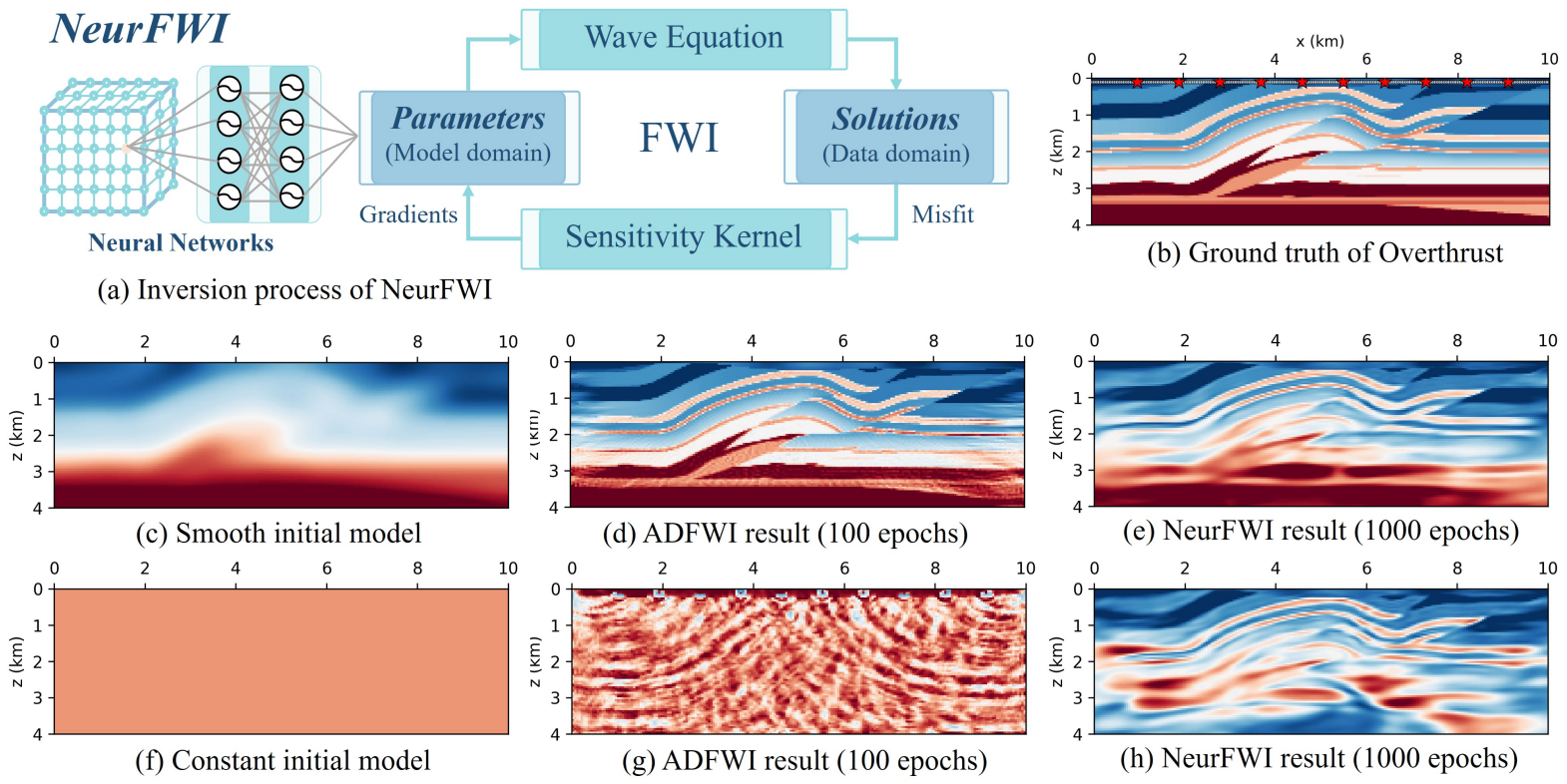

FIG. 3. *The process and performance of NeurFWI. (a) NeurFWI represents the physical parameters using networks and updates the neural parameters by minimizing the wavefield misfit. (b) Ground truth. (c) Smooth initial model. (d) ADFWI result (smooth). (e) NeurFWI result (smooth). (f) Constant initial model. (g) ADFWI result (constant). (h) NeurFWI result (constant).*

A deeper understanding of the underlying mechanisms within NeurFWI is essential for further enhancement. Existing studies attribute its success to various forms of implicit regularization, including initial model pre-training [57], sparse parameterization [22], and spatial correlation [71]. We previously validated that NeurFWI embeds smooth regularization and developed a low-rank NeurFWI (LR-FWI) method [11]. Regarding optimization dynamics, over-parameterization has been shown to increase degrees of freedom [25] and favor saddle points over local minima, thereby enhancing stability [40]. Furthermore, we have empirically validated the spectral bias in NeurFWI [27] (i.e., NeurFWI prioritizes fitting low-frequency data). Similarly, CNN-based FWI exhibits spectral bias [31] and the model updates are controlled by this bias [56]. However, existing literature offers primarily qualitative explanations, and there is currently no suitable theoretical framework for NeurFWI that adequately explains its convergence behavior. Here, we summarize some intriguing questions:

(**Model Domain**) Why does NeurFWI prioritize recovering low-wavenumber model components? Specifically, what changes occur in its gradient or sensitivity kernel to mitigate the unbalanced wavenumber distribution (Fig. 2 (a) and (b)), and how do different network architectures affect this modulated behavior?

(**Data Domain**) Why does NeurFWI exhibit slow convergence for high-frequency components? Is there a mathematical framework that characterizes residual evolution and explains cycle-skipping in FWI? How does the NeurFWI mitigate this issue?

In this paper, we investigate the above questions. A deeper understanding of both the model and data domains for NeurFWI can illuminate the role of neural representations and inspire the design of more tailored representations for FWI.

cal analysis. Our theoretical claims can be readily extended to other representations (e.g., CNNs, attention, recurrent neural networks) with appropriate neural tangent kernel formulations [6, 35, 4].

[4]Detailed settings are provided in the supplementary materials SM3.3.

**1.3. Our Works and Contributions.** In this study, we establish a theoretical foundation for NeurFWI to explain its inversion behaviors, analyzing existing methods from both model and data domains. Based on this, we propose new methodologies for NeurFWI and demonstrate its application across seismic and medical scenarios[5].

(**Theories in Model Domain**) Our theoretical analysis reveals that the neural tangent kernel (NTK) modulates the original sensitivity kernel to form the neural sensitivity kernel (NSK). The NSK induces smoother gradients via decomposing the original kernel along NTK eigenfunctions and weighting them by their eigenvalues, i.e., the spectral filtering effect (Fig. 4 and Proposition 3.3). Fourier analysis demonstrates that this modulation suppresses high-wavenumber components, i.e., the gradient wavenumber modulation (Fig. 5 and Theorem 3.8). We also explore the impact of activation functions on this modulation behavior, including tanh (NNFWI) [71], sinusoidal (IFWI) [40], and Gabor (WinFWI) [64] (Corollaries 3.10, 3.12, and 3.14).

(**Theories in Data Domain**) We establish the wave tangent kernel (WTK) to characterize wavefield residual evolution, where the WTK weights the data residual along different spectral directions by the corresponding eigenvalues to drive misfit reduction (Theorem 4.9). Under the Born approximation, conventional FWI inherently prioritizes high-frequency misfit reduction, rendering short-wavelength recovery susceptible to cycle-skipping [50]. In NeurFWI, the NTK modulates the WTK to form the wave neural tangent kernel (WNTK). We show that WNTK exhibits faster eigenvalue decay than WTK (Theorem 4.12), thereby preserving low-frequency evolution while suppressing high-frequency fitting. Specifically, the Fourier transform of WNTK contains a high-frequency attenuation term (Fig. 8 and Proposition 4.6).

(**Methodologies**) To achieve a more tailored spectral modulation and WTK's eigen-structure, we propose a systematic design methodology for NeurFWI (Theorems 4.10 and 5.3), such as a novel representation integrating INR with multi-resolution parametric encoding (MPE) [34] for FWI (IG-FWI) and low-rank INR [32] with MPE (LRG-FWI). These methods induce a more suitable eigenvalue decay of WTK and better gradient modulation tailored for FWI (Corollaries 5.6 and 5.8).

(**Experiments and Applications**) We numerically verify our theoretical findings, and we then evaluate the inversion performance of our proposed NeurFWI on several complex seismic models (i.e., BP salt, the Canadian Foothills, and the 2014 Chevron blind test models) and medical imaging (i.e., breast and brain).

**1.4. Outline of This Work.** This paper is structured as follows. Section 2 presents the preliminary notations. Section 3 establishes the NSK theory. Section 4 establishes the WTK theory. Section 5 provides a systematic design methodology. Applications are presented in Section 6. We summarize this paper in Section 7.

**2. Preliminaries and Background.** Let scalars, vectors, and matrices be $x$, $\mathbf{x}$, and $\mathbf{X}$. $(\delta\phi/\delta f)[f]$ denotes the Fréchet derivative of $\phi$ at $f \in L^p(\Omega)$. To avoid ambiguity, we denote $\langle\cdot,\cdot\rangle_{\mathbb{R}^n}$ by the standard Euclidean inner product on $\mathbb{R}^n$, and $\langle\cdot,\cdot\rangle$ by the $L^2$ inner product. We define $\mathcal{F}[\phi]$ (or $\hat{\phi}$) as the Fourier transform of the function $\phi$. The Fourier transform and its inverse transform are defined as follows:

$$
\begin{aligned}
\hat{\phi}(\boldsymbol{\xi}) = \mathcal{F}[\phi](\boldsymbol{\xi}) &= \mathcal{F}_{\mathbf{x}\to\boldsymbol{\xi}}[\phi](\boldsymbol{\xi}) = \int_{\mathbb{R}^d} \phi(\mathbf{x}) e^{-2\pi i \boldsymbol{\xi}\cdot\mathbf{x}}\, d\mathbf{x}, \\
\mathcal{F}^{-1}[\phi](\boldsymbol{\xi}) &= \mathcal{F}^{-1}_{\mathbf{x}\to\boldsymbol{\xi}}[\phi](\boldsymbol{\xi}) = \int_{\mathbb{R}^d} \phi(\mathbf{x}) e^{2\pi i \boldsymbol{\xi}\cdot\mathbf{x}}\, d\mathbf{x}.
\end{aligned}
\tag{2.1}
$$

[5] These theories and methodologies can be extended to other wave equation-based imaging problems, as well as offer potential solutions to a broader class of PDE-constrained inversion problems.

6 **2.1. Full-Waveform Inversion.** Without loss of generality, we only consider one source function to simplify the theoretical notation. Let $\mathcal{P}$ be the sampling operator, FWI estimates the true model $m^\dagger$ starting with an initial model $m_0$ by minimizing the data misfit between the observed data $d_{\text{obs}}$ and synthetic data $\mathcal{P}[u]$:

$$\begin{aligned}\min_m \mathcal{J}(m) &= \frac{1}{2}\int_T \int_{\Omega_r} \left|\mathcal{P}[u(m)](\mathbf{x},t) - d_{\text{obs}}(\mathbf{x},t)\right|^2 \mathrm{d}\mathbf{x}\mathrm{d}t, \\ &\text{s.t. wave equation (1.2) and } m^-(\mathbf{x}) \le m(\mathbf{x}) \le m^+(\mathbf{x}) \quad \text{for a.e. } \mathbf{x}\in\Omega,\end{aligned} \tag{2.2}$$

where $\Omega_r$ denotes the wavefield sampling domain, and $m^-, m^+$ denote the minimum and maximum velocity value, respectively. Let $(\delta\mathcal{J}/\delta m)[m]$ be the gradient and $m^{(k)}$ be the parameters at the $k$-th iteration, $\eta$ be the learning rate, and $\tau$ be the iteration pseudo-time. In the continuous time limit (i.e., as $\eta$ tends to zero) [24], the gradient descent-based FWI is formulated as the following iterative scheme [65]:

$$m^{(k+1)} = m^{(k)} - \eta\frac{\delta\mathcal{J}}{\delta m}[m^{(k)}] \xrightarrow{\eta\to 0} \frac{dm}{d\tau} = -\frac{\delta\mathcal{J}}{\delta m}[m], \tag{2.3}$$

where $m^{(0)} = m_0$ denotes the initial model, and the gradient can be computed via the adjoint-state method [50] or automatic differentiation [30]. Theoretically, the gradient can be expressed as the wavefield residual weighted by the sensitivity kernel [13] in equation (1.3) (see Fig. 4). The sensitivity kernel propagates wavefield errors back to parameter gradients. Analyzing its properties can reveal the intrinsic wavenumber balance of FWI gradients and explain the bias toward high-wavenumber updates [69].

**2.2. Neural Reparameterization.** Let the space of network parameters be denoted by $\boldsymbol{\Theta} \subseteq \mathbb{R}^n$ ($n$ is the parameter number). NeurFWI uses networks to represent the models, especially a coordinate-based network $F_{\boldsymbol{\theta}} : \Omega \to \mathbb{R}$ with learnable parameters $\boldsymbol{\theta} \in \boldsymbol{\Theta}$, which maps a coordinate $\mathbf{x} \in \Omega$ to the corresponding perturbation $F_{\boldsymbol{\theta}}(\mathbf{x})$. Hence, the model is reparameterized as $m_{\boldsymbol{\theta}}(\mathbf{x}) = F_{\boldsymbol{\theta}}(\mathbf{x}) + m_0(\mathbf{x})$, where $m_0 : \Omega \to \mathbb{R}$ is the initial model [12]. As a result, the optimization variables of NeurFWI change from the discrete parameter models $m$ in equation (2.2) to neural parameters $\boldsymbol{\theta}$, i.e.,

$$\begin{aligned}\min_{\boldsymbol{\theta}\in\boldsymbol{\Theta}} \mathcal{J}(\boldsymbol{\theta}) &= \frac{1}{2}\int_T \int_{\Omega_r} \left|\mathcal{P}[u(m_{\boldsymbol{\theta}})](\mathbf{x},t) - d_{\text{obs}}(\mathbf{x},t)\right|^2 \mathrm{d}\mathbf{x}\mathrm{d}t \\ &\text{s.t. wave equation (1.2) and } m^-(\mathbf{x}) \le m_{\boldsymbol{\theta}}(\mathbf{x}) \le m^+(\mathbf{x}) \quad \text{for a.e. } \mathbf{x}\in\Omega.\end{aligned} \tag{2.4}$$

Let $\boldsymbol{\theta}^{(k)}$ and $m_{\boldsymbol{\theta}}^{(k)}$ be the neural parameters and generated model at the $k$-th iteration. In the continuous time limit, the iterative scheme of NeurFWI is given by

$$\boldsymbol{\theta}^{(k+1)} = \boldsymbol{\theta}^{(k)} - \eta\frac{\partial\mathcal{J}}{\partial\boldsymbol{\theta}}[m_{\boldsymbol{\theta}}^{(k)}] \xrightarrow{\eta\to 0} \frac{d\boldsymbol{\theta}}{d\tau} = -\frac{\partial\mathcal{J}}{\partial\boldsymbol{\theta}}[m_{\boldsymbol{\theta}}], \tag{2.5}$$

where $\partial\mathcal{J}/\partial\boldsymbol{\theta}$ denotes the gradient of network parameters. To simplify the theoretical analysis, we consider a shallow fully-connected neural network with one hidden layer and a smooth nonlinear activation function $\sigma : \mathbb{R} \to \mathbb{R}$. Let $\boldsymbol{\theta} = \{W^1, W^0, b\}$ be trainable parameters, where $W^0 = \{\mathbf{w}_j\}_1^p \in \mathbb{R}^{d\times p}, W^1 = \{a_j\}_1^p \in \mathbb{R}^{p\times 1}$, and $\mathbf{b} = \{b_j\}_1^p \in \mathbb{R}^p$. For coordinates input $\mathbf{x} \in \Omega$, the network output is given by:

$$F_{\boldsymbol{\theta}}(\mathbf{x}) = \frac{1}{\sqrt{n}}\sigma^*(\mathbf{x}) = \frac{1}{\sqrt{n}}W^1 \cdot \sigma(W^0\mathbf{x} + \mathbf{b}) = \frac{1}{\sqrt{n}}\sum_{i=1}^{n} a_j\sigma(\mathbf{w}_j^\top\mathbf{x} + b_j), \tag{2.6}$$

where we define $\sigma^*(\mathbf{x}) = W^1 \cdot \sigma\big(W^0\mathbf{x} + \mathbf{b}\big)$ for simplifying the notation, and $\sigma$ can be tanh [71], sinusoidal [43], and Gabor wavelet [64] activation functions. In the theoretical analysis, we follow [24] for network initialization, i.e., all the weights and biases are initialized to be independent and identically distributed as the standard normal distribution $\mathcal{N}(0,1)$. Here, we employ the rescaling factor of $1/\sqrt{n}$ [24], which is crucial for ensuring the convergence of the initialized NTK [70]. Specifically, the NTK induced by networks (see Definition 2.1) governs the evolution of network output.

Definition 2.1 (NTK [24]). *For a neural network $F_{\boldsymbol{\theta}}(\mathbf{x})$, the NTK is defined as the Gram matrix of the network Jacobian w.r.t its learnable neural parameters:*

$$\Theta_n(\mathbf{x},\mathbf{x}';\boldsymbol{\theta}) = \left\langle \frac{\partial F_{\boldsymbol{\theta}}(\mathbf{x})}{\partial \boldsymbol{\theta}}, \frac{\partial F_{\boldsymbol{\theta}}(\mathbf{x}')}{\partial \boldsymbol{\theta}} \right\rangle_{\mathbb{R}^n} = \frac{1}{n}\sum_{i=1}^{n} \Big(\nabla_{\theta_i}\sigma^*(\mathbf{x},\boldsymbol{\theta})\Big)\cdot\Big(\nabla_{\theta_i}\sigma^*(\mathbf{x}',\boldsymbol{\theta})\Big), \tag{2.7}$$

*where $\langle\cdot,\cdot\rangle_{\mathbb{R}^n}$ denotes the inner product, and $n$ is the number of parameters.*

This NTK provides a mathematical framework for analyzing the training dynamics of networks [24]. We establish the main assumptions for the networks (2.6), which are relatively mild assumptions that could be realized through proper initializations.

Assumption 2.2. *For the network $F_{\boldsymbol{\theta}}$ (2.6), the parameters are mutually independent and remain uniformly bounded by a constant $C > 0$ independent of the hidden layer width $n$. Moreover, the parameter $b$ satisfies $\mathbb{E}[b] = 0$ and $\mathbb{E}[b^2] = \sigma_b < \infty$, and the distribution of the parameter $\mathbf{w}$ is radially symmetric (i.e., $\rho_{\mathbf{w}}(\mathbf{w}) = \rho_{\mathbf{w}}(||\mathbf{w}||\mathbf{e}_1)$, where $\mathbf{e}_1 = (1,0,\cdots,0)^T \in \mathbb{R}^d$) and satisfies $\mathbb{E}[\mathbf{w}] = 0$. The activation $\sigma$ is of class $C^{k+1}(\mathbb{R})$, and its derivatives up to order $k+1$ are uniformly bounded on $\mathbb{R}$.*

## 3. Neural Sensitivity Kernel Analysis.

This section elucidates how the NTK spectrally modulates the sensitivity kernel to yield smoother gradient components. Specifically, wavenumber-domain Fourier analysis reveals that NeurFWI inherently attenuates high-wavenumber components while enhancing low-wavenumber ones. Furthermore, we systematically compare this modulation behavior across different NeurFWI variants and validate the theoretical claims through numerical experiments.

### 3.1. Spectral Filtering Effect of NSK.

Consider the iterative scheme of conventional FWI (2.3) and the gradient (1.3). In the continuous time limit, the dynamic gradient flow of parameter models $m$ can be expressed as follows:

$$\frac{dm(\mathbf{y})}{d\tau} = -\frac{\partial \mathcal{J}(m)}{\partial m}(\mathbf{y}) = -\int_T\int_\Omega \frac{\partial u(\mathbf{x},t)}{\partial m(\mathbf{y})}\cdot \Delta D_m(\mathbf{x},t)\; d\mathbf{x}dt. \tag{3.1}$$

This expression indicates that the parameter update is governed by the gradients, which are derived by weighting the wavefield errors $\Delta D_m$ with the sensitivity kernel $\partial u/\partial m$. From the analysis of equation (1.4), the sensitivity kernel can induce more high-wavenumber gradient components, thereby hindering the update of the background model. The process of gradient generation is shown in Fig. 4. In contrast, NeurFWI prioritizes recovering the low-wavenumber background velocity model. The key to the underlying mechanism lies in the sensitive kernel of NeurFWI. The following proposition shows that NTK induced by networks acts as a low-pass filter.

Proposition 3.1 (Neural Sensitivity Kernel, NSK). *Consider the neural network (2.6), the objective function (2.4), and the gradient iterative scheme (2.5). Let $\Theta : \Omega\times\Omega \to \mathbb{R}$ be NTK and define the neural sensitivity kernel as follows:*

$$\frac{\partial u_\Theta(\mathbf{x},t)}{\partial m(\mathbf{y})} = \int_\Omega \frac{\partial u(\mathbf{x},t)}{\partial m(\mathbf{z})}\cdot\Theta(\mathbf{y},\mathbf{z};\boldsymbol{\theta})\; d\mathbf{z}, \tag{3.2}$$

*where the NSK can be regarded as a linear integral operator governed by NTK $\Theta$. As the learning rate $\eta$ tends to zero and the width of the network tends to infinity, the*

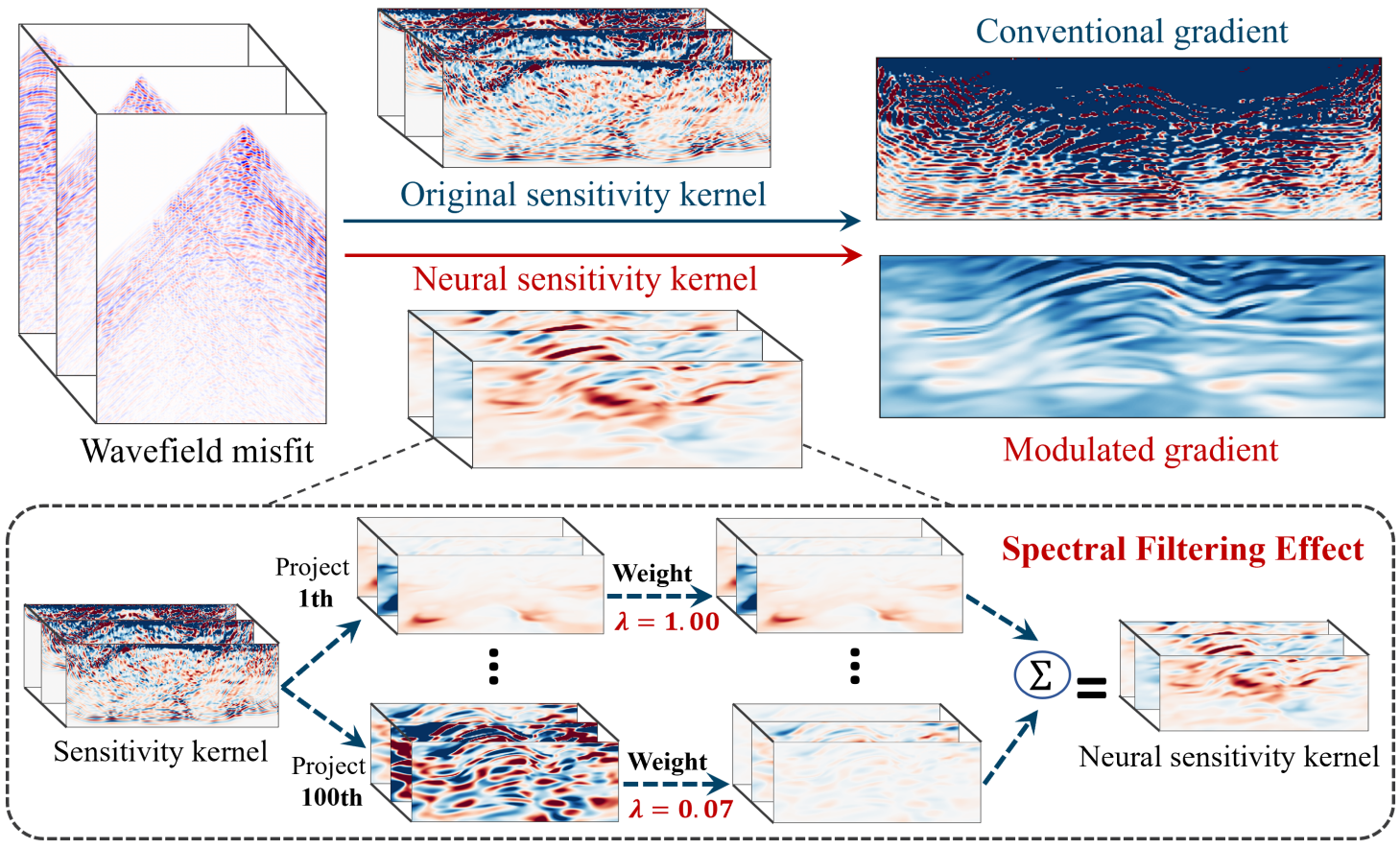

FIG. 4. *The gradient generation process of conventional FWI and NeurFWI. (a) The process for conventional FWI. (b) The spectral filtering mechanism of NeurFWI includes projecting the sensitivity kernel, weighting its components, and re-projecting them to the physical domain.*

*dynamic evolution flow of parameter model $m_{\boldsymbol{\theta}}$ can be expressed as follows:*

$$\frac{dm_{\boldsymbol{\theta}}(\mathbf{y})}{d\tau} = -\int_T \int_\Omega \frac{\partial u_\Theta(\mathbf{x},t)}{\partial m(\mathbf{y})} \Delta D_m(\mathbf{x},t)\ d\mathbf{x}dt. \tag{3.3}$$

*Proof.* The proof can be found in the supplementary materials SM1.2. ∎

*Remark* 3.2. This proposition shows that the parameter evolution is driven by the wavefield error $\Delta D_m$, weighted by the newly defined NSK in equation (3.2) (i.e., $\partial u_\Theta / \partial m$), where the structure-dependent NTK induced by neural networks modulates the original kernel. This modulation influences the mapping from data-domain errors to model-domain gradients, ultimately altering the optimization trajectory.

Furthermore, the NSK defined in Proposition 3.1 can be regarded as a spectrally filtered version of the sensitive kernel controlled by the NTK (see Fig. 4), i.e.,

PROPOSITION 3.3 (Spectral Filtering Effect). *Given Assumption 2.2 and the NTK spectral decomposition, the NSK projects the sensitivity kernel onto the NTK eigenbasis, weights it by eigenvalues, and reconstructs it in the spatial domain:*

$$\frac{\partial u_\Theta(\mathbf{x},t)}{\partial m(\mathbf{y})} = \sum_{k=1}^{\infty} \lambda_k \left\langle \frac{\partial u(\mathbf{x},t)}{\partial m(\cdot)}, \phi_k(\cdot) \right\rangle \phi_k(\mathbf{y}), \tag{3.4}$$

*where $\{\phi_k\}_{k=1}^{\infty}$ are the eigenbasis and $\{\lambda_k\}_{k=1}^{\infty}$ are the non-negative eigenvalues arranged in order, which can be obtained by eigenvalue decomposition of NTK based on its symmetry and semi-positive definiteness, i.e., $\Theta(\mathbf{y}, \mathbf{z}, \boldsymbol{\theta}) = \sum_{k=1}^{\infty} \lambda_k \phi_k(\mathbf{y}) \phi_k(\mathbf{z})$.*

*Proof.* The proof can be found in the supplementary materials SM1.3. ∎

*Remark* 3.4. The sensitivity kernel transforms into the NSK via filtering with the NTK, which supplies both the eigenbasis $\phi_k$ and the filtering coefficients $\lambda_k$. Because the smoothness of NTK and its eigenvalues decay rapidly with $k$ [24], the NTK acts as a low-pass filter for the sensitivity kernel, as shown in Fig. 4, i.e., low-wavenumber modes (large $\lambda_k$) are preserved, while high-wavenumber modes (small $\lambda_k$) are suppressed. This prevents overfitting to high-wavenumber noise, speeds up convergence of the background model, but slows recovery of fine details. Hence, a balanced NTK structure are required for NeurFWI. In Section 5, we propose hybrid neural representations to design a balanced NSK with controlled eigenvalue decay rate.

*Remark* 3.5. Due to the linearity property of the inner product, the spectral filtering effect can be regarded as acting on the original gradient directly. From the analysis of equation (1.4), three strategies can be used to generate low-wavenumber gradients: (1) Attenuating the energy of the sensitivity kernel $\omega^2$; (2) Reducing the wavenumber distribution $\exp(-i\frac{\omega}{c_0}(\hat{\mathbf{s}}+\hat{\mathbf{r}})\cdot\mathbf{y})$, e.g., changing the high-wavenumber modes to the low-wavenumber modes; (3) Limiting the high-frequency components of wavefield data residual, e.g., multiscale strategies. Which of the above is the reason why NeurFWI can generate low-wavenumber gradients? We consider the modulated gradient by NTK to unveil the mechanism, which can be expressed as follows:

$$\frac{\delta \mathcal{J}}{\delta m}[m_{\boldsymbol{\theta}}](\mathbf{y}) = -\sum_{\omega}\sum_{s}\sum_{r}\omega^2\Big(\int_{\Omega}\Re\{\exp(-i\frac{\omega}{c_0}(\hat{\mathbf{s}}+\hat{\mathbf{r}})\cdot\mathbf{z})\}\Theta(\mathbf{y},\mathbf{z};\boldsymbol{\theta})d\mathbf{z}\Big)\Delta D_m(\mathbf{x}_s,\mathbf{x}_r,\omega),$$

where NTK only modulates the spatial Fourier series while preserving the frequency energy distribution $\omega^2$, i.e., high-frequency NSK can generate more low-wavenumber gradients compared to original kernels. Fig. 5 compares the original kernel and NSK at 15 Hz, which empirically validates our theoretical claims. Hence, the novel neural sensitivity kernel can suppress the wrong high-wavenumber components and acheive the adaptive multiscale inversion process from coarse scale to fine scale. These claims are theoretically validated by the following section based on spatial Fourier analysis.

**3.2. Wavenumber Modulation of NSK.** Existing works have shown that NeurFWI preferentially inverts low-wavenumber background [43, 11, 27], whereas FWI tends to recover more high-wavenumber details (see Fig. 5). The NSK can induce the low-wavenumber gradient and achieve adapetive wavenumber modulation during the inversion process, as shown in Fig. 6. To elucidate the underlying theoretical mechanisms of gradient wavenumber modulation, we examine the following dynamic evolution (3.3) in the wavenumber domain via the Fourier transforms.

LEMMA 3.6 (NSK in Wavenumber Domain). *Consider the Fourier transform in equation* (2.1) *and let* $\frac{\partial u(\mathbf{x},t)}{\partial \hat{m}(\boldsymbol{\xi})} = \mathcal{F}_{\mathbf{y}\to\boldsymbol{\xi}}\big[\frac{\partial u(\mathbf{x},t)}{\partial m(\mathbf{y})}\big](\boldsymbol{\xi})$ *be the sensitivity kernel in the spatial wavenumber domain. Define the expression of NTK in the wavenumber domain as:*

$$\hat{\Theta}(\boldsymbol{\xi},\boldsymbol{\xi}') = \mathbb{E}_{\boldsymbol{\theta}}\Big(\mathcal{F}_{\mathbf{y}\to\boldsymbol{\xi}}\Big[\nabla_{\boldsymbol{\theta}}\sigma^*(\mathbf{y},\boldsymbol{\theta})\Big](\boldsymbol{\xi})\cdot\overline{\mathcal{F}_{\mathbf{z}\to\boldsymbol{\xi}}\Big[\nabla_{\boldsymbol{\theta}}\sigma^*(\mathbf{z},\boldsymbol{\theta})\Big](\boldsymbol{\xi}')}\Big), \tag{3.5}$$

*where the expectation* $\mathbb{E}_{\boldsymbol{\theta}}$ *is taken w.r.t. the initial distribution of parameters. Then, the dynamic evolution* (3.3) *in the spatial wavenumber domain has the expression:*

$$\mathcal{F}_{\mathbf{y}\to\boldsymbol{\xi}}\Big[\frac{dm_{\boldsymbol{\theta}}(\mathbf{y})}{d\tau}\Big](\boldsymbol{\xi}) = -\int_{\Omega}\int_{T}\Big(\int_{\mathbb{R}^d}\frac{\partial u(\mathbf{x},t)}{\partial \hat{m}(\boldsymbol{\xi}')}\cdot\hat{\Theta}(\boldsymbol{\xi},\boldsymbol{\xi}')\ d\boldsymbol{\xi}'\Big)\cdot\Delta D_m(\mathbf{x},t)\ d\mathbf{x}dt. \tag{3.6}$$

*Proof.* The proof can be found in the supplementary materials SM1.4. ∎

*Remark* 3.7. When the NTK $\hat{\Theta}(\boldsymbol{\xi},\boldsymbol{\xi}')$ reduces to a Dirac delta function, the dynamic evolution (3.6) degenerates to the evolution of conventional FWI. Thus, one fundamental distinction between FWI and NeurFWI lies in their sensitivity kernels, i.e., NTK modulates the wavenumber of the sensitivity kernel (see Theorem 3.8).

THEOREM 3.8. *Consider the network* (2.6) *and suppose Assumption 2.2 holds. The wavenumber-domain NTK decomposes according to* $\boldsymbol{\theta}=(a,\mathbf{w},b)$. *Then, we have*

$$\mathcal{F}_{\mathbf{y}\to\boldsymbol{\xi}}\Big[\frac{\partial u_{\Theta}}{\partial m(\mathbf{y})}\Big](\boldsymbol{\xi}) = \mathcal{F}_{\mathbf{y}\to\boldsymbol{\xi}}\Big[\frac{\partial u_{\Theta_a}}{\partial m(\mathbf{y})}\Big](\boldsymbol{\xi}) + \mathcal{F}_{\mathbf{y}\to\boldsymbol{\xi}}\Big[\frac{\partial u_{\Theta_{\mathbf{w}}}}{\partial m(\mathbf{y})}\Big](\boldsymbol{\xi}) + \mathcal{F}_{\mathbf{y}\to\boldsymbol{\xi}}\Big[\frac{\partial u_{\Theta_b}}{\partial m(\mathbf{y})}\Big](\boldsymbol{\xi}) \tag{3.7}$$

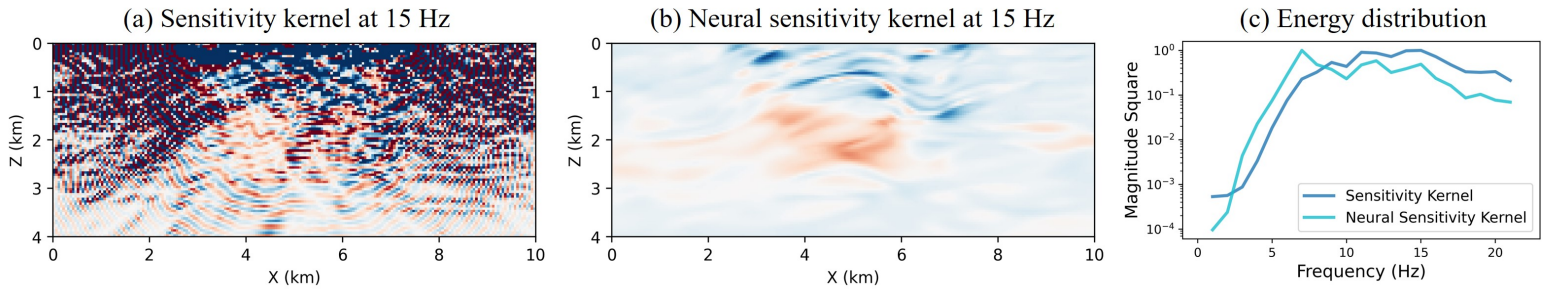

FIG. 5. *(a) The sensitivity kernel at 15 Hz. (b) Neural sensitivity kernel at 15 Hz. (c) Energy distribution of the sensitivity kernel and NSK with different frequencies.*

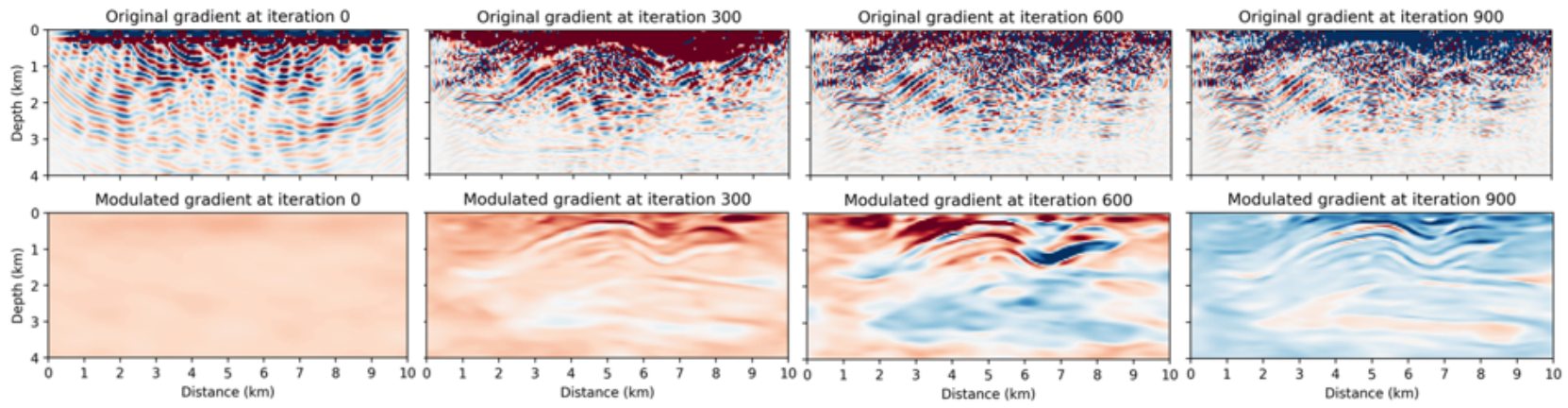

FIG. 6. *The gradient of FWI and modulated gradient of NeurFWI with different iterations.*

*For brevity, define the functional* $\mathcal{W}[g](\boldsymbol{\xi}) = \mathcal{F}[g](\frac{||\boldsymbol{\xi}||}{r}) \cdot \mathcal{F}[g](-\frac{||\boldsymbol{\xi}||}{r})$*, where* $r = ||\mathbf{w}||$ *and* $g$ *is an activation function. Then the wavenumber-domain NSK is given by*

$$\begin{aligned}
&\mathcal{F}_{\mathbf{y}\to\boldsymbol{\xi}}\Big[\frac{\partial u_{\Theta_a}}{\partial m(\mathbf{y})}\Big](\boldsymbol{\xi}) = \frac{C}{||\boldsymbol{\xi}||^{d-1}}\mathbb{E}_r\Big[\frac{1}{r}\mathcal{W}[\sigma](\boldsymbol{\xi})\Big]\mathcal{F}_{\mathbf{y}\to\boldsymbol{\xi}}\Big[\frac{\partial u}{\partial m(\mathbf{y})}\Big](\boldsymbol{\xi}),\\
&\mathcal{F}_{\mathbf{y}\to\boldsymbol{\xi}}\Big[\frac{\partial u_{\Theta_\mathbf{w}}}{\partial m(\mathbf{y})}\Big](\boldsymbol{\xi}) = \frac{C}{||\boldsymbol{\xi}||^{d-1}}\mathbb{E}_{a,r}\Big[\frac{1}{r}\mathcal{W}[a\sigma'](\boldsymbol{\xi})\Big]\mathcal{F}_{\mathbf{y}\to\boldsymbol{\xi}}\Big[\frac{\partial u}{\partial m(\mathbf{y})}\Big](\boldsymbol{\xi}),\\
&\mathcal{F}_{\mathbf{y}\to\boldsymbol{\xi}}\Big[\frac{\partial u_{\Theta_b}}{\partial m(\mathbf{y})}\Big](\boldsymbol{\xi}) = \frac{C}{4\pi^2||\boldsymbol{\xi}||^{d-1}}\nabla\cdot\Big(\mathbb{E}_{a,r}\Big[\frac{1}{r}\mathcal{W}[a\sigma'](\boldsymbol{\xi})\Big]\nabla\mathcal{F}_{\mathbf{y}\to\boldsymbol{\xi}}\Big[\frac{\partial u}{\partial m(\mathbf{y})}\Big](\boldsymbol{\xi})\Big),
\end{aligned} \tag{3.8}$$

*where the expectations are taken with respect to initial parameter distribution,* $C = p_b(0)\Gamma(d/2)/(2\pi^{d/2})$*,* $p_b(0)$ *is the density of* $b$*,* $\Gamma$ *denotes the Gamma distribution.*

*Proof.* The proof can be found in the supplementary materials SM1.5. This proof relies on the network's Fourier transform, the Laplace method, and the integral under radial symmetry, where the Laplace method can be used to approximate the expectation of bias $b$ and the radial symmetry can solve the expectation of weight $\mathbf{w}$. □

*Remark* 3.9. This theorem shows that the NSK is essentially the original kernel spectrally modulated by the NTK, with contributions derived from parameters $a$, $\mathbf{w}$, and $b$. All three components share a common decay factor $\mathcal{O}(\|\boldsymbol{\xi}\|^{-d+1})$, which acts as a low-pass filter by attenuating high-wavenumber modes. Specifically, the $a$- and $\mathbf{w}$-components weight the sensitivity kernel by the Fourier spectra of $\sigma$ and $\sigma'$, respectively, thereby encoding the activation function's smoothness and regularity. The bias $b$ introduces a divergence structure $\nabla_{\boldsymbol{\xi}} \cdot (\cdot)$, coupling spatial derivative information. This underlying mechanism of NSK can generate more low-wavenumber gradient by suppressing the sensitivity of high-wavenumber components, which can effectively mitigate cycle-skipping and reduce dependence on initial models.

Next, we will derive the expression of NSK for INR-based NeurFWI methods with different activation functions [71, 40, 64] (e.g., tanh, sinusoidal, and Gabor wavelet).

**3.2.1. NSK for NNFWI.** Zhu et al. [71] introduced NNFWI, which employs INR with a tanh activation function to parameterize the velocity models in FWI, which usually invert relative smooth velocity models. The following corollary gives the corresponding NSK expression for NNFWI to unveil the modulation mechanism.

COROLLARY 3.10 (NSK for NNFWI). *Suppose that Assumption 2.2 holds. If* $\sigma_b \gg 1$ *and* $\sigma(z) = \tanh(z)$*, then the NSK of NNFWI can be expressed as follows:*

$$\mathcal{F}[\frac{\partial u_\Theta}{\partial m}](\boldsymbol{\xi}) \simeq C \cdot \mathbb{E}_{a,r}\Big[\frac{\pi^2 r^2 + 4\pi^4 a^2 \|\boldsymbol{\xi}\|^2}{r^3 \|\boldsymbol{\xi}\|^{d-1}} \operatorname{csch}^2\Big(\frac{\pi^2\|\boldsymbol{\xi}\|}{r}\Big)\Big]\mathcal{F}[\frac{\partial u}{\partial m}](\boldsymbol{\xi}), \tag{3.9}$$

*where the expectations are taken w.r.t initial parameter distribution.*

*Proof.* The proof is provided in the supplementary materials SM1.6. ∎

*Remark* 3.11. This corollary reveals that the NNFWI sensitivity kernel acts as an intrinsic spectral filter. Although the numerator contains a wavenumber gain term $\|\boldsymbol{\xi}\|^2$, the $\operatorname{csch}^2(\pi^2\|\boldsymbol{\xi}\|/r)$ factor enforces exponential attenuation that overwhelmingly dominates the spectral response. Specifically, smaller $r$ values intensify this exponential suppression, severely limiting high-wavenumber updates and promoting smooth inversion. Conversely, larger $r$ values moderate the attenuation strength. Fig. 7 validates this sensitivity spectral filtering and wavenumber modulation behavior, which theoretically explains the inversion performance of NNFWI in existing works [71, 40].

**3.2.2. NSK for IFWI.** To obtain more high-resolution inversion results using NeurFWI, Sun et al. [40] introduced implicit FWI, which employs INR with a sinusoidal activation function [38] to parameterize the parameter model in FWI. The following corollary gives the corresponding NSK expression for IFWI[6].

Corollary 3.12 (NSK for IFWI). *Suppose that Assumption 2.2 holds. If $\sigma_b \gg 1$ and $\sigma(x) = \sin(\omega_0 x) \cdot \mathcal{X}_{[-T/2,T/2]}(x)$, then the NSK of IFWI is given by:*

$$\mathcal{F}[\frac{\partial u_\Theta}{\partial m}](\boldsymbol{\xi}) \simeq C \cdot \mathbb{E}_{a,r}\Big[\frac{T^2 + a^2\omega_0^2 T^2}{4\|\boldsymbol{\xi}\|^{d-1} r} sinc^2\Big(T\Big(\frac{\|\boldsymbol{\xi}\|}{r} - \frac{\omega_0}{2\pi}\Big)\Big)\Big]\mathcal{F}\Big[\frac{\partial u}{\partial m}\Big](\boldsymbol{\xi}), \tag{3.10}$$

*where the expectations are taken w.r.t initial parameter distribution.*

*Proof.* The proof is provided in the supplementary materials SM1.7. ∎

*Remark* 3.13. This NSK acts as a frequency-selective band-pass filter modulated by a polynomial envelope $\mathcal{O}(\|\boldsymbol{\xi}\|^{1-d})$. The $\operatorname{sinc}^2(\cdot)$ term concentrates gradient updates within a narrow spectral band centered at $\|\boldsymbol{\xi}\| = \omega_0 r/(2\pi)$. The intrinsic frequency $\omega_0$ directly governs the peak location, i.e., increasing $\omega_0$ shifts toward higher wavenumbers (high-resolution but unstable), whereas decreasing $\omega_0$ enforces a stronger low-wavenumber bias (stable but low-resolution). Furthermore, the weight norm $r$ serves as a neuron-specific scaling factor. A larger variance of weights accelerates the high-wavenumber convergence. Fig. 7 numerically validates the above claims and the inversion results of IFWI using different activation hyperparameters and neural network's weight variance are shown in supplementary materials SM3.4.

**3.2.3. NSK for WinFWI.** Yang et al. [64] introduced WinFWI, which employs INR with a Gabor wavelet activation [37] to parameterize the velocity models, which can also further mitigate the spectral bias in the IFWI method. The following corollary gives the corresponding NSK for WinFWI to explain its mechanism.

Corollary 3.14 (NSK for WinFWI). *Suppose that Assumption 2.2 holds. If $\sigma_b \gg 1$ and $\sigma(x) = \exp(-s_0^2 x^2)\cos(\omega_0 x)$, then the NSK of WinFWI is given by:*

$$\mathcal{F}[\frac{\partial u_\Theta}{\partial m}](\boldsymbol{\xi}) \simeq C \cdot \mathbb{E}_{a,\|\mathbf{w}\|}\Big[\frac{\pi r^3 + 4\pi^3 a^2 \|\boldsymbol{\xi}\|^2}{4 s_0^2 r^3 \|\boldsymbol{\xi}\|^{d-1}} \exp\Big(-\frac{(2\pi\|\boldsymbol{\xi}\|/r - \omega_0)^2}{2 s_0^2}\Big)\Big]\mathcal{F}\Big[\frac{\partial u}{\partial m}\Big](\boldsymbol{\xi}), \tag{3.11}$$

*where the expectations are taken w.r.t initial parameter distribution.*

[6]We consider the truncated sinusoidal activation here, since the product of delta functions is not well-defined when computing the NSK with the standard sinusoidal activation. In practice, because the parameters and outputs are bounded, this is equivalent to using a truncated sinusoidal function.

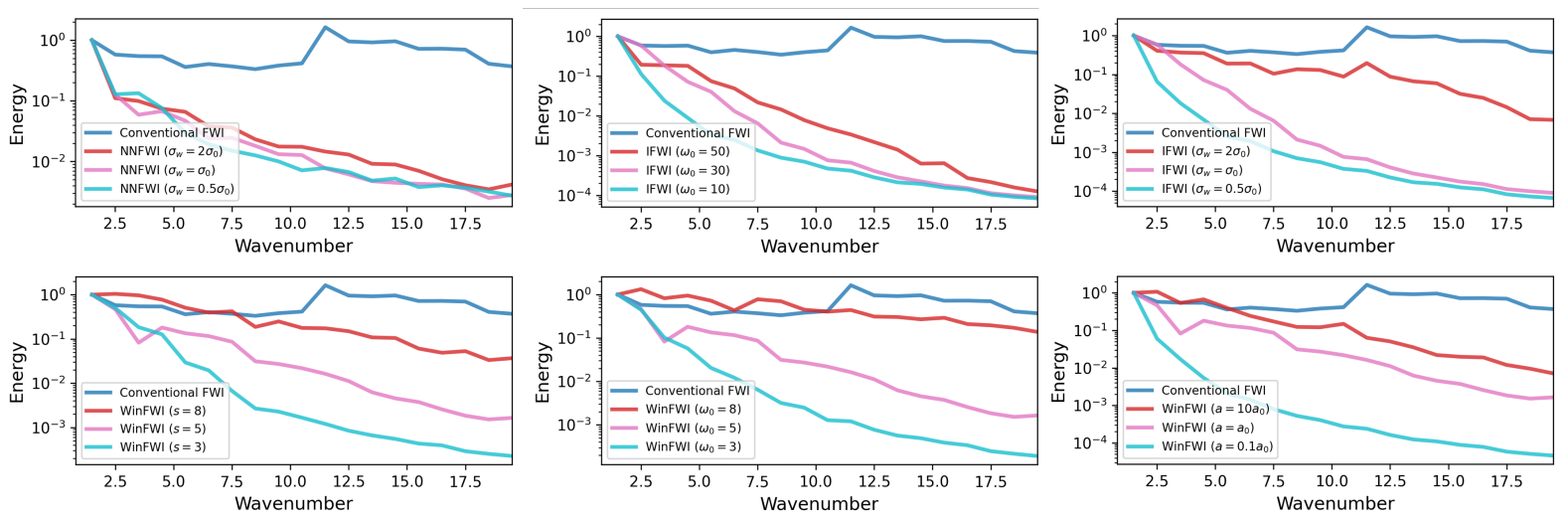

FIG. 7. *Gradient wavenumber distribution using FWI and different NeurFWI at the first epoch, where $a_0$ denotes a default output layer scaling and $\sigma_0$ is a default initial weight variance.*

*Proof.* The proof is provided in the supplementary materials SM1.8. ∎

*Remark* 3.15. The Gabor wavelet activation introduces three components, including a wavenumber attenuation term $\mathcal{O}(||\boldsymbol{\xi}||^{1-d})$, a wavenumber band-pass filtering term $\exp\big(-(2\pi||\boldsymbol{\xi}||-\omega_0 r)^2/(2s_0^2r^2)\big)$ centered at $\omega_0$ with bandwidth $s_0$, and a wavenumber enhancement term $\mathcal{O}(||\boldsymbol{\xi}||^{3-d})$. This modulation is controlled by bandwidth $s_0$ and frequency $\omega_0$. Specifically, larger $s_0$ (i.e., a wider frequency band range) and $\omega_0$ (i.e., the modulation frequency is concentrated at higher-frequency components) can enhance the high-wavenumber convergence. Moreover, the output layer scaling $a$ controls the amplitude of $\mathcal{O}(||\boldsymbol{\xi}||^{3-d})$, i.e., the larger $a$ leads to a faster convergence rate of high-wavenumber components. Fig. 7 illustrates the effect of these hyperparameters, and the inversion results of WinFWI using different activation hyperparameters and output scaling value are shown in supplementary materials SM3.4.

Overall, these theoretical analysis of NeurFWI within the model domain can unveil the sensitivity kernel and gradient modulation mechanism, which can be utilized to explain existing inversion performance using different NeurFWI methods. This NSK theory has the potential to further provide the methodological design framework.

## 4. Wave Tangent Kernel Analysis.

While sensitivity kernel analysis provides insights into model evolution, theoretical frameworks for investigating the convergence behavior of data misfit remain absent. In this section, we establish the wave tangent kernel (i.e., WTK) for FWI, enabling the analysis of data-fitting behavior.

### 4.1. WTK for FWI.

Consider the iterative scheme (2.3) and the objective function (2.2). Applying the chain rule, we prove the following proposition, which characterizes the evolution of the wavefield residuals during the inversion process.

PROPOSITION 4.1 (WTK). *Consider the wavefield data misfit (2.2) and the dynamic gradient flow (1.3), the synthetic data $u(\mathbf{x},t)$ evolves as follows:*

$$\frac{\partial u(\mathbf{x},t)}{\partial \tau} = -\int_T\int_\Omega\left(\int_\Omega \frac{\partial u(\mathbf{x},t)}{\partial m(\mathbf{y})}\frac{\partial u(\mathbf{x}',t')}{\partial m(\mathbf{y})}d\mathbf{y}\right)\Delta D_m(\mathbf{x}',t')\ d\mathbf{x}'dt', \tag{4.1}$$

*where we define $\mathbf{K}\big((\mathbf{x},t),(\mathbf{x}',t')\big) = \int_\Omega \frac{\partial u(\mathbf{x},t)}{\partial m(\mathbf{y})}\frac{\partial u(\mathbf{x}',t')}{\partial m(\mathbf{y})}d\mathbf{y}$ as WTK. Let $\mathcal{K}$ denote the WTK integral operator, defined by its action on any function $f \in L^2(\Omega\times T)$:*

$$\mathcal{K}[f](\mathbf{x},t) = \int_\Omega\int_T \mathbf{K}\big((\mathbf{x},t),(\mathbf{x}',t')\big)f(\mathbf{x}',t')d\mathbf{x}'dt'. \tag{4.2}$$

*With this operator, the evolution can be written as $\partial_\tau u(\mathbf{x},t) = -\mathcal{K}[\Delta D_m](\mathbf{x},t)$.*

*Proof.* The proof is provided in the supplementary materials SM2.1. ∎

13 The data fitting is driven by the wavefield residual $\Delta D_m$ (**data-driven**), weighted by WTK $\mathbf{K}$ (**physical guide**). Here, the WTK quantifies the coupling and correlation relationship between different wavefield data points $(\mathbf{x}, t)$ and $(\mathbf{x}', t')$ through point-wise model perturbations. To elucidate the mechanism of cycle-skipping, we examine the evolution of synthetic data in the frequency domain. Let $\Delta\hat{D}_m(\omega) = \mathcal{F}\big[\Delta D_m\big](\omega)$ be the frequency wavefield residual. Then, the evolution (4.1) is given by

$$\begin{aligned}\mathcal{F}_{t\to\omega}[\frac{\partial u(x,t)}{\partial \tau}](\mathbf{x},\omega) &= -\int_T\int_\Omega\Big(\int_\Omega \frac{\partial \hat{u}(\mathbf{x},\omega)}{\partial m(\mathbf{y})}\overline{\frac{\partial \hat{u}(\mathbf{x}',\omega')}{\partial m(\mathbf{y})}}\, d\mathbf{y}\Big)\Delta\hat{D}_m(\mathbf{x}',t')\, d\mathbf{x}'dt', \\ &= -\int_{\mathbb{R}}\int_\Omega \hat{\mathbf{K}}\big((\mathbf{x},\omega),(\mathbf{x},\omega')\big)\Delta\hat{D}_m(\mathbf{x}',\omega')\, d\mathbf{x}'d\omega',\end{aligned} \tag{4.3}$$

where $\partial\hat{u}/\partial m$ is the frequency sensitivity kernel and $\hat{\mathbf{K}}$ denotes the frequency WTK. Consider a homogeneous background medium, and the incident wavefield is a monochromatic plane wave propagating in direction $\hat{\mathbf{s}}$, and the scattered wavefield is observed in the far-field as a plane wave propagating in direction $\hat{\mathbf{r}}$ or $\hat{\mathbf{r}}'$. These assumptions enable explicit Green's functions and a scattering formulation. Especially, we have:

$$\frac{\partial \hat{u}(\mathbf{x}_r,\omega)}{\partial m(\mathbf{y})} \simeq \sum_s \omega^2 \hat{G}_0(\mathbf{x}_r,\mathbf{y};\omega)\hat{G}_0(\mathbf{y},\mathbf{x}_s;\omega) \simeq \sum_s \omega^2 \exp(iw\hat{\mathbf{r}}\cdot\mathbf{y})\exp(iw\hat{\mathbf{s}}\cdot\mathbf{y}), \tag{4.4}$$

where $\hat{G}_0(\mathbf{x}_r,\mathbf{y};\omega)$ is the Green's function from scattered point $\mathbf{y}$ to receiver $\mathbf{x}_r$, and $\hat{G}_0(\mathbf{y},\mathbf{x}_s;\omega)$ is the Green's function from source $\mathbf{x}_s$ to scattered point $\mathbf{y}$. Substituting the above sensitivity kernel expression into the frequency WTK (4.3), we have:

$$\hat{\mathbf{K}}\big((\mathbf{x}_r,\omega),(\mathbf{x}_r',\omega')\big) = \sum_s \omega^2\omega'^2\int_\Omega \exp\big[i\big(\omega(\hat{\mathbf{s}}+\hat{\mathbf{r}}) - \omega'(\hat{\mathbf{s}}+\hat{\mathbf{r}}')\big)\cdot\mathbf{y}\big]\, d\mathbf{y}, \tag{4.5}$$

where we define $\mathbf{k} = \omega(\hat{\mathbf{s}}+\hat{\mathbf{r}}) - \omega'(\hat{\mathbf{s}}+\hat{\mathbf{r}}')$ as a wave vector difference.

*Remark* 4.2. The WTK $\hat{\mathbf{K}}$ quantifies sensitivity correlations in wavefield data across space and frequency. The correlation between $u(\mathbf{x}_r,\omega)$ and $u(\mathbf{x}_r',\omega')$ depends on the frequencies $\omega,\omega'$ and the wave vector difference $\mathbf{k}$, which reflects the mismatch in scattering paths. This is described by the oscillatory integral $\int_\Omega e^{i\mathbf{k}\cdot\mathbf{y}}d\mathbf{y}$, whose magnitude decreases as $|\mathbf{k}|$ increases. For equal frequencies, $\mathbf{k} = \omega(\hat{\mathbf{r}} - \hat{\mathbf{r}}')$, so larger receiver angular separation increases $|\mathbf{k}|$ and reduces correlation. For identical receivers, $\mathbf{k} = (\omega-\omega')(\hat{\mathbf{s}}+\hat{\mathbf{r}})$, so a greater frequency difference also increases $|\mathbf{k}|$ and decreases correlation. Thus, larger receiver separations or frequency differences generally reduce data correlation during fitting, which is consistent with physical principles.

*Remark* 4.3. The frequency factors $\omega^2$ and $\omega'^2$ arise from the second-order time derivative after Fourier transformation [45], accelerating high-frequency wavefield data evolution. When $\mathbf{k} = 0$ (identical frequency and receiver direction), the correlation is strong for high frequencies and weak for low frequencies due to the amplitude $\omega^4$, i.e., strong self-coupling. Although a high $\omega'$ can increase the frequency factor when $\omega \neq \omega'$, the effect remains marginal because of the scattering-path mismatch ($|\mathbf{k}| \propto |\omega-\omega'|$), i.e., weak cross-coupling (Fig. 8 (c)). Consequently, the wavefield data favors point-by-point fitting due to the main diagonal of WTK dominates (i.e., strong self-coupling), which raises the risk of converging to a local minimum, thereby prioritizing local details over global fitting, and thus preferring high-frequency components. This is validated by MSE evolution of low- and high-frequency wavefield data residuals on the 2D Overthrust model with a constant initial model (Fig. 8 (a)).

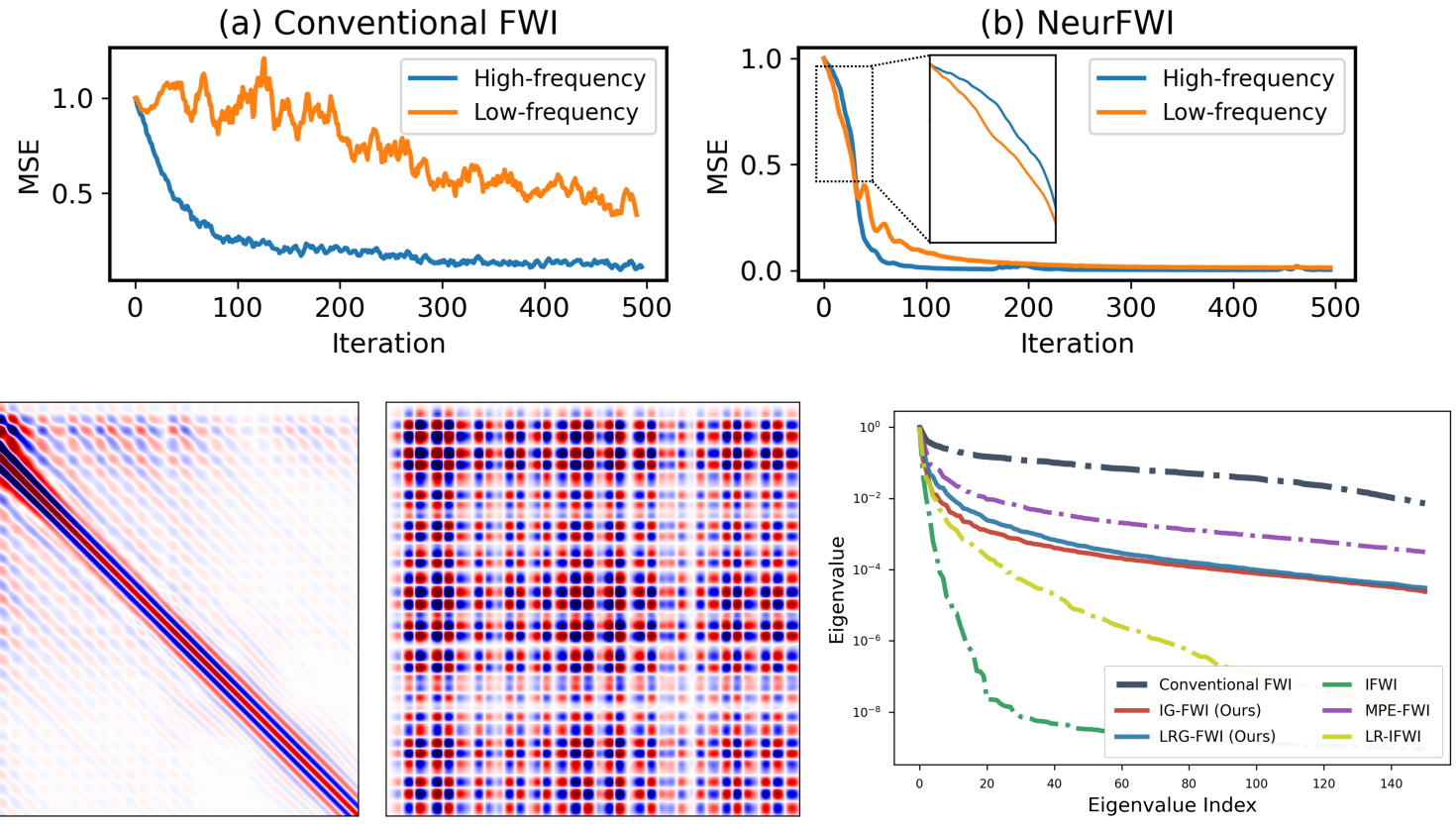

(c) WTK of FWI (d) WTK of NeurFWI (e) Eigenvalue decay rate

FIG. 8. *Data misfit convergence behavior using (a) conventional FWI and (b) NeurFWI methods. WTK for (c) FWI and (d) NeurFWI using the Overthrust model (one source at 2.5 km and receiver at 7.5 km). (e) Eigenvalue decay rate using FWI and different NeurFWI methods.*

**4.2. WTK for NeurFWI.** Next, we investigate the mechanism of NeurFWI from the perspective of WTK, i.e., how the NTK modulates the original WTK and changes the data fitting behavior. Consider the dynamic gradient flow (3.3). Applying the chain rule, we prove the wavefield evolution for NeurFWI as follows.

PROPOSITION 4.4 (WNTK for NeurFWI). *Suppose the Assumption 2.2 holds. Then the synthetic wavefield data $u$ evolves according to the following equation:*

$$\frac{\partial u(\mathbf{x},t)}{\partial \tau} = -\int_T \int_\Omega \Big( \int_\Omega \int_\Omega \frac{\partial u(\mathbf{x},t)}{\partial m(\mathbf{y})} \frac{\partial u(\mathbf{x}',t')}{\partial m(\mathbf{z})} \Theta(\mathbf{y},\mathbf{z};\boldsymbol{\theta})\ d\mathbf{z}d\mathbf{y} \Big) \Delta D_m(\mathbf{x}',t')\ d\mathbf{x}'dt'. \tag{4.6}$$

*where we define $\mathbf{K}_N\big((\mathbf{x},t),(\mathbf{x}',t')\big) = \int_\Omega \int_\Omega \frac{\partial u(\mathbf{x},t)}{\partial m(\mathbf{y})} \frac{\partial u(\mathbf{x}',t')}{\partial m(\mathbf{z})} \Theta(\mathbf{y},\mathbf{z};\boldsymbol{\theta})\ d\mathbf{z}d\mathbf{y}$ as the neural wave tangent kernel (termed WNTK). Let $\mathcal{K}_N$ denote the integral operator associated with the WNTK, defined by its action on a function $f \in L^2(\Omega \times T)$, i.e.,*

$$\mathcal{K}_N[f](\mathbf{x},t) = \int_\Omega \int_T \mathbf{K}_N\big((\mathbf{x},t),(\mathbf{x}',t')\big) f(\mathbf{x}',t') d\mathbf{x}'dt'. \tag{4.7}$$

*With this operator, the evolution can be written as $\partial_\tau u(\mathbf{x},t) = -\mathcal{K}_N[\Delta D_m](\mathbf{x},t)$.*

*Proof.* The proof is consistent with the Proposition 4.1. ∎

*Remark* 4.5. The WNTK $\mathbf{K}_N$ degrades into the WTK $\mathbf{K}$ without neural representation (the proof is provided in SM2.1). The WNTK exhibits strong off-diagonal dominance due to spatial correlations induced by NTK. Hence, WNTK incorporates a joint guidance derived from the product of sensitivity kernels across different velocity model points, rather than relying on point-wise computations in the original WTK, thereby enabling global correlations. The WTK and WNTK are shown in Fig. 8 (c) and (d), respectively, using one source at 2.5 km and receiver at 7.5 km on the 2D modified Overthrust model. This result shows that NeurFWI can mitigate the self-coupling term and enhance the cross-coupling item, which can focus on global wavefield fitting and may suppress the high-frequency fitting. The following theorem can unveil the frequency modulation mechanism of NeurFWI.

THEOREM 4.6 (Wave Frequency Bias). *Consider the frequency dynamic evolution (4.3). Let $\mathbf{n} = (\hat{\mathbf{s}} + \hat{\mathbf{x}})$ and $\mathbf{n}' = (\hat{\mathbf{s}} + \hat{\mathbf{x}}')$ be the fixed direction vector. Assume that*

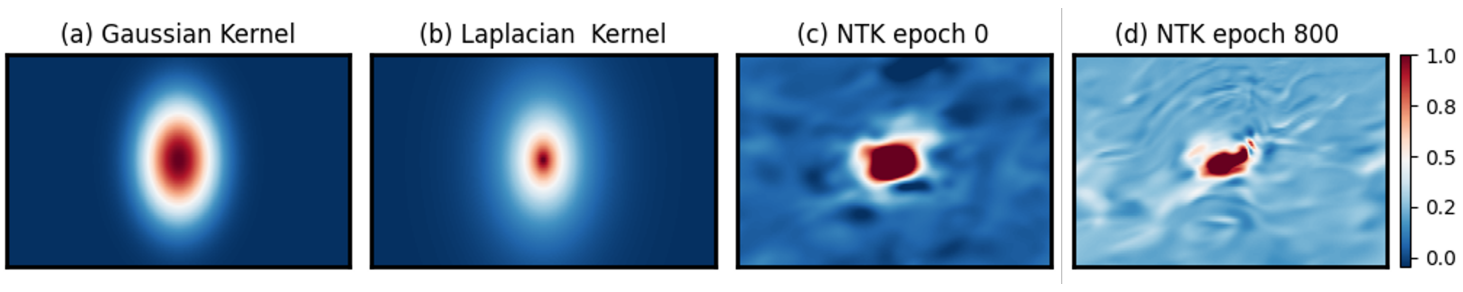

FIG. 9. *(a) Gaussian kernel. (b) Laplacian kernel. (c) Initialized NTK. (d) NTK at 800 th epoch. Here, the NTK captures center-surround correlation on the Overthrust model.*

*the NTK is a translation invariant kernel*[7]*, the WNTK can be expressed as follows:*

$$\hat{\mathbf{K}}_N\big((\mathbf{x},\omega),(\mathbf{x}',\omega')\big) = \hat{\mathbf{K}}\big((\mathbf{x},\omega),(\mathbf{x}',\omega')\big)\cdot\hat{\kappa}\big(\omega'\mathbf{n}'\big), \tag{4.8}$$

*where we define* $\hat{\kappa}(\omega'\mathbf{n}') = \big(\int_\Omega \exp(i\omega'\mathbf{n}'\cdot\mathbf{u})\kappa(\mathbf{u})d\mathbf{u}\big)$ *as frequency NTK.*

*Proof.* The proof is provided in the supplementary materials SM2.2. ∎

*Remark* 4.7. The above theorem demonstrates that the NTK modulates the original WTK by controlling the frequency factor, i.e., the $\hat{\kappa}$ acts as an adaptive low-pass filter to attenuate the high-frequency correlation in WTK, thereby prioritizing low-frequency data fitting. If the kernel is a Gaussian kernel $\kappa(\mathbf{u}) = \exp\big(-||\mathbf{u}||^2/(2\sigma^2)\big)$, it induces the following exponential low-pass frequency filtering effect, i.e.,

$$\hat{\kappa}(\omega'\mathbf{n}') = (2\pi\sigma^2)^{d/2}\exp\big(-\sigma^2\|\omega'\mathbf{n}'\|^2/2\big), \tag{4.9}$$

where $d$ is the dimension of NTK's input. Moreover, if the kernel is a Laplacian kernel $\kappa(\mathbf{u}) = \exp(-||\mathbf{u}||/\sigma)$, it induceds the following polynomial low-pass filtering effect

$$\hat{\kappa}(\omega'\mathbf{n}') = \frac{C_d}{\big(1+\sigma^2\|\omega'\mathbf{n}'\|^2\big)^{(d+1)/2}}, \tag{4.10}$$

where $C_d$ is a constant depending on dimension $d$ and scaling $\sigma$. When considering the NTK, the filtering effect depends on the smoothness of activation functions. The initialized NTK shares the similar function properties with Gaussian and Laplacian kernels, as shown in Fig. 9. The main difference is the learnable mechanism of NTK, which can capture the structural information. Fig. 8 (b) validates our theoretical claims (i.e., NeurFWI can accelerate the reduction of low-frequency data residuals).

**4.3. Spectral Analysis for WTK.** To quantitatively characterize the data fitting behavior, we analyze the WTK via spectral decomposition, following a similar approach to the spectral analysis of the NSK. Within a sufficiently small training window, the model changes slightly, keeping the WTK nearly constant [10], and its eigenvalue spectrum thus provides a quantitative estimate of local convergence rates.

PROPOSITION 4.8 (Spectral Decomposition of WTK). *The WTK and WNTK integral defined in equations (4.2) and (4.7) operators are self-adjoint, non-negative definite, and compact operators. Then, for any square-integrable function $f$, these integral operators have the spectral decomposition $\mathcal{K}[f] = \sum_{k=1}^{\infty}\lambda_k\langle\phi_k, f\rangle\otimes\phi_k$, where $\{\phi_k\}_{k=1}^{\infty}$ are eigenfunctions and $\{\lambda_k\}_{k=1}^{\infty}$ represent non-negative eigenvalues.*

*Proof.* The proof is provided in the supplementary materials SM2.3. ∎

The subsequent substitution of this decomposition into equation (4.6) leads to the following data misfit evolution equation along different spectral directions.

[7]This translation invariance of neural tangent kernel can be achieved by encoding the coordinate input, e.g., using random Fourier features or positional encoding [44], which is a common technique to make NTKs translation-invariant.

THEOREM 4.9 (Spectral Data Convergence). *Consider the wavefield evolution equation (4.6) and spectral decomposition of WTK in Proposition 4.8. Then the dynamic evolution of the wavefield data residual in the spectral domain satisfies:*

$$\langle \Delta D_m^\tau, \phi_m \rangle = -\exp(-\lambda_m \tau)\langle \Delta D_m^0, \phi_m \rangle,\ \forall\ m \in [0, 1, \cdots], \tag{4.11}$$

*where $\phi_m$ is the spectral eigenfunction, $\lambda_m \geq 0$ is the corresponding eigenvalue, $\Delta D_m^\tau$ denotes the data residual at iterative time $\tau$, and $\Delta D_m^0$ denotes the initial residual.*

*Proof.* The proof is provided in the supplementary materials SM2.4. ∎

The eigenvalue spectrum of the WTK governs the convergence rate of wavefield data across distinct spectral components, i.e., larger eigenvalues accelerate error reduction along their corresponding eigen-directions, whereas smaller eigenvalues impede corresponding spectral convergence. Consequently, the overall convergence behavior is dictated by the spectral decay rate, i.e., a flatter spectrum enables faster global convergence. To elucidate the performance disparities between FWI and NeurFWI methods from the spectral convergence perspective, the following theorem establishes a unified theoretical framework for comparing their WTK eigenvalue decay rates.

THEOREM 4.10 (Spectral Decay Comparison). *Consider any two NTK $\Theta_1, \Theta_2 : \Omega \times \Omega \to \mathbb{R}$, and the WTK integral operator $\mathcal{K}1$ and $\mathcal{K}_2$ in equation (4.7) with NTK $\Theta_1$ and $\Theta_2$. If the NTK satisfies the dominance condition $\Theta_1 \succeq \Theta_2$ (i.e., $\Theta_1 - \Theta_2$ is a positive semi-definite kernel), then $\mathcal{K}_1 \succeq \mathcal{K}_2$, i.e., the eigenvalues of the corresponding integral operators of WTK are similarly ordered $\lambda_j(\mathcal{K}_1) \geq \lambda_j(\mathcal{K}_2)$ for all $j = 1, 2, \ldots,$ where $\{\lambda_j(\mathcal{K}_i)\}_{j=1}^\infty$ denote the eigenvalues of $\mathcal{K}_i$ arranged in non-increasing order.*

*Proof.* The proof is provided in the supplementary materials SM2.5. ∎

*Remark* 4.11. This theorem enables a practical spectral comparison framework for WTK and WNTK with different neural architectures, i.e, we can compare the eigenvalue decay of different NTK instead of directly analyzing WTK decay rates. A slower eigenvalue decay implies faster convergence while losing inversion robustness (i.e., premature fitting of high-frequency data). A faster eigenvalue decay can enhance the robustness, while slowing down the convergence of high-frequency components.

Next, we demonstrate that NeurFWI exhibits a faster eigenvalue decay rate compared to FWI, which prioritizes low-frequency data fitting to enhance the robustness.

THEOREM 4.12. *Consider the WTK $\mathbf{K}$ in (4.1) and the normalized WNTK $\mathbf{K}_N$ in (4.6) with eigenvalues $\{\lambda_j\}_{j=1}^\infty$ and $\{\mu_j\}_{j=1}^\infty$ (arranged in non-increasing order), respectively. Then, the eigenvalue of WTK satisfies $\lambda_j \leq \mu_j$ for all $j \in \mathbb{N}$.*

*Proof.* Consider the Dirac delta kernel $\delta(\mathbf{y} - \mathbf{z})$ for FWI and the normalized NTK $\widetilde{\Theta}(\mathbf{y}, \mathbf{z}; \boldsymbol{\theta})$ for NeurFWI. The eigenvalue of the Dirac kernel is equal to 1, and the maximal eigenvalue of NTK is 1 due to the normalization. By construction, the NTK satisfies $\delta \preceq \widetilde{\Theta}$. Hence, applying Theorem 4.10 yields $\lambda_j \leq \mu_j$ for all $j \in \mathbb{N}$. ∎

*Remark* 4.13. The above theorem is validated in Fig. 8 (e). The rapid decay of eigenvalues in WNTK implies that the low-frequency components (corresponding to larger eigenvalues) are optimized rapidly, thereby alleviating the risk of cycle-skipping and decreasing reliance on an accurate initial model (see the Fig. 3). In contrast, high-frequency components, associated with the sharply decaying tail of the eigenvalue spectrum, converge more slowly due to their small eigenvalues. By connecting the eigenvalue decay behavior of WNTK to optimization behavior, it becomes essential to develop novel NeurFWI methods with deliberately tailored eigenvalue decay.

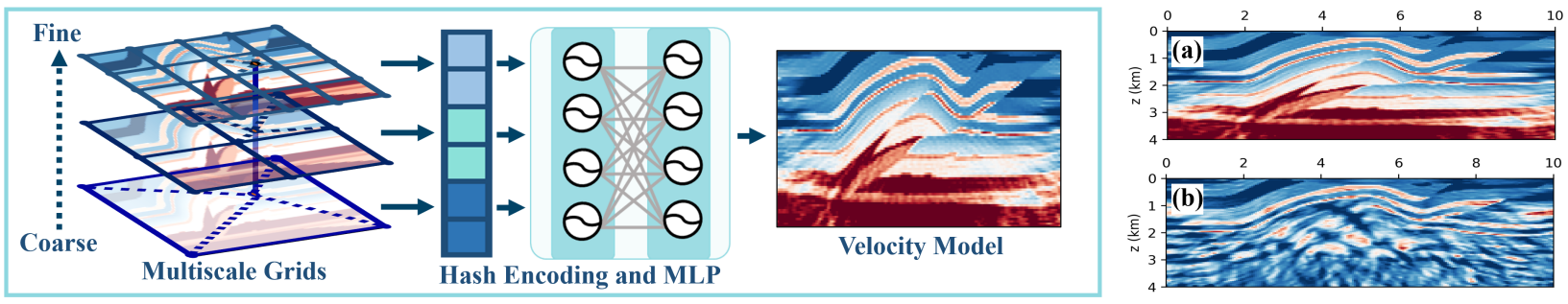

Fig. 10. *The velocity model generation process of the proposed MPE-FWI and its inversion results using (a) the smooth initial model and (b) the constant initial model.*

**5. Hybrid Neural Representation.** In this section, we propose a novel multi-grid parametric encoding [34] based NeurFWI method (i.e., MPE-FWI), which can improve the high-frequency convergence rate by reducing the attenuation rate of eigenvalues. To further achieve a more tailored eigen-structure for FWI, we establish a hybrid neural representation paradigm to design a suitable eigen-structure, and further propose two novel neural representations for FWI, i.e., hybrid INR [38, 40] and multi-grid representation [34] (termed IG-FWI) and hybrid low-rank tensor decomposition [11, 32] and multi-grid representation [34] (termed LRG-FWI). These novel methods can achieve a trade-off between robustness and convergence.

**5.1. Multi-Grid Parametric NeurFWI.** Multi-grid parametric encoding employs trainable auxiliary data structures to construct higher-dimensional embedding spaces, which can be used to represent the velocity model, as shown in Fig. 10. Here, the hash encoding function with learnable parameters maps a coordinate point to a feature vector via multi-resolution grid interpolation and concatenation, where $n_g$ denotes the number of multi-grid levels, and $n_f$ is the number of features per grid level. For each query point $\mathbf{x}$, the feature is obtained by interpolating the embeddings from adjacent grid vertices across multiple resolution levels [5, 33]. Then, these interpolated features are passed through a lightweight INR to produce the velocity value. The overall neural network representation can be expressed as follows:

$$\mathbf{F}_{\boldsymbol{\theta}}(\mathbf{x}) = \mathrm{MLP}\big(h_{\boldsymbol{\theta}_1}(\mathbf{x}); \boldsymbol{\theta}_0\big), \text{ where } h_{\boldsymbol{\theta}_1}(\mathbf{x}) = \bigoplus_{l=1}^{n_g} E_l(\mathbf{x}, T_l; \boldsymbol{\theta}_1) \oplus \mathbf{x}, \tag{5.1}$$

where $\oplus$ denotes the vector concatenation operator, and $E_l(\cdot, T_l)$ is a feature query function that retrieves via bilinear interpolation, i.e., an encoding vector from the learnable feature grid $T_l$ at level $l$. By integrating the representational flexibility of INR with the efficient spatial lookup of multi-grid hash encoding, the MPE approach leverages a grid structure to accelerate convergence and mitigate spectral bias [5].

Theorem 5.1. *Consider the $\mathbf{K}_n^{inr}$ in (4.6) for IFWI and the $\mathbf{K}_n^{mpe}$ for MPE-FWI with eigenvalues $\{\lambda_j\}_{j=1}^{\infty}$ and $\{\mu_j\}_{j=1}^{\infty}$ (arranged in non-increasing order), respectively. Then, the eigenvalue of the WTK satisfies $\lambda_j \leq \mu_j$ for all $j \in \mathbb{N}$.*

*Proof.* The proof can be found in the supplementary materials SM2.6. ∎

*Remark* 5.2. This theorem ensures a significant elevation of the entire eigenvalue spectrum (Fig. 8 (e)), thereby accelerating high-frequency convergence (Fig. 10 (a)). However, the eigenvalue decay rate is inadequate to address the ill-posedness when using an inaccurate initial model (Fig. 10 (b)). Therefore, a novel neural representation tailored for FWI needs to be designed, one that incorporates a suitable decay rate to achieve a more balanced trade-off between convergence and robustness.

**5.2. Hybrid Representation Methodologies.** Existing NeurFWI methods differ in WTK eigenvalue decay rates. In practice, the choice of neural networks should be adapted to the initial model and seismic data. For instance, when an accurate initial model is available, MPE-FWI can accelerate the high-wavenumber convergence. In contrast, IFWI [40] and LR-FWI [11] can stably invert satisfactory results even using inaccurate initial models. Hence, we propose a hybrid neural representation with a suitable and tailored eigen-structure. The following theorem demonstrates that such hybrid representation achieves a trade-off in eigenvalue decay rate.

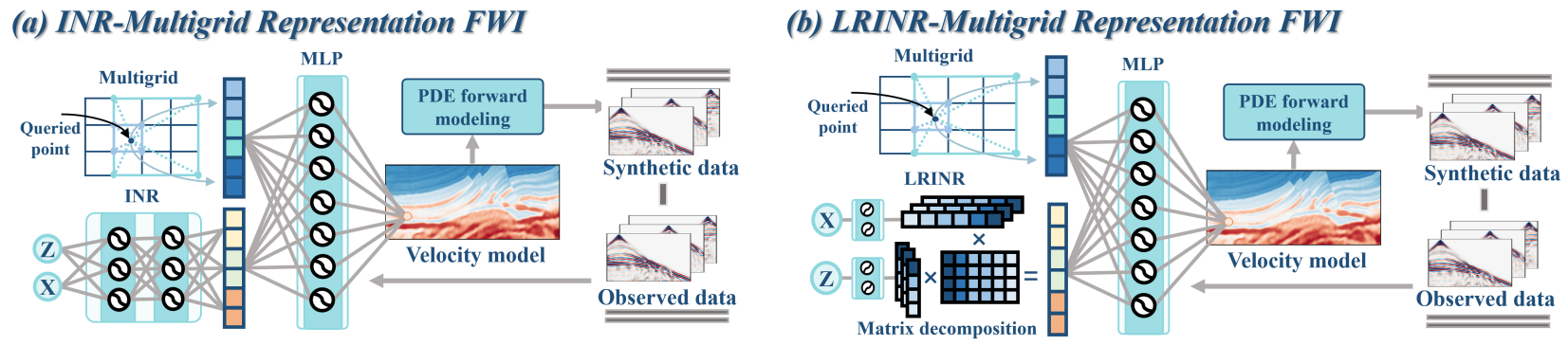

FIG. 11. *The pipeline of our proposed hybrid neural representation. (a) INR-Multigrid hybrid representation for FWI. (b) LRINR-Multigrid hybrid representation for FWI.*

THEOREM 5.3. *Consider two WNTKs $\mathbf{K}_N^{(1)}$ and $\mathbf{K}_N^{(2)}$ induced by neural representations $\mathbf{F}_{\boldsymbol{\theta}}^1(\mathbf{x}) = MLP\big(h_{\boldsymbol{\theta}_1}(\mathbf{x}); \boldsymbol{\theta}_0\big)$ and $\mathbf{F}_{\boldsymbol{\theta}}^2 = MLP\big(h_{\boldsymbol{\theta}_2}(\mathbf{x}); \boldsymbol{\theta}_0\big)$, where $\boldsymbol{\theta}_0$ are MLP parameters and $\boldsymbol{\theta}_1, \boldsymbol{\theta}_2$ are parameters of latent encoding functions $h_{\boldsymbol{\theta}_1}, h_{\boldsymbol{\theta}_2}$. Assume the NTK eigenvalue decay rate of $\mathbf{F}_{\boldsymbol{\theta}}^1(\mathbf{x})$ is slower than $\mathbf{F}_{\boldsymbol{\theta}}^2(\mathbf{x})$, i.e., $\Theta^{(1)} \succeq \Theta^{(2)}$. Consider the hybrid neural representation combining these two latent features, i.e.,*

$$\mathbf{F}_{\boldsymbol{\theta}}^{(HR)}(\mathbf{x}) = MLP\Big(\big(\sqrt{\alpha}\cdot h_{\boldsymbol{\theta}_1}(\mathbf{x})\big) \otimes \big(\sqrt{(1-\alpha)}\cdot h_{\boldsymbol{\theta}_2}(\mathbf{x})\big); \boldsymbol{\theta}_0\Big),\ \boldsymbol{\theta} = \{\boldsymbol{\theta}_0, \boldsymbol{\theta}_1, \boldsymbol{\theta}_2\}, \tag{5.2}$$

*where $\otimes$ denotes the vector concatenation operator, $\alpha$ is a scaling parameter, and let $\mathbf{K}_N^{(HR)}$ be the WTK for hybrid NeurFWI. Then, let $\lambda_j(\mathbf{K}_N^{(1)}), \lambda_j(\mathbf{K}_N^{(2)}), \lambda_j(\mathbf{K}_N^{(HR)})$ be the WTK for representation $\mathbf{F}_{\boldsymbol{\theta}}^1, \mathbf{F}_{\boldsymbol{\theta}}^2, \mathbf{F}_{\boldsymbol{\theta}}^{(HR)}$, respectively. Then, we have*

$$\lambda_j(\mathbf{K}_N^{(1)}) \geq \lambda_j(\mathbf{K}_N^{(HR)}) \geq \lambda_j(\mathbf{K}_N^{(2)}),\ \forall\ j \in [1, 2, \cdots], \tag{5.3}$$

*where the decay rate of $\lambda_j(\mathbf{K}_{wtk}^{(HR)})$ lies between those of $\lambda_j(\mathbf{K}_{wtk}^{(1)})$ and $\lambda_j(\mathbf{K}_{wtk}^{(2)})$.*

*Proof.* The proof can be found in the supplementary materials SM2.7. □

*Remark* 5.4. By balancing the eigenvalue decay in this manner, the hybrid neural representation can inherit the robustness and superior convergence rate from different neural representations, thereby achieving more accurate and reliable inversion results.

*Remark* 5.5. The scaling parameter $\alpha$ controls the eigenvalue decay rate of the hybrid NeurFWI, i.e., a larger $\alpha$ accelerates the high-frequency fitting. A suitable scaling $\alpha$ depends on the frequency distribution of wavefield data residuals. In this study, we propose an automatic criterion to select a suitable parameter based on the energy ratio of low- and high-frequency data weighted by a distribution factor:

$$\sqrt{\alpha} = 1 - \frac{\int_0^{\omega_c} \sum_{i=1}^{N_s} \left|\mathcal{F}_{t\to\omega}[\Delta D_m(\mathbf{x}_s^{(i)})](\omega)\right| \cdot W_i(\omega) d\omega}{\int_0^{\infty} \sum_{i=1}^{N_s} \left|\mathcal{F}_{t\to\omega}[\Delta D_m(\mathbf{x}_s^{(i)})](\omega)\right| \cdot W_i(\omega) d\omega},\ W_i(\omega) = \left|\mathcal{F}_{t\to\omega}[d_{obs}(t)](\omega)\right|, \tag{5.4}$$

where $\omega_c$ denotes the main frequency of the wavefield data, $N_s$ denotes the source number, $\mathbf{x}_s^{(i)}$ represents the location for the $i^{th}$ source. We compare the performance using different $\alpha$ and the selection criterion in the supplementary materials SM3.6.

19 **5.2.1. INR-Multigrid Hybrid Representation.** To achieve a performance trade-off, we propose a novel neural representation by integrating the INR and MPE, termed IG-FWI, as shown in Fig. 11 (a). IG-FWI employs a tiny INR to encode smooth features, which are then combined with multi-grid encoding features. Next, these features are combined using a MLP. The reparameterized model is given by:

$$F_{\boldsymbol{\theta}}^{ig}(\mathbf{x}) = \text{MLP}\big(\mathbf{v}(\mathbf{x};\boldsymbol{\theta}_1,\boldsymbol{\theta}_2);\boldsymbol{\theta}_0\big), \text{ where } \mathbf{v}(\mathbf{x}) = \sqrt{(1-\alpha)}\cdot I_{\boldsymbol{\theta}_1}(\mathbf{x}) \oplus \sqrt{\alpha}\cdot h_{\boldsymbol{\theta}_2}(\mathbf{x}), \tag{5.5}$$

where $I_{\boldsymbol{\theta}_1}(\cdot)$ is a tiny INR, $h_{\boldsymbol{\theta}_2}(\cdot)$ is a hash multi-resolution encoding, and hybrid latent features $\mathbf{v} \in \mathbb{R}^{(n_g n_f + n_r)}$ are formed by splicing $h_{\boldsymbol{\theta}_2}(\cdot)$ and $I_{\boldsymbol{\theta}_1}(\cdot)$ with scaling $\sqrt{\alpha}$ and $\sqrt{1-\alpha}$, respectively, where $n_r$ denotes the output dimension of tiny INR and $n_g n_f$ represents the output dimension of hash encoding. The following corollary indicates that the eigenvalue decay of IG-FWI is between that of IFWI and MPE-FWI.

Corollary 5.6. *Let $\mathbf{K}_N^{ig}, \mathbf{K}_N^{inr}, \mathbf{K}_N^{mpe}$ be the WTK of IG-FWI, IFWI, and MPE-FWI with eigenvalues $\lambda(\mathbf{K}_N^{ig}), \lambda(\mathbf{K}_N^{inr}), \lambda(\mathbf{K}_N^{mpe})$, respectively. When INR and MPE features are normalized such that comparable gradient norms, then for all $i \in \mathbb{N}$,*

$$\lambda_i(\mathbf{K}_N^{inr}) \le \lambda_i(\mathbf{K}_N^{ig}) \le \lambda_i(\mathbf{K}_N^{mpe}),\ i \in \mathbb{N}, \tag{5.6}$$

*where the spectral decay rate of $\lambda(\mathbf{K}_N^{ig})$ lies between those of $\lambda(\mathbf{K}_N^{inr})$ and $\lambda(\mathbf{K}_N^{mpe})$.*

*Proof.* Let $\mathbf{F}_{\boldsymbol{\theta}}^1(\mathbf{x})$ and $\mathbf{F}_{\boldsymbol{\theta}}^2(\mathbf{x})$ be INR and MPE in Theorem 4.10, respectively. By the Theorem 4.10 and 5.1, we can obtain the target eigenvalue inequalities. ∎

*Remark* 5.7. By balancing the WTK eigenvalue decay rate in this manner (see Fig. 8 (c)), IG-FWI inherits the robustness of INR and the superior high-wavenumber convergence of MPE, thereby achieving more accurate and reliable results.

**5.2.2. LRINR-Multigrid Hybrid Representation.** The geophysical parameters typically exhibit inherent structural priors (e.g., low-rank) [28]. Our previous work [11] proposes the LR-FWI method, which decomposes the velocity model using matrix factorization and represents the low-dimensional matrices separately using multiple 1D INRs [32]. The formulation of 2D LR-FWI can be expressed as follows:

$$F_{\boldsymbol{\theta}}^{lr}(\mathbf{x}) = [\mathbf{C}; F_{\boldsymbol{\theta}_1}, F_{\boldsymbol{\theta}_2}](\mathbf{x}) = F_{\boldsymbol{\theta}_1}(x_1) \times \mathbf{C} \times F_{\boldsymbol{\theta}_2}(x_2)^\top, \tag{5.7}$$

where $\mathbf{x} = (x_1, x_2)$ denotes spatial coordinates, $F_{\boldsymbol{\theta}_1} : \mathbb{R} \to \mathbb{R}^{1\times r_1}$, $F_{\boldsymbol{\theta}_2} : \mathbb{R} \to \mathbb{R}^{1\times r_2}$ are 1-D INRs, and $\mathbf{C} \in \mathbb{R}^{r_1 \times r_2}$ is the core matrix. Hence, the neural parameters are $\boldsymbol{\theta} = \{\boldsymbol{\theta}_1, \boldsymbol{\theta}_2, \mathbf{C}\}$. The LR-FWI method enhances the convergence rate for high-frequency components due to an appropriate eigenvalue decay rate, which we have empirically verified through numerical experiments (see Fig. 8). To further enhance the inversion performance, we integrate the multigrid hash encoding into LR-FWI (termed LRG-FWI) to improve the high-wavenumber convergence, as shown in Fig. 11 (b). Hence, the reparameterized model using LRG-FWI is given by:

$$\mathbf{F}_{\boldsymbol{\theta}}^{lrg}(\mathbf{x}) = \text{MLP}\Big(\mathbf{v}(\mathbf{x};\boldsymbol{\theta}_1,\boldsymbol{\theta}_2);\boldsymbol{\theta}_0\Big), \text{ where } \mathbf{v}(\mathbf{x}) = \sqrt{1-a}\cdot \mathbf{F}_{\boldsymbol{\theta}_1}^{lr}(\mathbf{x}) \otimes \sqrt{a}\cdot h_{\boldsymbol{\theta}_2}(\mathbf{x}), \tag{5.8}$$

where $\mathbf{F}_{\boldsymbol{\theta}_1}^{lr}$ denotes the low-rank implicit representation [32] and $h_{\boldsymbol{\theta}_2}$ represents the hash multi-resolution encoding [34]. The hybrid latent features $\mathbf{v} \in \mathbb{R}^{(n_g n_f + n_r)}$ are formed by splicing $h_{\boldsymbol{\theta}_2}(\cdot)$ and $\mathbf{F}_{\boldsymbol{\theta}_1}^{lr}(\cdot)$ with scaling $\sqrt{\alpha}$ and $\sqrt{1-\alpha}$, respectively, where $n_r$ denotes the output dimension of LRINR. The following corollary indicates that the eigenvalue decay of LRG-FWI is between that of LR-FWI and MPE-FWI.

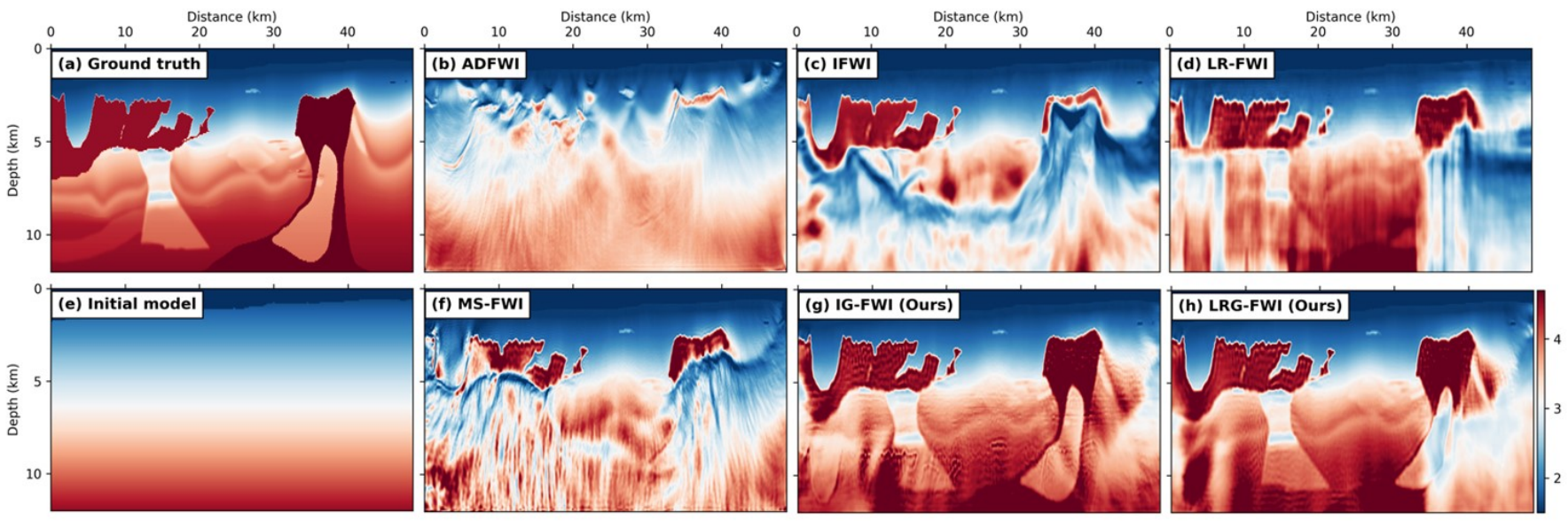

FIG. 12. *Inversion results on the BP model using different FWI methods under same iterations. (a) Ground truth. (d) Conventional FWI. (c) IFWI. (d) Low-rank FWI. (e) 1-D initial model. (f) Frequency multiscale FWI. (g) Our proposed IG-FWI. (h) Our proposed LRG-FWI.*

COROLLARY 5.8. *Let $\mathbf{K}_N^{lrg}, \mathbf{K}_N^{lr}, \mathbf{K}_N^{mpe}$ be the WTK of LRG-FWI, LR-FWI, and MPE-FWI with eigenvalues $\lambda(\mathbf{K}_N^{lrg}), \lambda(\mathbf{K}_N^{lr}), \lambda(\mathbf{K}_N^{mpe})$, respectively. When LRINR and MPE features are normalized with comparable gradient norms, for all $i \in \mathbb{N}$,*

$$\min\{\lambda_i(\mathbf{K}_N^{lr}), \lambda_i(\mathbf{K}_N^{mpe})\} \leq \lambda_i(\mathbf{K}_N^{lrg}) \leq \max\{\lambda_i(\mathbf{K}_N^{lr}), \lambda_i(\mathbf{K}_N^{mpe})\}, \ i \in \mathbb{N} \tag{5.9}$$

*where the decay rate of $\lambda_j(\mathbf{K}_{wtk}^{lrg})$ lies between those of $\lambda_j(\mathbf{K}_{wtk}^{lr})$ and $\lambda_j(\mathbf{K}_{wtk}^{mpe})$.*

*Proof.* Let $\mathbf{F}_{\boldsymbol{\theta}}^1(\mathbf{x})$ and $\mathbf{F}_{\boldsymbol{\theta}}^2(\mathbf{x})$ be LRINR and MPE in Theorem 4.10, respectively. Assume that $\Theta_{mpe} \succeq \Theta_{lr}$ or $\Theta_{lr} \succeq \Theta_{mpe}$, we can obtain the eigenvalue inequalities. ∎

**6. Applications.** We carefully select widely used FWI methods, including automatic differentiable FWI (i.e., ADFWI) [30], frequency multiscale FWI (i.e., MS-FWI) [18, 8], and existing state-of-the-art NeurFWI methods, including INR-based FWI with sine activation function (i.e., IFWI) [40] and LR-FWI method [11] as baselines, to validate the performance of our proposed novel NeurFWI methods, including INR-Grid based FWI (i.e., IG-FWI) and low-rank multigrid based FWI (i.e., LRG-FWI), on seismic exploration and medical imaging. To streamline the article, the NeurFWI implementation details are provided in the supplementary materials SM3.1.

**6.1. Marine Seismic Exploration.** Firstly, we validate our proposed methods on the challenging BP model (see Fig. 12 (a)), which includes two salt index masks and inserts salt in the model with a constant velocity of 4510 m/s for the left salt body and 4790 m/s for the right salt body. In this region, the large velocity contrasts give rise to significant multiple scattering, making it difficult to invert the resulting wavefields for subsalt imaging. Spanning 49 km in width and 12 km in depth, we discretized the subsurface model using a 75 m × 75 m grid and utilized a 1-D initial velocity model (see Fig. 12 (e)) for the inversion. The ultra-long-offset OBN acquisition employs 49 seabed hydrophones at 1 km intervals and 325 sources (150 m spacing, 25 m depth). A 3 Hz Ricker wavelet is used as the source signature with 10 ms sampling interval and 2000 time steps, and a free-surface boundary condition is applied at the top, while PML absorbing boundaries are imposed on the others. For the multiscale FWI and NeurFWI, we adopt ten multi-frequency bands ranging from 0.5 Hz to 5 Hz with a 0.5 Hz interval and 200 iterations based on the deepwave procedure.

The inversion results are shown in Fig. 12. These results seem to indicate that ADFWI method tends to be affected by local minima issues, and the frequency multiscale FWI can only recover the shallow salt body velocity structure. IFWI can successfully invert the left salt body structure, although the deep velocity and right salt body need further improvement. LR-FWI can further enhance the inversion performance of deep velocity structure using low-rank tensor function decomposition. Moreover, combining multi-grid parametric hash encoding into IFWI and LR-FWI methods (i.e., IG-FWI and LRG-FWI) can consistently improve the performance of large-scale inversion scenarios, especially in the salt body and deep velocity structure, which can be attributed to their suitable and balanced eigenvalue decay rate of WTK.

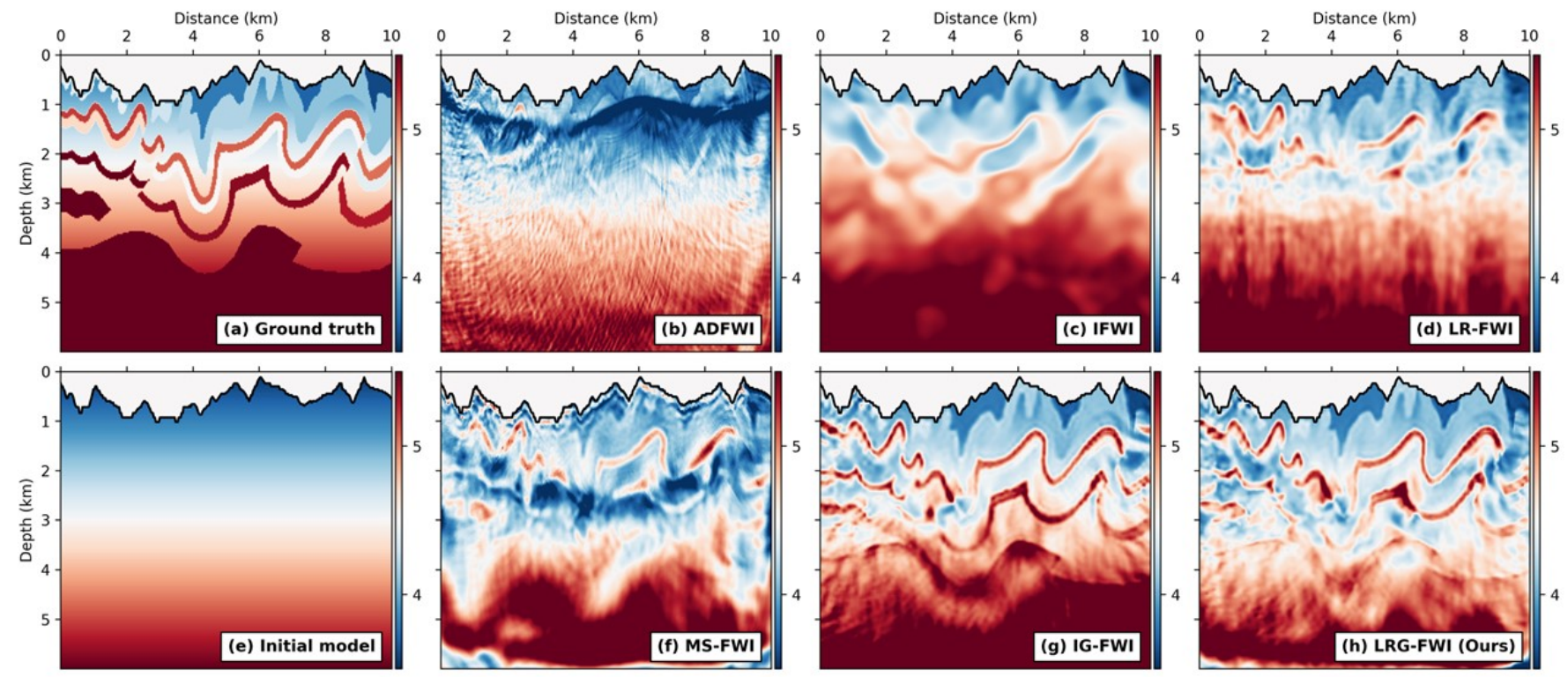

FIG. 13. *Inversion results on the Foothills model using different FWI methods under same iterations. (a) Ground truth. (b) Conventional ADFWI. (c) IFWI. (d) Low-rank FWI. (e) 1-D initial model. (f) Multiscale FWI. (g) Our proposed IG-FWI. (h) Our proposed LRG-FWI.*

**6.2. Land Seismic Exploration.** In land seismic exploration, the complex multiple reflected waves due to the irregular surface further increase the difficulty of inversion. We validate our proposed FWI methods on the challenging Canadian Foothills model (see Fig. 13 (a)) with an irregular surface. Spanning 10 km in width and 6 km in depth, we discretized the subsurface model using a 30 m $\times$ 30 m grid and utilized a 1-D initial velocity model (see Fig. 13 (b)) for the inversion. Synthetic data are generated from 30 shot gathers, with sources spaced 300 m apart at a depth of 30 m, while all receivers were fixed at a depth of 60 m from left to right. A Ricker wavelet with a dominant frequency of 8 Hz with 2 ms sampling interval and 1500 time steps. A free-surface boundary condition based on the immersive boundary method is applied at the top, while PML absorbing boundaries are imposed on the others.

The inversion results are shown in Fig. 13. These results show that the ADFWI method may falls into the local minima due to its slow eigenvalue decay rate, and the frequency multiscale FWI can only recover the shallow velocity structure. IFWI and LR-FWI can recover the overall background at the cost of poor details recovery due to their fast eigenvalue decay rate. Moreover, combining multi-grid parametric hash encoding into IFWI and LR-FWI methods (i.e., IG-FWI and LRG-FWI) can consistently improve the performance of land seismic imaging with irregular surface, which can be attributed to their suitable and balanced eigenvalue decay rate of WTK.

**6.3. More Realistic Chevron Blind Test Model.** To further validate our proposed IG-FWI and LRG-FWI methods on more realistic field data and consider the inversion crime problem, we compare our methods with ADFWI [30], multiscale FWI [8], IFWI [40], and LR-FWI [11] on the 2014 Chevron blind test model. This 2D marine towed-streamer dataset, released by Chevron for FWI applications, features a maximum offset of 8 km. The observed data were generated by simulating 2D isotropic elastic wave propagation with a free surface and absorbing boundary conditions. The resulting records contain both low- and high-frequency noise, with a useful signal predominantly above 3 Hz. Acquisition parameters include 1600 shots spaced at 15 m intervals and receivers deployed at 15 m spacing, all positioned at a water depth of 15 m. Data were recorded for 8 s at a sampling interval of 4 ms. The velocity model spans approximately 45 km in length and 6 km in depth, discretized on a 25 m grid, with a water layer velocity of 1510 m/s. A 1D starting model (see Fig. 14 (a)), source wavelet, and a well log were provided, while the ground truth model was withheld.

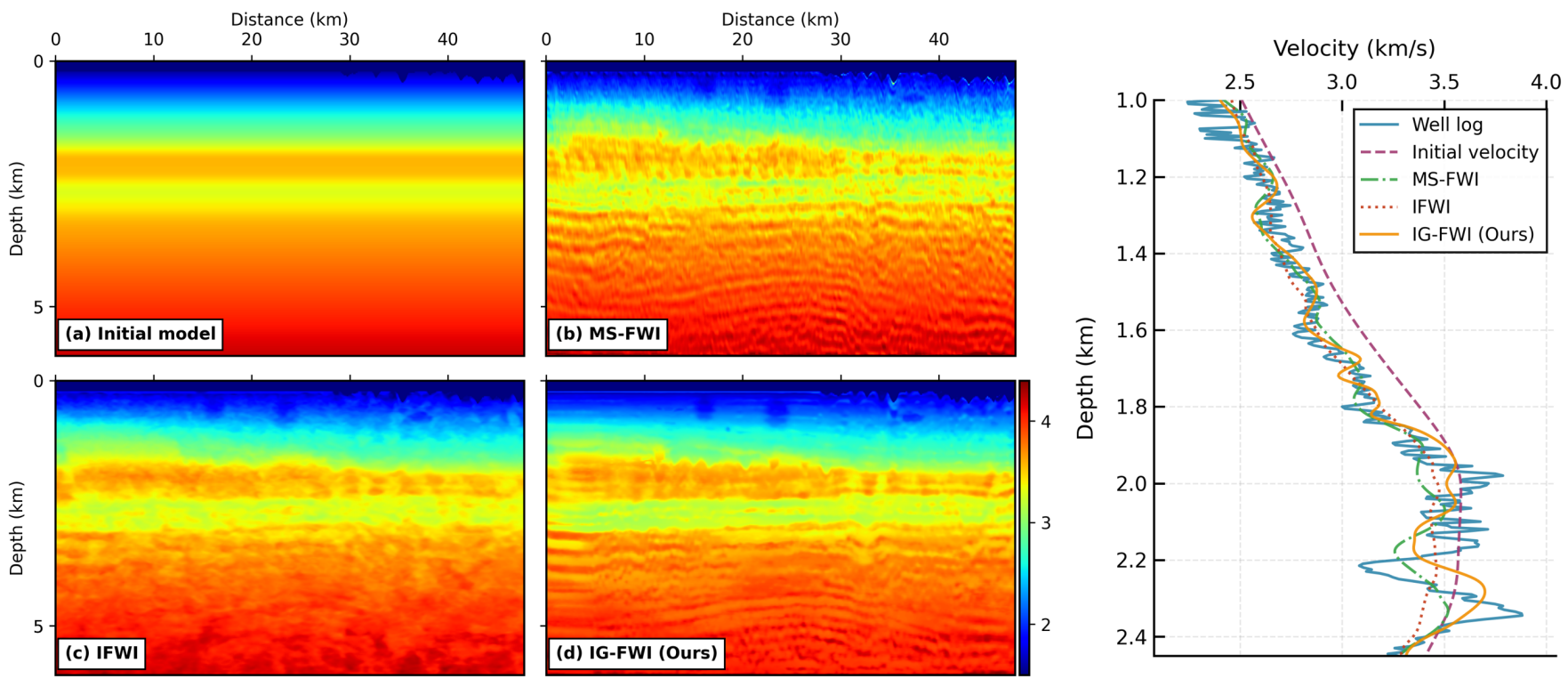

FIG. 14. *Inversion results on the Chevron model under same iterations. (a) 1-D initial model. (b) MS-FWI. (c) IFWI. (d) IG-FWI. The right image denotes the well log of the Chevron model.*

Multiscale FWI and all NeurFWI methods adopt the frequency multiscale strategies, which are divided into five frequency bands ranging from 3 Hz to 11 Hz with a 2 Hz interval and 50 iterations. Here, we produce the synthetic seismic data using the acoustic wave equation rather than the isotropic and elastic wave equation, which avoids the inversion crime problem. The inversion results on the Chevron model using MS-FWI, IFWI, and the proposed IG-FWI methods are shown in Fig. 14 (ADFWI, LR-FWI, and the proposed LRG-FWI results are shown in the supplemental materials SM3.5), showing that these methods successfully recover the shallow structures. However, the inversion becomes more challenging at greater depths. Velocity profiles (the right image of Fig. 14) indicate good agreement with well-log data from 1 km to 2.5 km, though some deviations occur below 2.0 km. Despite these difficulties, our proposed IG-FWI delivers better performance compared to other FWI methods.

**6.4. Medical Imaging.** Beyond seismic imaging, NeurFWI and our proposed FWI methods are also applicable to other scientific imaging domains (e.g., medical imaging). Guasch et al. [20] introduce the FWI approach into human brain imaging, which validates the high-resolution and effectiveness of FWI. Recently, Yan et al. [59] first introduced untrained neural networks into FWI medical imaging, thereby enhancing robustness by representing human brain velocity using the DIP. In this study, we first extend INR-based NeurFWI (e.g., IFWI and LR-FWI) to medical imaging and validate that our proposed methods further improve inversion performance.

Unlike the surface boundary receivers in seismic FWI, medical FWI can obtain a comprehensive boundary observation wavefield information to recover the interior structure, such as the breast and brain. The representative wavefield data are shown in Fig. 15 (c) and (f). The breast model is discretized as 544 $\times$ 544 with a 0.5 mm grid interval, and the brain model is discretized as 472 $\times$ 472 with a 0.5 mm grid interval. The acquisition system consists of 20 shots and 499 receivers per shot for both breast and brain models. Both sources and receivers are uniformly distributed on two concentric ellipses (same center, different semi-axis lengths). The source wavelet is a Ricker wavelet with a center frequency of 0.7 MHz and 0.3 MHz for breast and brain, respectively, sampling over 2000 and 4000 time steps with a time interval of 0.1 $\mu$s, respectively. The initial models are constant, as shown in Fig. 15 (b) and (e).

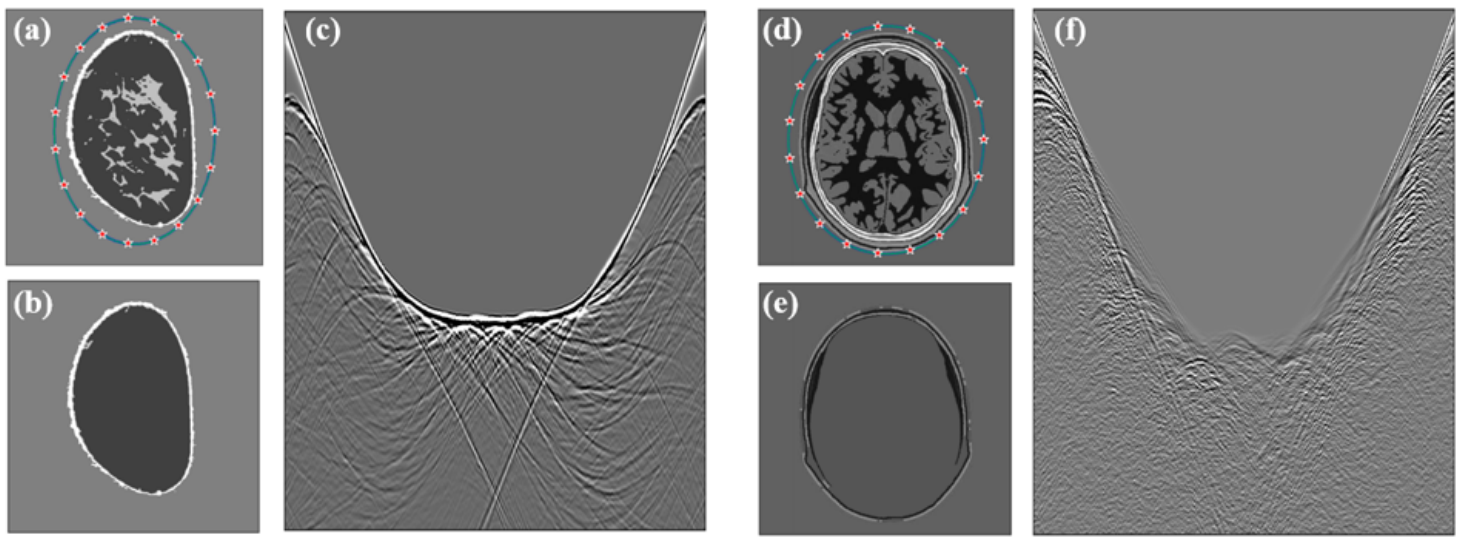

FIG. 15. *Breast and brain models. (a) Ground truth of breast. (b) Breast initial model. (c) Observed data. (d) Ground truth of the brain. (e) Brain initial model. (f) Observed data.*

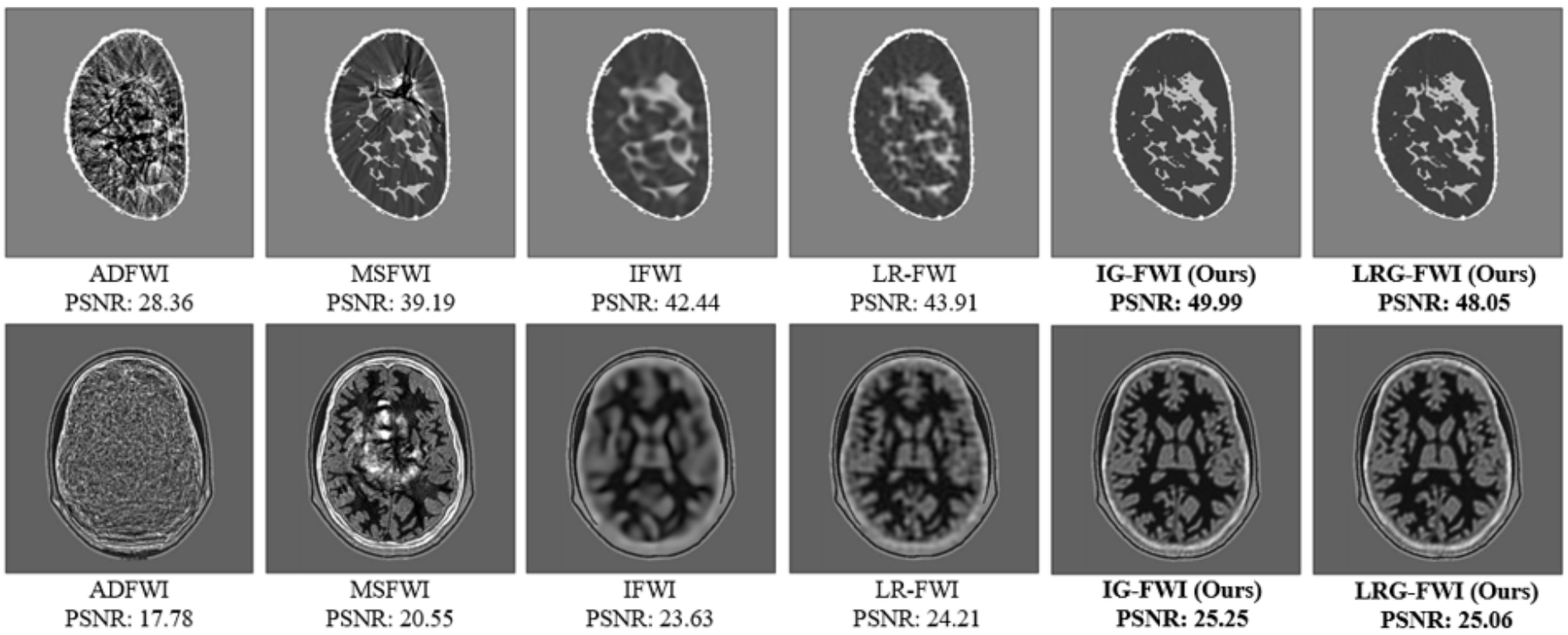

FIG. 16. *Inversion results using different FWI methods under same iterations.*

These inversion results are shown in Fig. 16 using different FWI methods. Conventional FWI based on automatic differentiation can not recover the effective structure due to the high-frequency wavefield data and the barrier of the high-speed brain boundary. Multiscale FWI can recover the vast majority of the structure, while some areas have fallen into local minima. NeurFWI can enhance the inversion stability due to the gradient wavenumber modulation and wave frequency bias. Furthermore, our proposed IG-FWI and LRG-FWI methods can further enhance the imaging resolution, which can be attributed to their more balanced eigenvalue decay rate of WTK.

**7. Conclusions.** This work establishes a theoretical framework for neural reparameterized FWI (NeurFWI) by introducing the neural sensitivity kernel and wave tangent kernel. We demonstrate that the NTK induced by neural representation modulates both gradient spectra and data residual evolution, providing an explanation for the enhanced stability and slower high-resolution convergence observed in NeurFWI. Guided by these insights, we propose a design methodology and instantiated through IG-FWI and LRG-FWI, which explicitly tailors spectral filtering and eigenvalue decay for more balanced inversion performance. Numerical validations across complex seismic and medical imaging benchmarks confirm our theoretical claims and the proposed methods. Beyond FWI, these theories and methods may have the potential to be extended to more PDE-constrained inversion problems and unveil the mechanism of parameter crosstalk decoupling in multi-parameter FWI.

# SUPPLEMENTARY MATERIALS: DECIPHERING NEURAL REPARAMETERIZED FULL-WAVEFORM INVERSION WITH NEURAL SENSITIVITY KERNEL AND WAVE TANGENT KERNEL

RUIHUA CHEN*, BANGYU WU*, YISI LUO‡, XILE ZHAO†, AND DEYU MENG‡

## SM1. Theoretical Proof of NSK.

This section provides the detailed proof of the proposed Theorems, Propositions, and Corollaries about NSK theories.

### SM1.1. Proof of Equation (1.4).

Consider the data misfit functional

$$\mathcal{J}(m) = \frac{1}{2} \sum_{s,\,r,\,\omega} \left| d_{\text{syn}}(\mathbf{x}_s, \mathbf{x}_r, \omega) - d_{\text{obs}}(\mathbf{x}_s, \mathbf{x}_r, \omega) \right|^2, \tag{SM1.1}$$

where the summation is taken over sources $s$, receivers $r$, and angular frequencies $\omega$. Assume a homogeneous background medium with constant wave speed $c_0$. Decompose the total wavefield $u$ into a background field $u_0$ and a scattered field $u_s$ (i.e., $u = u_0 + u_s$) under the Born approximation. Then, for each source–receiver pair $(\mathbf{x}_s, \mathbf{x}_r)$ and frequency $\omega$, the sensitivity kernel can be expressed as follows [SM11]:

$$\frac{\partial u(\mathbf{x}_s, \mathbf{x}_r, \omega)}{\partial m(\mathbf{y})} = \Re\Big\{ \frac{-2\omega^2 f(\omega)}{c_0^3(\mathbf{y})} G(\mathbf{x}_s, \mathbf{y}, \omega) G(\mathbf{y}, \mathbf{x}_r, \omega) \Big\}, \tag{SM1.2}$$

where $f(\omega)$ denotes the source function, $G(\mathbf{x}_s, \mathbf{y}, \omega)$ denotes the Green's function at location $\mathbf{y}$ with the source at $\mathbf{x}_s$, and $\Re\{\cdot\}$ denotes the real part. Using a reciprocity principle, the following relation between the source Green's function and receiver Green's function holds $G(\mathbf{y}, \mathbf{x}_s, \omega) = G(\mathbf{x}_s, \mathbf{y}, \omega)$. Then, the gradient of the wavefield data misfit function with respect to parameter models is given by:

$$\begin{aligned} \frac{\delta \mathcal{J}(m)}{\delta m}(\mathbf{y}) &= \sum_{\omega} \sum_{s} \sum_{r} \left( \frac{\partial u(\mathbf{x}_s, \mathbf{x}_r, \omega)}{\partial m(\mathbf{y})} \right)^* \Delta D_m(\mathbf{x}_s, \mathbf{x}_r, \omega) \\ &= \sum_{\omega} \sum_{s} \sum_{r} \Re\Big\{ \frac{-2\omega^2 f(\omega)}{c_0^3(\mathbf{y})} G^*(\mathbf{y}, \mathbf{x}_s, \omega) G^*(\mathbf{y}, \mathbf{x}_r, \omega) \Delta D_m(\mathbf{x}_s, \mathbf{x}_r, \omega) \Big\}. \end{aligned} \tag{SM1.3}$$

Let the incident field be a monochromatic plane wave propagating in direction $\hat{\mathbf{s}}$ and let the scattered field be recorded as a plane wave propagating in direction $\hat{\mathbf{r}}$ under the far-field approximation and homogeneous background velocity model, the Green's functions $G(\mathbf{y}, \mathbf{x}_s, \omega)$ and $G(\mathbf{y}, \mathbf{x}_r, \omega)$ can be expressed as follows:

$$G(\mathbf{y}, \mathbf{x}_s, \omega) \simeq \exp(i \frac{\omega}{c_0} \hat{\mathbf{s}} \cdot \mathbf{y}), \;\; G(\mathbf{y}, \mathbf{x}_r, \omega) \simeq \exp(i \frac{\omega}{c_0} \hat{\mathbf{r}} \cdot \mathbf{y}). \tag{SM1.4}$$

Inserting the expression of the incident and scattered plane waves into the gradient of the misfit function, we have the following gradient expression of FWI:

$$\frac{\delta \mathcal{J}(m)}{\delta m}(\mathbf{y}) \simeq \sum_{\omega} \sum_{s} \sum_{r} \omega^2 \, \Re\Big\{ \exp\Big( -i \frac{\omega}{c_0} (\hat{\mathbf{s}} + \hat{\mathbf{r}}) \cdot \mathbf{y} \Big) \Big\} \, \Delta D_m(\mathbf{x}_s, \mathbf{x}_r, \omega), \tag{SM1.5}$$

*School of Mathematics and Statistics, Xi'an Jiaotong University, Xi'an, Shaanxi 710049, China (Ruihua.Chen@stu.xjtu.edu.cn, bangyuwu@xjtu.edu.cn).

†School of Mathematical Sciences, University of Electronic Science and Technology of China, Chengdu 610000, China (xlzhao122003@163.com).

‡Corresponding author. School of Mathematics and Statistics, Xi'an Jiaotong University, Xi'an, Shaanxi, China (yisiluo1221@foxmail.com, dymeng@mail.xjtu.edu.cn)

where we ignore the constant term $-\frac{2f(\omega)}{c_0^3(\mathbf{y})}$ for simplified representation.

**SM1.2. Proof of Proposition 3.1.** Under the continuous-time gradient flow $\frac{d\boldsymbol{\theta}}{d\tau} = -\nabla_{\boldsymbol{\theta}}\mathcal{J}$, the evolution of $m_{\boldsymbol{\theta}}(\mathbf{y})$ is governed by the chain rule:

$$\frac{\mathrm{d}m_{\boldsymbol{\theta}}(\mathbf{y})}{\mathrm{d}\tau} = \left\langle \nabla_{\boldsymbol{\theta}} m_{\boldsymbol{\theta}}(\mathbf{y}), \frac{\mathrm{d}\boldsymbol{\theta}}{\mathrm{d}\tau} \right\rangle_{\mathbb{R}^n} = -\left\langle \nabla_{\boldsymbol{\theta}} m_{\boldsymbol{\theta}}(\mathbf{y}), \nabla_{\boldsymbol{\theta}}\mathcal{J} \right\rangle_{\mathbb{R}^n}. \tag{SM1.6}$$

Applying the chain rule for functional derivatives, we expand the inner product as

$$\begin{aligned} \frac{\mathrm{d}m_{\boldsymbol{\theta}}(\mathbf{y})}{\mathrm{d}\tau} &= -\sum_{i=1}^{n} \frac{\partial m_{\boldsymbol{\theta}}(\mathbf{y})}{\partial \theta_i} \left( \int_\Omega \frac{\delta \mathcal{J}}{\delta m}[m_{\boldsymbol{\theta}}](\mathbf{z}) \frac{\partial m_{\boldsymbol{\theta}}(\mathbf{z})}{\partial \theta_i}\,\mathrm{d}\mathbf{z} \right) \\ &= -\int_\Omega \frac{\delta \mathcal{J}}{\delta m}[m_{\boldsymbol{\theta}}](\mathbf{z}) \underbrace{\left[ \sum_{i=1}^{n} \frac{\partial m_{\boldsymbol{\theta}}(\mathbf{y})}{\partial \theta_i} \frac{\partial m_{\boldsymbol{\theta}}(\mathbf{z})}{\partial \theta_i} \right]}_{\Theta(\mathbf{y},\mathbf{z};\boldsymbol{\theta})} \mathrm{d}\mathbf{z}, \end{aligned} \tag{SM1.7}$$

where $\Theta(\mathbf{y},\mathbf{z};\boldsymbol{\theta})$ denotes the neural tangent kernel (NTK). Substituting the functional gradient expression (1.3) into the evolution equation yields

$$\begin{aligned} \frac{\mathrm{d}m_{\boldsymbol{\theta}}(\mathbf{y})}{\mathrm{d}\tau} &= -\int_\Omega \left( \int_T \int_\Omega \frac{\partial u(\mathbf{x},t)}{\partial m(\mathbf{z})} \Delta D_m(\mathbf{x},t)\,\mathrm{d}\mathbf{x}\,\mathrm{d}t \right) \Theta(\mathbf{y},\mathbf{z};\boldsymbol{\theta})\,\mathrm{d}\mathbf{z} \\ &= -\int_T \int_\Omega \left( \int_\Omega \frac{\partial u(\mathbf{x},t)}{\partial m(\mathbf{z})} \Theta(\mathbf{y},\mathbf{z};\boldsymbol{\theta})\,\mathrm{d}\mathbf{z} \right) \Delta D_m(\mathbf{x},t)\,\mathrm{d}\mathbf{x}\,\mathrm{d}t, \end{aligned} \tag{SM1.8}$$

where the interchange of integrals is justified by Fubini's theorem under standard regularity assumptions. Defining the *neural sensitivity kernel* (NSK) as

$$\frac{\partial u_\Theta(\mathbf{x},t)}{\partial m(\mathbf{y})} = \int_\Omega \frac{\partial u(\mathbf{x},t)}{\partial m(\mathbf{z})} \Theta(\mathbf{y},\mathbf{z};\boldsymbol{\theta})\,\mathrm{d}\mathbf{z}, \tag{SM1.9}$$

we finally obtain the desired evolution dynamics. This completes the proof.

**SM1.3. Proof of Proposition 3.3.** Before proceeding with the proof of the main proposition, we introduce a fundamental lemma regarding the convergence and structural properties of the neural tangent kernel. The proofs of these properties are well-established in the literature; see, e.g., [SM4, SM1, SM6, SM5].

LEMMA SM1.1. *Let $\Theta_n(\mathbf{x},\mathbf{x}';\boldsymbol{\theta})$ denote the NTK of a network $f(\boldsymbol{\theta})$ as defined in Definition 2.1. For any fixed parameters $\boldsymbol{\theta}$, the finite-width NTK is symmetric and positive semi-definite. Furthermore, in the infinite-width limit (as hidden layer widths tend to infinity), under standard initialization and scaling regimes, the initialized NTK converges in probability to a deterministic limiting kernel $\Theta(\mathbf{x},\mathbf{x}')$. Moreover, under gradient flow training on the squared loss, this limiting kernel remains constant throughout the training dynamics, i.e., $\Theta_\infty(\mathbf{x},\mathbf{x}';\boldsymbol{\theta}_\tau) = \Theta(\mathbf{x},\mathbf{x}')$ for all $\tau \geq 0$.*

We now present the proof of Proposition 3.3 utilizing the spectral decomposition.

*Proof.* By Lemma SM1.1, the NTK $\Theta(\mathbf{x},\mathbf{x}')$ is symmetric and positive semi-definite. By Mercer's Theorem [SM7], $\Theta$ admits an eigen-decomposition. There exists an orthonormal basis $\{\phi_k\}_{k=1}^\infty$ of $L^2(\Omega)$ and a sequence of non-negative eigenvalues $\{\lambda_k\}_{k=1}^\infty$ (sorted in non-increasing order, i.e., $\lambda_1 \geq \lambda_2 \geq \cdots \geq 0$) such that:

$$\Theta(\mathbf{y},\mathbf{z};\boldsymbol{\theta}) = \sum_{k=1}^{\infty} \lambda_k \phi_k(\mathbf{y}) \phi_k(\mathbf{z}), \tag{SM1.10}$$

Substituting the expansion (SM1.10) into the definition of NSK, we obtain:

$$\begin{aligned}\frac{\partial u_\Theta(\mathbf{x},t)}{\partial m(\mathbf{y})} &= \int_\Omega \frac{\partial u(\mathbf{x},t)}{\partial m(\mathbf{z})}\left[\sum_{k=1}^{\infty}\lambda_k\phi_k(\mathbf{y})\phi_k(\mathbf{z})\right]\mathrm{d}\mathbf{z}\\ &= \sum_{k=1}^{\infty}\lambda_k\left\langle\frac{\partial u(\mathbf{x},t)}{\partial m(\cdot)},\phi_k(\cdot)\right\rangle_{L^2(\Omega)}\phi_k(\mathbf{y}),\end{aligned} \tag{SM1.11}$$

where $\langle\cdot,\cdot\rangle_{L^2(\Omega)}$ denotes the standard inner product in the Hilbert space $L^2(\Omega)$. ∎

**SM1.4. Proof of Lemma 3.6.** Applying the spatial Fourier transform defined in (2.1) to the model dynamics, the evolution in the wavenumber domain is given by

$$\begin{aligned}\mathcal{F}_{\mathbf{y}\to\boldsymbol{\xi}}\left[\frac{\mathrm{d}m_{\boldsymbol{\theta}}(\mathbf{y})}{\mathrm{d}\tau}\right](\boldsymbol{\xi}) &= \int_\Omega \frac{\mathrm{d}m_{\boldsymbol{\theta}}(\mathbf{y})}{\mathrm{d}\tau}e^{-2\pi i\mathbf{y}\cdot\boldsymbol{\xi}}\,\mathrm{d}\mathbf{y}\\ &= \int_\Omega\left(\int_T\int_\Omega\frac{\partial u_\Theta(\mathbf{x},t)}{\partial m(\mathbf{y})}\Delta D_m(\mathbf{x},t)\,\mathrm{d}\mathbf{x}\,\mathrm{d}t\right)e^{-2\pi i\mathbf{y}\cdot\boldsymbol{\xi}}\,\mathrm{d}\mathbf{y}\\ &= \int_T\int_\Omega\left(\int_\Omega\frac{\partial u_\Theta(\mathbf{x},t)}{\partial m(\mathbf{y})}e^{-2\pi i\mathbf{y}\cdot\boldsymbol{\xi}}\,\mathrm{d}\mathbf{y}\right)\Delta D_m(\mathbf{x},t)\,\mathrm{d}\mathbf{x}\,\mathrm{d}t\\ &= \int_T\int_\Omega\mathcal{F}_{\mathbf{y}\to\boldsymbol{\xi}}\left[\frac{\partial u_\Theta(\mathbf{x},t)}{\partial m(\mathbf{y})}\right](\boldsymbol{\xi})\Delta D_m(\mathbf{x},t)\,\mathrm{d}\mathbf{x}\,\mathrm{d}t.\end{aligned} \tag{SM1.12}$$

Next, we expand the Fourier transform of NSK using the definition of the NTK:

$$\begin{aligned}\mathcal{F}_{\mathbf{y}\to\boldsymbol{\xi}}\left[\frac{\partial u_\Theta(\mathbf{x},t)}{\partial m(\mathbf{y})}\right](\boldsymbol{\xi}) &= \int_\Omega e^{-2\pi i\mathbf{y}\cdot\boldsymbol{\xi}}\left(\int_\Omega\Theta(\mathbf{y},\mathbf{z};\boldsymbol{\theta})\frac{\partial u(\mathbf{x},t)}{\partial m(\mathbf{z})}\,\mathrm{d}\mathbf{z}\right)\mathrm{d}\mathbf{y}\\ &= \mathbb{E}_{\boldsymbol{\theta}}\left[\int_\Omega\int_\Omega e^{-2\pi i\mathbf{y}\cdot\boldsymbol{\xi}}\left(\nabla_{\boldsymbol{\theta}}\sigma^*(\mathbf{y},\boldsymbol{\theta})\cdot\nabla_{\boldsymbol{\theta}}\sigma^*(\mathbf{z},\boldsymbol{\theta})\right)\frac{\partial u(\mathbf{x},t)}{\partial m(\mathbf{z})}\,\mathrm{d}\mathbf{z}\,\mathrm{d}\mathbf{y}\right]\\ &= \mathbb{E}_{\boldsymbol{\theta}}\left[\left(\int_\Omega\nabla_{\boldsymbol{\theta}}\sigma^*(\mathbf{y},\boldsymbol{\theta})e^{-2\pi i\mathbf{y}\cdot\boldsymbol{\xi}}\,\mathrm{d}\mathbf{y}\right)\cdot\left(\int_\Omega\nabla_{\boldsymbol{\theta}}\sigma^*(\mathbf{z},\boldsymbol{\theta})\frac{\partial u(\mathbf{x},t)}{\partial m(\mathbf{z})}\,\mathrm{d}\mathbf{z}\right)\right].\end{aligned} \tag{SM1.13}$$

Applying Parseval's identity to the inner product over $\mathbf{z}$, we obtain

$$\mathcal{F}_{\mathbf{y}\to\boldsymbol{\xi}}\left[\frac{\partial u_\Theta(\mathbf{x},t)}{\partial m(\mathbf{y})}\right](\boldsymbol{\xi}) = \mathbb{E}_{\boldsymbol{\theta}}\left[\int_{\mathbb{R}^d}\mathcal{F}_{\mathbf{y}\to\boldsymbol{\xi}}[\nabla_{\boldsymbol{\theta}}\sigma^*(\cdot,\boldsymbol{\theta})](\boldsymbol{\xi})\cdot\overline{\mathcal{F}_{\mathbf{z}\to\boldsymbol{\xi}'}[\nabla_{\boldsymbol{\theta}}\sigma^*(\cdot,\boldsymbol{\theta})](\boldsymbol{\xi}')}\cdot\frac{\partial u(\mathbf{x},t)}{\partial\hat{m}(\boldsymbol{\xi}')}\,\mathrm{d}\boldsymbol{\xi}'\right],$$

where $\overline{(\cdot)}$ denotes the complex conjugate. Defining the *spectral NTK* as

$$\hat{\Theta}(\boldsymbol{\xi},\boldsymbol{\xi}') = \mathbb{E}_{\boldsymbol{\theta}}\left(\mathcal{F}_{\mathbf{y}\to\boldsymbol{\xi}}[\nabla_{\boldsymbol{\theta}}\sigma^*(\cdot,\boldsymbol{\theta})](\boldsymbol{\xi})\cdot\overline{\mathcal{F}_{\mathbf{z}\to\boldsymbol{\xi}'}[\nabla_{\boldsymbol{\theta}}\sigma^*(\cdot,\boldsymbol{\theta})](\boldsymbol{\xi}')}\right), \tag{SM1.14}$$

and $\frac{\partial u(\mathbf{x},t)}{\partial\hat{m}(\boldsymbol{\xi}')}$ as the functional derivative in the wavenumber domain, then we have

$$\mathcal{F}_{\mathbf{y}\to\boldsymbol{\xi}}\left[\frac{\partial u_\Theta(\mathbf{x},t)}{\partial m(\mathbf{y})}\right](\boldsymbol{\xi}) = \int_{\mathbb{R}^d}\hat{\Theta}(\boldsymbol{\xi},\boldsymbol{\xi}')\frac{\partial u(\mathbf{x},t)}{\partial\hat{m}(\boldsymbol{\xi}')}\,\mathrm{d}\boldsymbol{\xi}'. \tag{SM1.15}$$

Substituting this result back into the first equation completes the derivation of (3.6).

**SM1.5. Proof of Theorem 3.8.** First, we present several essential lemmas required for the subsequent derivation. Lemma SM1.2 provides the Fourier transform of the neural network. Lemma SM1.3 (the Laplace method) is utilized to approximate the expectation over the bias term $b$, while Lemma SM1.5 addresses the expectation over the weight vector $\mathbf{w}$ under radial symmetry. These results collectively facilitate the simplification of the explicit NSK to analyze the modulation performance.

LEMMA SM1.2 (Fourier Transforms of Networks). *For any unit vector $\mathbf{v} \in \mathbb{R}^d$, any nonzero vector $\mathbf{w} \in \mathbb{R}^d$ with $\hat{\mathbf{w}} = \frac{\mathbf{w}}{||\mathbf{w}||}$, and an activation function $\sigma(\cdot) : \mathbb{R} \to \mathbb{R}$ such that $\mathcal{F}[\sigma] \in C(\mathbb{R})$, the following equations hold in the sense of distributions:*

$$\begin{aligned}
&(a) \quad \mathcal{F}_{\mathbf{x}\to\boldsymbol{\xi}}[\sigma(\mathbf{v}^\top \mathbf{x})](\boldsymbol{\xi}) = \delta_{\mathbf{v}}(\boldsymbol{\xi})\mathcal{F}[\sigma](\boldsymbol{\xi}^\top \mathbf{v}),\\
&(b) \quad \mathcal{F}_{\mathbf{x}\to\boldsymbol{\xi}}[\sigma(\mathbf{w}^\top \mathbf{x} + b)](\boldsymbol{\xi}) = \delta_{\mathbf{w}}(\boldsymbol{\xi})\mathcal{F}[\sigma]\Big((\boldsymbol{\xi}^\top \hat{\mathbf{w}})/||\mathbf{w}||\Big) e^{(2\pi i \frac{b}{||\mathbf{w}||}\boldsymbol{\xi}^\top \hat{\mathbf{w}})},\\
&(c) \quad \mathcal{F}_{\mathbf{x}\to\boldsymbol{\xi}}[\mathbf{x}\sigma(\mathbf{w}^\top \mathbf{x} + b)](\boldsymbol{\xi}) = \frac{i}{2\pi}\nabla_{\boldsymbol{\xi}}\Big[\delta_{\mathbf{w}}(\boldsymbol{\xi})\mathcal{F}[\sigma]\Big((\boldsymbol{\xi}^\top \hat{\mathbf{w}})/||\mathbf{w}||\Big)\Big] e^{(2\pi i \frac{b}{||\mathbf{w}||}\boldsymbol{\xi}^\top \hat{\mathbf{w}})},
\end{aligned}$$

*where $\delta_{\mathbf{w}} : \mathcal{S}(\mathbb{R}^d) \to \mathbb{R}$ is defined such that for any test function $\phi \in \mathcal{S}(\mathbb{R}^d)$ (the Schwartz space on $\mathbb{R}^d$), we have $\langle \delta_{\mathbf{w}}, \phi\rangle = \int_{\mathbb{R}} \phi(y\mathbf{w})dy$.*

*Proof.* The proof can be found in [Lemma 2 of [SM5]]. □

For the two-layer neural network defined in equation (2.6), the learnable parameter set is $\Theta = \{a, \mathbf{w}, b\}$. The NTK $\hat{\Theta}(\boldsymbol{\xi}, \boldsymbol{\xi}')$ can be decomposed into three components corresponding to the parameters $a$, $\mathbf{w}$, and $b$, respectively:

$$\begin{aligned}
\hat{\Theta}(\boldsymbol{\xi},\boldsymbol{\xi}') &= \mathbb{E}_{\boldsymbol{\theta}}\hat{\Theta}_{\boldsymbol{\theta}}(\boldsymbol{\xi},\boldsymbol{\xi}') = \mathbb{E}_{\boldsymbol{\theta}}\Big(\mathcal{F}_{\mathbf{x}\to\boldsymbol{\xi}}\Big[\nabla_{\boldsymbol{\theta}}\sigma^*(\mathbf{x},\boldsymbol{\theta})\Big](\boldsymbol{\xi}) \cdot \overline{\mathcal{F}_{\mathbf{x}'\to\boldsymbol{\xi}}\Big[\nabla_{\boldsymbol{\theta}}\sigma^*(\mathbf{x}',\boldsymbol{\theta})\Big](\boldsymbol{\xi}')}\Big)\\
&= \mathbb{E}_a\Big(\mathcal{F}\big[\nabla_a\sigma^*\big](\boldsymbol{\xi}) \cdot \overline{\mathcal{F}\big[\nabla_a\sigma^*\big]}(\boldsymbol{\xi}')\Big) + \mathbb{E}_{\mathbf{w}}\Big(\mathcal{F}\big[\nabla_{\mathbf{w}}\sigma^*\big](\boldsymbol{\xi}) \cdot \overline{\mathcal{F}\big[\nabla_{\mathbf{w}}\sigma^*\big]}(\boldsymbol{\xi}')\Big)\\
&+ \mathbb{E}_b\Big(\mathcal{F}\big[\nabla_b\sigma^*\big](\boldsymbol{\xi}) \cdot \overline{\mathcal{F}\big[\nabla_b\sigma^*\big]}(\boldsymbol{\xi}')\Big) = \mathbb{E}_{\boldsymbol{\theta}}\Big(\hat{\Theta}_a(\boldsymbol{\xi},\boldsymbol{\xi}') + \hat{\Theta}_{\mathbf{w}}(\boldsymbol{\xi},\boldsymbol{\xi}') + \hat{\Theta}_b(\boldsymbol{\xi},\boldsymbol{\xi}')\Big).
\end{aligned}$$

Furthermore, the neural sensitivity kernel $\partial u_\Theta/\partial m$ in the wavenumber domain given by equation (3.6) can be decomposed into three analogous parts:

$$\begin{aligned}
\mathcal{F}_{\mathbf{y}\to\boldsymbol{\xi}}\Big[\frac{\partial u_\Theta(\mathbf{x},t)}{\partial m(\mathbf{y})}\Big](\boldsymbol{\xi}) &= \int_{\mathbb{R}^d} \frac{\partial u(\mathbf{x},t)}{\partial \hat{m}(\boldsymbol{\xi}')} \cdot \hat{\Theta}(\boldsymbol{\xi},\boldsymbol{\xi}')\, d\boldsymbol{\xi}'\\
&= \mathbb{E}_{\boldsymbol{\theta}}\Big(\int_{\mathbb{R}^d} \frac{\partial u(\mathbf{x},t)}{\partial \hat{m}(\boldsymbol{\xi}')} \cdot \Big(\hat{\Theta}_a(\boldsymbol{\xi},\boldsymbol{\xi}') + \hat{\Theta}_b(\boldsymbol{\xi},\boldsymbol{\xi}') + \hat{\Theta}_{\mathbf{w}}(\boldsymbol{\xi},\boldsymbol{\xi}')\Big)\, d\boldsymbol{\xi}'\Big)\\
&= \mathcal{F}_{\mathbf{y}\to\boldsymbol{\xi}}\Big[\frac{\partial u_{\Theta_a}}{\partial m(\mathbf{y})}\Big](\boldsymbol{\xi}) + \mathcal{F}_{\mathbf{y}\to\boldsymbol{\xi}}\Big[\frac{\partial u_{\Theta_{\mathbf{w}}}}{\partial m(\mathbf{y})}\Big](\boldsymbol{\xi}) + \mathcal{F}_{\mathbf{y}\to\boldsymbol{\xi}}\Big[\frac{\partial u_{\Theta_b}}{\partial m(\mathbf{y})}\Big](\boldsymbol{\xi}),
\end{aligned} \tag{SM1.16}$$

where $\mathcal{F}\big[\frac{\partial u_{\Theta_a}}{\partial m(\mathbf{y})}\big](\boldsymbol{\xi}) = \mathbb{E}_a\big(\int_{\mathbb{R}^d} \frac{\partial u(\mathbf{x},t)}{\partial \hat{m}(\boldsymbol{\xi}')} \cdot \hat{\Theta}_a(\boldsymbol{\xi},\boldsymbol{\xi}')\, d\boldsymbol{\xi}'\big)$, and others are defined analogously. By applying Lemma SM1.2, the decomposed NTK components are given by:

$$\begin{aligned}
\hat{\Theta}_a(\boldsymbol{\xi},\boldsymbol{\xi}') &= \mathcal{F}_{\mathbf{x}\to\boldsymbol{\xi}}\big[\sigma(\mathbf{w}^\top\mathbf{x}+b)\big](\boldsymbol{\xi}) \cdot \overline{\mathcal{F}_{\mathbf{x}'\to\boldsymbol{\xi}'}\big[\sigma(\mathbf{w}^\top\mathbf{x}'+b)\big]}(\boldsymbol{\xi}')\\
&= \delta_{\mathbf{w}}(\boldsymbol{\xi})\delta_{\mathbf{w}}(\boldsymbol{\xi}')\mathcal{F}[\sigma]\Big(\frac{\boldsymbol{\xi}^\top\hat{\mathbf{w}}}{||\mathbf{w}||}\Big)\overline{\mathcal{F}[\sigma]\Big(\frac{\boldsymbol{\xi}'^\top\hat{\mathbf{w}}}{||\mathbf{w}||}\Big)} e^{\frac{2\pi i b(\boldsymbol{\xi}-\boldsymbol{\xi}')\hat{\mathbf{w}}}{||\mathbf{w}||}},\\
\hat{\Theta}_b(\boldsymbol{\xi},\boldsymbol{\xi}') &= \mathcal{F}_{\mathbf{x}\to\boldsymbol{\xi}}\big[a\sigma'(\mathbf{w}^\top\mathbf{x}+b)\big](\boldsymbol{\xi}) \cdot \overline{\mathcal{F}_{\mathbf{x}'\to\boldsymbol{\xi}'}\big[a\sigma'(\mathbf{w}^\top\mathbf{x}'+b)\big]}(\boldsymbol{\xi}')\\
&= \delta_{\mathbf{w}}(\boldsymbol{\xi})\delta_{\mathbf{w}}(\boldsymbol{\xi}')\mathcal{F}[a\sigma']\Big(\frac{\boldsymbol{\xi}^\top\hat{\mathbf{w}}}{||\mathbf{w}||}\Big)\overline{\mathcal{F}[a\sigma']\Big(\frac{\boldsymbol{\xi}'^\top\hat{\mathbf{w}}}{||\mathbf{w}||}\Big)} e^{\frac{2\pi i b(\boldsymbol{\xi}-\boldsymbol{\xi}')\hat{\mathbf{w}}}{||\mathbf{w}||}},\\
\hat{\Theta}_{\mathbf{w}}(\boldsymbol{\xi},\boldsymbol{\xi}') &= \mathcal{F}_{\mathbf{x}\to\boldsymbol{\xi}}\big[a\mathbf{x}\sigma'(\mathbf{w}^\top\mathbf{x}+b)\big](\boldsymbol{\xi}) \cdot \overline{\mathcal{F}_{\mathbf{x}'\to\boldsymbol{\xi}'}\big[a\mathbf{x}'\sigma'(\mathbf{w}^\top\mathbf{x}'+b)\big]}(\boldsymbol{\xi}')\\
&= -\frac{1}{4\pi^2}\nabla_{\boldsymbol{\xi}}\Big[\delta_{\mathbf{w}}(\boldsymbol{\xi})\mathcal{F}[a\sigma']\Big(\frac{\boldsymbol{\xi}^\top\hat{\mathbf{w}}}{||\mathbf{w}||}\Big)e^{\frac{2\pi i b\boldsymbol{\xi}^\top\hat{\mathbf{w}}}{||\mathbf{w}||}}\Big]\nabla_{\boldsymbol{\xi}'}\Big[\delta_{\mathbf{w}}(\boldsymbol{\xi}')\overline{\mathcal{F}[a\sigma']\Big(\frac{\boldsymbol{\xi}'^\top\hat{\mathbf{w}}}{||\mathbf{w}||}\Big)}e^{\frac{2\pi i b\boldsymbol{\xi}'^\top\hat{\mathbf{w}}}{||\mathbf{w}||}}\Big].
\end{aligned}$$

For any two test functions $\phi, \psi \in \mathcal{S}(\mathbb{R}^d)$, the decomposed neural sensitivity kernel can be recast into the form given in equation (SM1.6), i.e.,

$$\big\langle \hat{\Theta}, \phi \otimes \psi \big\rangle = \int_{\mathbb{R}^d} \hat{\Theta}(\boldsymbol{\xi},\boldsymbol{\xi}')\phi(\boldsymbol{\xi})\psi(\boldsymbol{\xi}')d\boldsymbol{\xi}d\boldsymbol{\xi}'. \tag{SM1.17}$$

Consider the general distributions for bias $b$ with zero-mean and variance $\sigma_b \gg 1$. Let $\eta = \frac{\boldsymbol{\xi}^\top \hat{\mathbf{w}}}{|\mathbf{w}|}$ and $\eta' = \frac{\boldsymbol{\xi}'^\top \hat{\mathbf{w}}}{|\mathbf{w}|}$, by the method of integration by parts and the test function whose decay at infinity is zero, for any test functions $\phi, \psi \in \mathcal{S}(\mathbb{R}^d)$, we have:

$$\begin{aligned}
\left\langle \hat{\Theta}_a, \phi \otimes \psi \right\rangle &= \int_{\mathbb{R}^{2d}} \phi(\eta \mathbf{w}) \psi(\eta' \mathbf{w}) \mathcal{F}[\sigma](\eta) \overline{\mathcal{F}[\sigma](\eta')} e^{2\pi i b(\eta - \eta')} d\eta d\eta' \\
\left\langle \hat{\Theta}_b, \phi \otimes \psi \right\rangle &= \int_{\mathbb{R}^{2d}} \phi(\eta \mathbf{w}) \psi(\eta' \mathbf{w}) \mathcal{F}[a\sigma'](\eta) \overline{\mathcal{F}[a\sigma'](\eta')} e^{2\pi i b(\eta - \eta')} d\eta d\eta' \\
\left\langle \hat{\Theta}_{\mathbf{w}}, \phi \otimes \psi \right\rangle &= \frac{1}{4\pi^2} \int_{\mathbb{R} \times \mathbb{R}} \nabla \phi(\eta \mathbf{w}) \cdot \nabla \psi(\eta' \mathbf{w}) \mathcal{F}[a\sigma'](\eta) \overline{\mathcal{F}[a\sigma'](\eta')} e^{2\pi i b(\eta - \eta')} d\eta d\eta'.
\end{aligned}$$

To deal with the expectation of neural parameters in equation (SM1.16) (e.g., $\mathbb{E}_b$ and $\mathbb{E}_{\mathbf{w}}$), we introduce the Laplace method and the integral under radial symmetry.

LEMMA SM1.3 (Laplace Method). *Let $b$ be a real-valued random variable such that $\mathbb{E}[b] = 0$ and $Var(b) = \sigma_b^2$ with $\sigma_b \gg 1$ and even probability density function $p_b$. Then, for Schwartz function $f \in \mathcal{S}(\mathbb{R}^2)$ and large variance $\sigma_b \gg 1$, we have*

$$\text{(SM1.18)} \quad I[f] = \mathbb{E}_b \left[ \int_{\mathbb{R}^2} f(\eta, \eta') e^{2\pi i b(\eta - \eta')} \, d\eta \, d\eta' \right] = p_b(0) \int_{\mathbb{R}} f(\eta, \eta) \, d\eta + \mathcal{O}(\sigma_b^{-3}).$$

*Proof.* Since $f \in \mathcal{S}(\mathbb{R}^2)$ and $|\varphi_b| \leq 1$, Fubini's theorem applies. Write $b = \sigma_b Z$ with $\mathbb{E}[Z] = 0$, $\mathrm{Var}(Z) = 1$, and even density $p_Z$, so that $\varphi_b(t) = \varphi_Z(\sigma_b t)$. Applying the change of variables $x = \eta$, $t = \sigma_b(\eta - \eta')$ (Jacobian $= 1/\sigma_b$) gives

$$\text{(SM1.19)} \quad \begin{aligned}
I[f] &= \int_{\mathbb{R}^2} f(\eta, \eta') \varphi_b\big(2\pi(\eta - \eta')\big) d\eta d\eta' \\
&= \frac{1}{\sigma_b} \int_{\mathbb{R}} \int_{\mathbb{R}} f\left(x, x - \frac{t}{\sigma_b}\right) \varphi_b\left(\frac{2\pi t}{\sigma_b}\right) dt \, dx \\
&= \frac{1}{\sigma_b} \int_{\mathbb{R}} \int_{\mathbb{R}} f\left(x, x - \frac{t}{\sigma_b}\right) \varphi_Z(2\pi t) \, dt \, dx.
\end{aligned}$$

By Taylor's theorem with Lagrange remainder, for each $x \in \mathbb{R}$ and $t \in \mathbb{R}$, Then, expanding $f(x, x - t/\sigma_b)$ in a Taylor series around $t = 0$, we have:

$$\text{(SM1.20)} \quad I[f] = \frac{1}{\sigma_b} \int_{\mathbb{R}} \int_{\mathbb{R}} \left( f(x,x) - \frac{t}{\sigma_b} \partial_2 f(x,x) + \frac{t^2}{2\sigma_b^2} \partial_2^2 f(x,x) + R \right) \varphi_b(\frac{2\pi t}{\sigma_b}) \, dx \, dt,$$

where let $I[f] = I_0 + I_1 + I_2 + I_R$. Then, we estimate each term as follows:
**(Leading-order term $I_0$)** Using the substitution $u = 2\pi t/\sigma_b$ (so $dt = \frac{\sigma_b}{2\pi} du$) and the Fourier inversion formula evaluated at zero, $p_b(0) = \frac{1}{2\pi} \int_{\mathbb{R}} \varphi_b(u) \, du$, we have

$$\text{(SM1.21)} \quad I_0 = \frac{1}{\sigma_b} \left( \int_{\mathbb{R}} f(x,x) \, dx \right) \left( \frac{\sigma_b}{2\pi} \int_{\mathbb{R}} \varphi_b(u) \, du \right) = p_b(0) \int_{\mathbb{R}} f(x,x) \, dx.$$

**(First-order term $I_1$)** Since $p_b$ is even, its characteristic function $\varphi_b$ is also even. Consequently, the integrand $t\varphi_b(2\pi t/\sigma_b)$ is an odd function of $t$. Thus, we have:

$$\text{(SM1.22)} \quad I_1 = -\frac{1}{\sigma_b^2} \left( \int_{\mathbb{R}} \partial_2 f(x,x) \, dx \right) \left( \int_{\mathbb{R}} t \, \varphi_b \left( \frac{2\pi t}{\sigma_b} \right) dt \right) = 0.$$

**(Second-order term $I_2$)** For $\sigma_b \gg 1$, we assume that $b$ belongs to a scale family, i.e., $b = \sigma_b Z$ where $Z$ is a standardized random variable with $\mathbb{E}[Z] = 0$, $\mathrm{Var}(Z) = 1$, and a fixed even density $p_Z$. Then, we have $\varphi_b(s) = \varphi_Z(\sigma_b s)$ and $\varphi_b(2\pi t/\sigma_b) = \varphi_Z(2\pi t)$ becomes independent of $\sigma_b$. Since both integrals are finite constants. Therefore,

$$I_2 = \frac{1}{2\sigma_b^3}\left(\int_{\mathbb{R}} \partial_2^2 f(x,x)\,dx\right)\left(\int_{\mathbb{R}} t^2\varphi_Z(2\pi t)\,dt\right) = \mathcal{O}(\sigma_b^{-3}). \tag{SM1.23}$$

**(Remainder term $I_R$)** Taylor's theorem with integral remainder gives $|R(x,t)| \leq \frac{|t|^3}{6\sigma_b^3}\sup_{\xi\in\mathbb{R}}|\partial_2^3 f(x,\xi)|$. Given the Schwartz property of $f$ and $\varphi_Z$, we have

$$|I_R| \leq \frac{1}{6\sigma_b^4}\int_{\mathbb{R}}\int_{\mathbb{R}} |t|^3\|\partial_2^3 f(x,\cdot)\|_{L^\infty}|\varphi_Z(2\pi t)|\,dtdx = \mathcal{O}(\sigma_b^{-4}). \tag{SM1.24}$$

Combining the estimates, we obtain $I[f] = p_b(0)\int_{\mathbb{R}} f(\eta,\eta)\,d\eta + \mathcal{O}(\sigma_b^{-3})$. ∎

*Remark* SM1.4. The diagonal concentration $\eta = \eta'$ stems from rapid oscillation cancellation away from the diagonal when $\sigma_b$ is large. The symmetry of $p_b$ is essential: it eliminates the $\mathcal{O}(\sigma_b^{-2})$ term in the expansion. If $p_b$ is not even, the linear term survives and the error degrades to $\mathcal{O}(\sigma_b^{-2})$. The lemma applies to any zero-mean distribution with a smooth, rapidly decaying density (e.g., Gaussian, symmetric uniform, or heavy-tailed symmetric laws truncated at large $\sigma_b$).

LEMMA SM1.5 (Radial Integral Identity). *Let $\mathbf{w} \in \mathbb{R}^d$ have a radially symmetric density $\rho(\mathbf{w}) = \phi(\|\mathbf{w}\|)$, and let $r = \|\mathbf{w}\|$ with density $\rho(r) = |S^{d-1}|r^{d-1}\phi(r)$, where $|S^{d-1}| = 2\pi^{d/2}/\Gamma(d/2)$, and $\eta = ||\boldsymbol{\xi}||/||\mathbf{w}||$. For any measurable $f$, we have*

$$\mathbb{E}_{\mathbf{w}}\left[\int_0^\infty f(\eta,\mathbf{w})\,d\eta\right] = \mathbb{E}_r\left[\frac{1}{|S^{d-1}|}\int_{\mathbb{R}^d}\frac{1}{r\|\boldsymbol{\xi}\|^{d-1}} f\left(\frac{\|\boldsymbol{\xi}\|}{r}, \frac{r}{\|\boldsymbol{\xi}\|}\boldsymbol{\xi}\right)d\boldsymbol{\xi}\right]. \tag{SM1.25}$$

*Proof.* Consider the change of variables $\boldsymbol{\xi} = \eta\mathbf{w}$ and $r = \|\mathbf{w}\|$. Then, we have $\eta = \|\boldsymbol{\xi}\|/r$ and $\mathbf{w} = (r\boldsymbol{\xi})/\|\boldsymbol{\xi}\|$. Parameterize $\mathbf{w} = (r\mathbf{w})/||\mathbf{w}||$ so that $d\mathbf{w} = r^{d-1}drd\sigma(\mathbf{w}/||\mathbf{w}||)$. The radial component of $\boldsymbol{\xi}$ satisfies $\|\boldsymbol{\xi}\| = \eta r$, hence $d\|\boldsymbol{\xi}\| = rd\eta$. Using the polar volume element $d\boldsymbol{\xi} = \|\boldsymbol{\xi}\|^{d-1}d\|\boldsymbol{\xi}\|d\sigma(\mathbf{w}/||\mathbf{w}||)$, we obtain

$$d\eta\,d\mathbf{w} = \frac{d\|\boldsymbol{\xi}\|}{r}\,r^{d-1}drd\sigma(\mathbf{w}/||\mathbf{w}||) = r^{d-2}\|\boldsymbol{\xi}\|^{-(d-1)}d\boldsymbol{\xi}\,dr. \tag{SM1.26}$$

Substituting this into the expectation and using the radial symmetry $\rho(\mathbf{w}) = \phi(r)$:

$$\begin{aligned}\mathbb{E}_{\mathbf{w}}\left[\int_0^\infty f(\eta,\mathbf{w})\,d\eta\right] &= \int_{\mathbb{R}^d}\int_0^\infty f(\eta,\mathbf{w})\rho(\mathbf{w})\,d\eta\,d\mathbf{w} \\ &= \int_0^\infty\int_{\mathbb{R}^d} f\left(\frac{\|\boldsymbol{\xi}\|}{r}, \frac{r}{\|\boldsymbol{\xi}\|}\boldsymbol{\xi}\right)\phi(r)\,\frac{r^{d-2}}{\|\boldsymbol{\xi}\|^{d-1}}\,d\boldsymbol{\xi}\,dr \\ &= \int_0^\infty\left(\frac{1}{|S^{d-1}|}\int_{\mathbb{R}^d}\frac{1}{r\|\boldsymbol{\xi}\|^{d-1}} f\left(\frac{\|\boldsymbol{\xi}\|}{r}, \frac{r}{\|\boldsymbol{\xi}\|}\boldsymbol{\xi}\right)d\boldsymbol{\xi}\right)\rho_r(r)\,dr,\end{aligned} \tag{SM1.27}$$

where we substituted $\phi(r) = \rho_r(r)/(|S^{d-1}|r^{d-1})$ [SM5]. ∎

Based on the above lemmas, we have the following key proposition. This proposition can be utilized to derive the explicit expression of NSK in the wavenumber domain, thereby unveiling the wavenumber modulation mechanism of NeurFWI.

LEMMA SM1.6. *Let $\boldsymbol{\xi}$ be the wavenumber, $\mathbf{w}$ be the neural weights with $\hat{\mathbf{w}} = \mathbf{w}/||\mathbf{w}||$, $b$ be the neural bias with $\sigma_b \gg 1$, and $L$ a the self-adjoint operator. Define $\eta = \boldsymbol{\xi}^\top\hat{\mathbf{w}}/||\mathbf{w}||$, $\eta' = \boldsymbol{\xi}'\hat{\mathbf{w}}/||\mathbf{w}||$, and $r = ||\mathbf{w}||$ be the norm of neural weights. Under the*

*Assumption 2.2, for any test functions $\phi, \psi \in \mathcal{S}(\mathbb{R}^d)$ and any temporal distribution $g \in \mathcal{S}'(\mathbb{R}^d)$, the integral formula defined below satisfies the following equality:*

$$
\begin{aligned}
I =&\mathbb{E}_{\boldsymbol{\theta}}\Big(\int_{\mathbb{R}^2} L[\phi(\eta\mathbf{w})]\cdot L[\psi(\eta'\mathbf{w})]\mathcal{F}[g](\eta)\overline{\mathcal{F}[g](\eta')}\exp(2\pi i b(\eta-\eta'))\Big)\\
=&\frac{p_b(0)}{|S^{d-1}|}\mathbb{E}_{a,r}\Big(\int_{\mathbb{R}^d}\frac{\phi(\boldsymbol{\xi})}{r||\boldsymbol{\xi}||^{d-1}}L\Big[\mathcal{F}[g](\frac{||\boldsymbol{\xi}||}{r})\cdot\mathcal{F}[g](-\frac{||\boldsymbol{\xi}||}{r})L[\psi(\boldsymbol{\xi})]\Big]d\boldsymbol{\xi}\Big)+\mathcal{O}(\sigma_b^{-3}),
\end{aligned}
$$

*where $p_b(0)$ denotes the density of $b$ at zero and $|S^{d-1}| = (2\pi^{d/2})/\,\Gamma(d/2)$ represents the surface area of the unit sphere in $\mathbb{R}^d$ (where $\Gamma$ is the Gamma distribution).*

*Proof.* Let $f(\eta,\eta') = L[\phi(\eta\mathbf{w})]\cdot L[\psi(\eta'\mathbf{w})]\mathcal{F}[g](\eta)\overline{\mathcal{F}[g](\eta')}$ and the change of variables $\eta\mathbf{w} = \boldsymbol{\xi}$ and $r = ||\mathbf{w}||$. By the Lemmas SM1.3 and SM1.5 we have

$$
\begin{aligned}
I &= \mathbb{E}_{\mathbf{w},a,\mathbf{b}}\Big(\int_{\mathbb{R}^2} L[\phi(\eta\mathbf{w})]\cdot L[\psi(\eta'\mathbf{w})]\mathcal{F}[g](\eta)\overline{\mathcal{F}[g](\eta')}\exp(2\pi i b(\eta-\eta'))\Big)\\
&= p_b(0)\cdot\mathbb{E}_{\mathbf{w},a}\Big(\int_{\mathbb{R}+} L[\phi(\eta\mathbf{w})]\cdot L[\psi(\eta\mathbf{w})]\mathcal{F}[g](\eta)\overline{\mathcal{F}[g](\eta)}d\eta\Big)+\mathcal{O}(\sigma_b^{-3})\\
&= \frac{p_b(0)}{|S^{d-1}|}\cdot\mathbb{E}_{a,r}\Big(\int_{\mathbb{R}^d}\frac{1}{r||\boldsymbol{\xi}||^{d-1}}\mathcal{F}[g](\frac{||\boldsymbol{\xi}||}{r})\overline{\mathcal{F}[g](\frac{||\boldsymbol{\xi}||}{r})}L[\phi(\boldsymbol{\xi})]L[\psi(\boldsymbol{\xi})]d\boldsymbol{\xi}\Big)+\mathcal{O}(\sigma_b^{-3})\\
&= \frac{p_b(0)}{|S^{d-1}|}\cdot\mathbb{E}_{a,r}\Big(\int_{\mathbb{R}^d}\phi(\boldsymbol{\xi})L\Big[\frac{1}{r||\boldsymbol{\xi}||^{d-1}}\mathcal{F}[g](\frac{||\boldsymbol{\xi}||}{r})\overline{\mathcal{F}[g](\frac{||\boldsymbol{\xi}||}{r})}L[\psi(\boldsymbol{\xi})]\Big]d\boldsymbol{\xi}\Big)+\mathcal{O}(\sigma_b^{-3}),
\end{aligned}
$$

where $|S^{d-1}|$ is the surface area of the unit sphere and the last equality holds since $L$ is a self-adjoint operator (i.e., $\int L[\phi](\boldsymbol{\xi})f(\boldsymbol{\xi})d\boldsymbol{\xi} = \int\phi(\boldsymbol{\xi})L[f](\boldsymbol{\xi})d\boldsymbol{\xi}$). ∎

By the Lemma SM1.6 and let $\mathcal{W}[g](\boldsymbol{\xi}) = \mathcal{F}[g](\frac{||\boldsymbol{\xi}||}{r})\cdot\mathcal{F}[g](-\frac{||\boldsymbol{\xi}||}{r})$, we have:

$$
\begin{aligned}
\mathbb{E}_{\boldsymbol{\theta}}\Big(\langle\hat{\Theta}_a,\phi\otimes\psi\rangle\Big) &=\frac{p_b(0)}{|S^{d-1}|\cdot||\boldsymbol{\xi}||^{d-1}}\mathbb{E}_r\Big(\int_{\mathbb{R}^d}\frac{\phi(\boldsymbol{\xi})}{r}\mathcal{W}[\sigma](\boldsymbol{\xi})\psi(\boldsymbol{\xi})d\boldsymbol{\xi}\Big)+\mathcal{O}(\sigma_b^{-3})\\
\mathbb{E}_{\boldsymbol{\theta}}\Big(\langle\hat{\Theta}_b,\phi\otimes\psi\rangle\Big) &=\frac{p_b(0)}{|S^{d-1}|\cdot||\boldsymbol{\xi}||^{d-1}}\mathbb{E}_{a,r}\Big(\int_{\mathbb{R}^d}\frac{\phi(\boldsymbol{\xi})}{r}\mathcal{W}[a\sigma'](\boldsymbol{\xi})\psi(\boldsymbol{\xi})d\boldsymbol{\xi}\Big)+\mathcal{O}(\sigma_b^{-3})\\
\mathbb{E}_{\boldsymbol{\theta}}\Big(\langle\hat{\Theta}_{\mathbf{w}},\phi\otimes\psi\rangle\Big) &=\frac{p_b(0)}{|S^{d-1}|\cdot||\boldsymbol{\xi}||^{d-1}}\mathbb{E}_{a,r}\Big(\int_{\mathbb{R}^d}\frac{\phi(\boldsymbol{\xi})}{r}\nabla\Big[\mathcal{W}[a\sigma'](\boldsymbol{\xi})\nabla\psi(\boldsymbol{\xi})\Big]d\boldsymbol{\xi}\Big)+\mathcal{O}(\sigma_b^{-3}),
\end{aligned}
$$

where $p_b(0)$ be the density of $b$ at zero, $|S^{d-1}| = \frac{2\pi^{d/2}}{\Gamma(d/2)}$ be the surface area of the unit sphere in $\mathbb{R}^d$, $r = ||\mathbf{w}||$ with the probability density $\rho_r(r) = |S^{d-1}|\rho_{\mathbf{w}}(r\mathbf{e}_1)r^{d-1}$, and the expectations are taken w.r.t. initial parameter distribution. Furthermore, taking $\psi$ to be the sensitivity kernel in the wavenumber domain, i.e., $\psi = \partial u/\partial m$. Hence,

$$
\begin{aligned}
\mathbb{E}_{\boldsymbol{\theta}}\langle\hat{\Theta},\phi\otimes\psi\rangle &= \frac{p_b(0)}{|S^{d-1}|\,\|\boldsymbol{\xi}\|^{d-1}}\int_{\mathbb{R}^d}\phi(\boldsymbol{\xi})\Bigg\{\mathbb{E}_{a,r}\Big[\frac{\mathcal{W}[\sigma](\boldsymbol{\xi})+\mathcal{W}[a\sigma'](\boldsymbol{\xi})}{r}\Big]\psi(\boldsymbol{\xi})\\
&\qquad\mathbb{E}_{a,r}\Big[\frac{1}{r}\nabla\big(\mathcal{W}[a\sigma'](\boldsymbol{\xi})\nabla\psi(\boldsymbol{\xi})\big)\Big]\Bigg\}d\boldsymbol{\xi}+\mathcal{O}(\sigma_b^{-3})\\
&= \frac{C}{\|\boldsymbol{\xi}\|^{d-1}}\Bigg\{\mathbb{E}_{a,r}\Big[\frac{\mathcal{W}[\sigma](\boldsymbol{\xi})+\mathcal{W}[a\sigma'](\boldsymbol{\xi})}{r}\Big]\mathcal{I}+\mathcal{L}_{\boldsymbol{\xi}}\Bigg\}\mathcal{F}\Big[\frac{\partial u}{\partial m(\mathbf{y})}\Big](\boldsymbol{\xi}),
\end{aligned}
\tag{SM1.28}
$$

where the expectations are taken with respect to initial parameter distribution, $C = p_b(0)\Gamma(d/2)/(2\pi^{d/2})$, $p_b(0)$ is the density of $b$, $\mathcal{I}$ is the identity operator, and $\mathcal{L}_{\boldsymbol{\xi}}[\cdot] = \nabla_{\boldsymbol{\xi}}\cdot\big(\mathbb{E}_{a,r}[\mathcal{W}[a\sigma']/r]\nabla_{\boldsymbol{\xi}}(\cdot)\big)$ is the wavenumber-domain weighted Laplacian operator.

**SM1.6. Proof of Corollary 3.10.** Consider the NNFWI with tanh activation function $\sigma(x) = \frac{e^x - e^{-x}}{e^x + e^{-x}}$ and its Fourier transform can be expressed as follows:

$$\begin{aligned} \mathcal{F}_{x\to\xi}\big[\sigma(x)\big](\xi) &= -\, i\pi \text{csch}(\pi^2\xi), \\ \mathcal{F}_{x\to\xi}\big[\sigma'(\mathbf{x})\big](\xi) &= 2\pi^2\xi\text{csch}(\pi^2\xi), \end{aligned} \tag{SM1.29}$$

where $\text{csch}(x) = \frac{2}{e^x - e^{-x}}$. Since $\delta(\boldsymbol{\xi}) = 0$ when $\boldsymbol{\xi} > 0$, we have

$$\begin{aligned} \mathcal{W}[\sigma] &= \mathcal{F}[\sigma](\frac{||\boldsymbol{\xi}||}{r}) \cdot \mathcal{F}[\sigma](-\frac{||\boldsymbol{\xi}||}{r}) = \pi^2\text{csch}^2(\pi^2\frac{||\boldsymbol{\xi}||}{r}), \\ \mathcal{W}[a\sigma'] &= \mathcal{F}[a\sigma'](\frac{||\boldsymbol{\xi}||}{r}) \cdot \mathcal{F}[a\sigma'](-\frac{||\boldsymbol{\xi}||}{r}) = 4a^2\pi^4\frac{||\boldsymbol{\xi}||^2}{r^2}\text{csch}^2(\pi^2\frac{||\boldsymbol{\xi}||}{r}), \end{aligned} \tag{SM1.30}$$

Substituting the above equation into the equation (3.8), we have

$$\begin{aligned} \mathcal{F}[\frac{\partial u_{\Theta_a}}{\partial m}](\boldsymbol{\xi}) &= \frac{\pi^2 p_b(0)}{|S^{d-1}|\cdot||\boldsymbol{\xi}||^{d-1}}\mathbb{E}_r\Big[\frac{1}{r}\text{csch}^2(\pi^2\frac{||\boldsymbol{\xi}||}{r})\Big]\mathcal{F}[\frac{\partial u}{\partial m}](\boldsymbol{\xi}), \\ \mathcal{F}[\frac{\partial u_{\Theta_b}}{\partial m}](\boldsymbol{\xi}) &= \frac{4\pi^4 p_b(0)}{|S^{d-1}|\cdot||\boldsymbol{\xi}||^{d-1}}\mathbb{E}_{a,r}\Big[\frac{a^2||\boldsymbol{\xi}||^2}{r^3}\text{csch}^2(\pi^2\frac{||\boldsymbol{\xi}||}{r})\Big]\mathcal{F}[\frac{\partial u}{\partial m}](\boldsymbol{\xi}), \\ \mathcal{F}[\frac{\partial u_{\Theta_\mathbf{w}}}{\partial m}](\boldsymbol{\xi}) &= \frac{4\pi^4 p_b(0)}{|S^{d-1}|\cdot||\boldsymbol{\xi}||^{d-1}}\nabla\cdot\Big(\mathbb{E}_{a,r}\Big[\frac{a^2||\boldsymbol{\xi}||^2}{r^3}\text{csch}^2(\pi^2\frac{||\boldsymbol{\xi}||}{r})\Big]\nabla\mathcal{F}[\frac{\partial u}{\partial m}](\boldsymbol{\xi})\Big), \end{aligned} \tag{SM1.31}$$

Substituting (SM1.31) into (SM1.16) and ignoring the complicated last term for simplifying expression, which arises from the evolution of parameter $\mathbf{w}$ and without which we can also explain the modulation behavior of NeurFWI. Then, we have

$$\mathcal{F}[\frac{\partial u_\Theta}{\partial m}](\boldsymbol{\xi}) \simeq C\cdot\mathbb{E}_{a,r}\left[\frac{\pi^2 r^2 + 4\pi^4 a^2\|\boldsymbol{\xi}\|^2}{r^3||\boldsymbol{\xi}||^{d-1}}\text{csch}^2\left(\frac{\pi^2\|\boldsymbol{\xi}\|}{r}\right)\right]\mathcal{F}[\frac{\partial u}{\partial m}](\boldsymbol{\xi}), \tag{SM1.32}$$

where $C = p_b(0)/|S^{d-1}|$. Then, we obtain equation (3.9).

**SM1.7. Proof of Corollary 3.12.** Consider the sinusoidal activation $\sigma(x) = \sin(\omega_0 x)\cdot\mathcal{X}_{[-T/2,T/2]}(x)$ and its Fourier transform can be expressed as follows:

$$\begin{aligned} \mathcal{F}_{\mathbf{x}\to\xi}[\sigma(x)](\xi) &= \frac{T}{2i}\Big[\text{sinc}\big(T(\xi - \frac{\omega_0}{2\pi})\big) - \text{sinc}\big(T(\xi + \frac{\omega_0}{2\pi})\big)\Big], \\ \mathcal{F}_{\mathbf{x}\to\xi}[\sigma'(x)](\xi) &= \frac{\omega_0 T}{2}\Big[\text{sinc}\big(T(\xi - \frac{\omega_0}{2\pi})\big) + \text{sinc}\big(T(\xi + \frac{\omega_0}{2\pi})\big)\Big], \end{aligned} \tag{SM1.33}$$

Since $\boldsymbol{\xi} > 0$, we can ignore $\text{sinc}(\frac{||\boldsymbol{\xi}||}{r} + \frac{\omega_0}{2\pi})$ and $\text{sinc}(-\frac{||\boldsymbol{\xi}||}{r} - \frac{\omega_0}{2\pi})$. Then, we have

$$\begin{aligned} \mathcal{W}[\sigma] &= \mathcal{F}[\sigma](\frac{||\boldsymbol{\xi}||}{r}) \cdot \mathcal{F}[\sigma](-\frac{||\boldsymbol{\xi}||}{r}) = \frac{T^2}{4}\text{sinc}^2\Big(T\Big(\frac{||\boldsymbol{\xi}||}{r} - \frac{\omega_0}{2\pi}\Big)\Big), \\ \mathcal{W}[a\sigma'] &= \mathcal{F}[a\sigma'](\frac{||\boldsymbol{\xi}||}{r}) \cdot \mathcal{F}[a\sigma'](-\frac{||\boldsymbol{\xi}||}{r}) = \frac{a^2\omega_0^2 T^2}{4}\text{sinc}^2\Big(T\Big(\frac{||\boldsymbol{\xi}||}{r} - \frac{\omega_0}{2\pi}\Big)\Big). \end{aligned} \tag{SM1.34}$$

Then, substituting the above equation into equation (3.8), we have

$$\begin{aligned} \mathcal{F}[\frac{\partial u_{\Theta_a}}{\partial m}](\boldsymbol{\xi}) &= \frac{p_b(0)}{|S^{d-1}|}\mathbb{E}_r\Big[\frac{T^2}{4r||\boldsymbol{\xi}||^{d-1}}\text{sinc}^2\Big(T\Big(\frac{||\boldsymbol{\xi}||}{r} - \frac{\omega_0}{2\pi}\Big)\Big)\Big]\mathcal{F}[\frac{\partial u}{\partial m}](\boldsymbol{\xi}), \\ \mathcal{F}[\frac{\partial u_{\Theta_b}}{\partial m}](\boldsymbol{\xi}) &= \frac{p_b(0)}{|S^{d-1}|}\mathbb{E}_{a,r}\Big[\frac{a^2\omega_0^2T^2}{4r||\boldsymbol{\xi}||^{d-1}}\text{sinc}^2\Big(T\Big(\frac{||\boldsymbol{\xi}||}{r} - \frac{\omega_0}{2\pi}\Big)\Big)\Big]\mathcal{F}[\frac{\partial u}{\partial m}](\boldsymbol{\xi}), \\ \mathcal{F}[\frac{\partial u_{\Theta_\mathbf{w}}}{\partial m}](\boldsymbol{\xi}) &= \frac{p_b(0)}{|S^{d-1}|}\nabla\cdot\Big(\mathbb{E}_{a,r}\Big[\frac{a^2\omega_0^2T^2}{4r||\boldsymbol{\xi}||^{d-1}}\text{sinc}^2\Big(T\Big(\frac{||\boldsymbol{\xi}||}{r} - \frac{\omega_0}{2\pi}\Big)\Big)\Big]\nabla\mathcal{F}[\frac{\partial u}{\partial m}](\boldsymbol{\xi})\Big), \end{aligned}$$

Substituting the above equation into (SM1.16) and ignoring the complicated last term for simplifying expression, which arises from the evolution of parameter $\mathbf{w}$ and without which we can also explain the modulation behavior of NeurFWI. Then, we have

$$\mathcal{F}\Big[\frac{\partial u_\Theta}{\partial m}\Big](\boldsymbol{\xi}) \simeq C\cdot\mathbb{E}_{a,r}\Big[\frac{T^2+a^2\omega_0^2T^2}{4||\boldsymbol{\xi}||^{d-1}r}\mathrm{sinc}^2\Big(T\Big(\frac{||\boldsymbol{\xi}||}{r}-\frac{\omega_0}{2\pi}\Big)\Big)\Big]\mathcal{F}\Big[\frac{\partial u}{\partial m}\Big](\boldsymbol{\xi}), \tag{SM1.35}$$

where $C = p_b(0)/|S^{d-1}|$. This completes the proof.

**SM1.8. Proof of Corollary 3.14.** Consider the real Gabor wavelet activation $\sigma(x) = \exp(-s_0^2x^2)\cos(\omega_0 x)$ and its Fourier transform can be expressed as follows:

$$\begin{aligned}\mathcal{F}_{x\to\xi}[\sigma(x)](\xi) =&\frac{\sqrt{\pi}}{2s_0}\left[\exp\Big(-\frac{(2\pi\xi-\omega_0)^2}{4s_0^2}\Big)+\exp\Big(-\frac{(2\pi\xi+\omega_0)^2}{4s_0^2}\Big)\right],\\ \mathcal{F}_{x\to\xi}[\sigma'(x)](\xi) =&\frac{j\pi^{3/2}\xi}{s_0}\left[\exp\Big(-\frac{(2\pi\xi-\omega_0)^2}{4s_0^2}\Big)+\exp\Big(-\frac{(2\pi\xi+\omega_0)^2}{4s_0^2}\Big)\right].\end{aligned}$$

For sufficiently large $\omega_0$, the two Gaussian peaks are well separated. Thus, for any fixed $\|\boldsymbol{\xi}\|/r$, only one of the exponentials is significant. Without loss of generality, we retain the term with $\exp\big(-(2\pi\xi_1-\omega_0)^2/(4s_0^2)\big)$ (the other term contributes negligibly for $\omega_0 > 0$ and $\xi_1 > 0$). Then, we have the following equalities hold:

$$\begin{aligned}\mathcal{W}[\sigma] =&\mathcal{F}[\sigma](\frac{||\boldsymbol{\xi}||}{r})\cdot\mathcal{F}[\sigma](-\frac{||\boldsymbol{\xi}||}{r}) \approx \frac{\pi}{4s_0^2}\exp\Big(-\frac{(2\pi||\boldsymbol{\xi}||/r-\omega_0)^2}{2s_0^2}\Big),\\ \mathcal{W}[a\sigma'] =&\mathcal{F}[a\sigma'](\frac{||\boldsymbol{\xi}||}{r})\cdot\mathcal{F}[a\sigma'](-\frac{||\boldsymbol{\xi}||}{r}) \approx \frac{\pi^3a^2||\boldsymbol{\xi}||^2}{s_0^2r^2}\exp\Big(-\frac{(2\pi||\boldsymbol{\xi}||/r-\omega_0)^2}{2s_0^2}\Big).\end{aligned}$$

Then, substituting the above equations into equation (3.8), we have

$$\begin{aligned}\mathcal{F}\Big[\frac{\partial u_{\Theta_a}}{\partial m}\Big](\boldsymbol{\xi}) =&\frac{p_b(0)}{|S^{d-1}|}\mathbb{E}_r\Big[\frac{\pi}{4s_0^2r||\boldsymbol{\xi}||^{d-1}}\exp\Big(-\frac{(2\pi||\boldsymbol{\xi}||/r-\omega_0)^2}{2s_0^2}\Big)\Big]\mathcal{F}\Big[\frac{\partial u}{\partial m}\Big](\boldsymbol{\xi}),\\ \mathcal{F}\Big[\frac{\partial u_{\Theta_b}}{\partial m}\Big](\boldsymbol{\xi}) =&\frac{p_b(0)}{|S^{d-1}|}\mathbb{E}_{a,r}\Big[\frac{\pi^3a^2||\boldsymbol{\xi}||^2}{s_0^2r^3||\boldsymbol{\xi}||^{d-1}}\exp\Big(-\frac{(2\pi||\boldsymbol{\xi}||/r-\omega_0)^2}{2s_0^2}\Big)\Big]\mathcal{F}\Big[\frac{\partial u}{\partial m}\Big](\boldsymbol{\xi}),\\ \mathcal{F}\Big[\frac{\partial u_{\Theta_\mathbf{w}}}{\partial m}\Big](\boldsymbol{\xi}) =&\frac{p_b(0)}{|S^{d-1}|}\nabla\cdot\Big(\mathbb{E}_{a,r}\Big[\frac{\pi^3a^2||\boldsymbol{\xi}||^2}{s_0^2r^3||\boldsymbol{\xi}||^{d-1}}\exp\Big(-\frac{(2\pi||\boldsymbol{\xi}||/r-\omega_0)^2}{2s_0^2}\Big)\Big]\nabla\mathcal{F}\Big[\frac{\partial u}{\partial m}\Big](\boldsymbol{\xi})\Big).\end{aligned}$$

Substituting the above equation into (SM1.16) and ignoring the complicated last term for simplifying expression, which arises from the evolution of parameter $\mathbf{w}$ and without which we can also explain the modulation behavior of NeurFWI. Then, we have

$$\mathcal{F}\Big[\frac{\partial u_\Theta}{\partial m}\Big](\boldsymbol{\xi}) \simeq C\cdot\mathbb{E}_{a,r}\Big[\frac{\pi r^3+4\pi^3a^2||\boldsymbol{\xi}||^2}{4s_0^2r^3||\boldsymbol{\xi}||^{d-1}}e^{\Big(-\frac{(2\pi||\boldsymbol{\xi}||/r-\omega_0)^2}{2s_0^2}\Big)}\Big]\mathcal{F}\Big[\frac{\partial u}{\partial m}\Big](\boldsymbol{\xi}), \tag{SM1.36}$$

where $C = p_b(0)/|S^{d-1}|$. This completes the proof.

**SM2. Theoretical Proof of WTK.** This section provides the detailed proof of the proposed Theorems, Propositions, and Corollaries about WTK theories.

**SM2.1. Proof of Proposition 4.1.** Consider the fixed wavefield data point $u(\mathbf{x}, t)$, the wavefield dynamic evolution is given by the chain rule:

$$\frac{\partial u(\mathbf{x},t)}{\partial\tau} = \int_\Omega \frac{\partial u(\mathbf{x},t)}{\partial m(\mathbf{y})}\frac{\partial m(\mathbf{y})}{\partial\tau}d\mathbf{y}. \tag{SM2.1}$$

Substituting the gradient flow dynamics from (2.3) into (SM2.1) yields:

$$\begin{aligned}\frac{\partial u(\mathbf{x},t)}{\partial\tau} &= -\int_\Omega \frac{\partial u(\mathbf{x},t)}{\partial m(\mathbf{y})}\Big[\int_T\int_\Omega \frac{\partial u(\mathbf{x}',t')}{\partial m(\mathbf{y})}\cdot\Delta D_m(\mathbf{x}',t')\,d\mathbf{x}'dt'\Big]d\mathbf{y}\\ &= -\int_T\int_\Omega\Big(\int_\Omega\frac{\partial u(\mathbf{x},t)}{\partial m(\mathbf{y})}\frac{\partial u(\mathbf{x}',t')}{\partial m(\mathbf{y})}d\mathbf{y}\Big)\Delta D_m(\mathbf{x}',t')\,d\mathbf{x}'dt',\end{aligned} \tag{SM2.2}$$

where the last step of the equation holds due to Fubini's theorem [SM9].

COROLLARY SM2.1. *The WTK $\mathbf{K}_N$ defined in Proposition 4.4 degrades into the WTK $\mathbf{K}$ defined in Proposition 4.1 without neural representation.*

*Proof.* Observed that change in the parameter $\boldsymbol{\theta}$ (i.e., the velocity model $m$) at the point $\mathbf{z}$ directly and solely affects the model $m_{\boldsymbol{\theta}}$ at the point $\mathbf{y} = \mathbf{z}$. Hence, the Fréchet derivative of the model $m_{\boldsymbol{\theta}}$ w.r.t. $\boldsymbol{\theta}$ at a point $\mathbf{z}$ is given by

$$\frac{dm_{\boldsymbol{\theta}}(\mathbf{y})}{d\boldsymbol{\theta}(\mathbf{z})} = \frac{d\boldsymbol{\theta}(\mathbf{y})}{d\boldsymbol{\theta}(\mathbf{z})} = \delta(\mathbf{y}-\mathbf{z}). \tag{SM2.3}$$

Substituting the above equality into the expression of NTK, i.e.,

$$\Theta(\mathbf{y},\mathbf{z};\boldsymbol{\theta}) = \int_\Omega \frac{\partial m_{\boldsymbol{\theta}}(\mathbf{y})}{\partial \boldsymbol{\theta}(\mathbf{x})} \cdot \frac{\partial m_{\boldsymbol{\theta}}(\mathbf{z})}{\partial \boldsymbol{\theta}(\mathbf{x})}\, d\mathbf{x} = \delta(\mathbf{y}-\mathbf{z}), \tag{SM2.4}$$

where the inner product is taken over the parameter space, resulting in the Dirac delta function, i.e., a perturbation at $\mathbf{x}$ affects the output only at $\mathbf{x}$. Hence,

$$\begin{aligned} \mathbf{K}_N\big((\mathbf{x},t),(\mathbf{x}',t')\big) &= \int_\Omega\int_\Omega \frac{\partial u(\mathbf{x},t)}{\partial m(\mathbf{y})}\frac{\partial u(\mathbf{x}',t')}{\partial m(\mathbf{z})}\delta(\mathbf{y}-\mathbf{z})\, d\mathbf{z}d\mathbf{y} \\ &= \int_\Omega \frac{\partial u(\mathbf{x},t)}{\partial m(\mathbf{y})}\frac{\partial u(\mathbf{x}',t')}{\partial m(\mathbf{y})} d\mathbf{y} = \mathbf{K}\big((\mathbf{x},t),(\mathbf{x}',t')\big), \end{aligned} \tag{SM2.5}$$

This completes the proof. ∎

**SM2.2. Proof of Theorem 4.6.** Starting from the definition in proposition 4.4, consider the variable substitution $\mathbf{u} = \mathbf{y}-\mathbf{z}, \mathbf{v} = \mathbf{y}$, we have the expression:

$$\begin{aligned} \hat{\mathbf{K}}_N\big((\mathbf{x},\omega),(\mathbf{x}',\omega')\big) &= \int_\Omega\int_\Omega \frac{\partial \hat{u}(\mathbf{x},t)}{\partial m(\mathbf{y})}\overline{\frac{\partial u(\mathbf{x}',t')}{\partial m(\mathbf{z})}}\Theta(\mathbf{y},\mathbf{z};\boldsymbol{\theta})\, d\mathbf{z}d\mathbf{y}, \\ &= \omega^2\omega'^2 \int_\Omega\int_\Omega \exp\big[i\big(\omega\mathbf{n}\cdot\mathbf{y} - \omega'\mathbf{n}'\cdot\mathbf{z}\big)\big]\kappa(\mathbf{y}-\mathbf{z})d\mathbf{z}d\mathbf{y} \\ &= \omega^2\omega'^2 \int_\Omega\int_\Omega \exp\big[i\big(\omega\mathbf{n}\cdot\mathbf{v} - \omega'\mathbf{n}'\cdot(\mathbf{v}-\mathbf{u})\big)\big]\kappa(\mathbf{u})d\mathbf{v}d\mathbf{u} \\ &= \underbrace{\omega^2\omega'^2\Big(\int_\Omega \exp\big[i\big(\omega\mathbf{n}\cdot\mathbf{v} - \omega'\mathbf{n}'\cdot\mathbf{v}\big)\big]d\mathbf{v}\Big)}_{\text{Wave tangent kernel for FWI}} \cdot \underbrace{\Big(\int_\Omega \exp(i\omega'\mathbf{n}'\cdot\mathbf{u})\kappa(\mathbf{u})d\mathbf{u}\Big)}_{\text{Frequency Modulation}} \\ &= \hat{\mathbf{K}}\big((\mathbf{x},\omega),(\mathbf{x}',\omega')\big)\cdot\hat{\kappa}(\omega'\mathbf{n}'), \end{aligned} \tag{SM2.6}$$

where $\hat{\kappa}$ denotes the frequency modulation term in the frequency domain.

**SM2.3. Proof of Proposition 4.8.** Since WNTK in equation 4.7 includes WTK in equation 4.2 when $\Theta(\mathbf{x},\mathbf{y};\boldsymbol{\theta}) = \delta(\mathbf{x}-\mathbf{y})$, we only consider WNTK in the next derivation and prove the properties of the integral operator as follows.

**(Symmetry).** Observe that the expression for $\Theta$ is symmetric in $(\mathbf{x},t)$ and $(\mathbf{x}',t')$ in Lemma SM1.1. Explicitly, interchanging $(\mathbf{x},t)$ and $(\mathbf{x}',t')$ for WTK yields:

$$\begin{aligned} \mathbf{K}_N\big((\mathbf{x},t),(\mathbf{x}',t')\big) &= \int_\Omega\int_\Omega \frac{\partial u(\mathbf{x},t)}{\partial m(\mathbf{y})}\frac{\partial u(\mathbf{x}',t')}{\partial m(\mathbf{z})}\Theta(\mathbf{y},\mathbf{z};\boldsymbol{\theta})d\mathbf{y}d\mathbf{z} \\ &= \int_\Omega\int_\Omega \frac{\partial u(\mathbf{x}',t')}{\partial m(\mathbf{z})}\frac{\partial u(\mathbf{x},t)}{\partial m(\mathbf{y})}\Theta(\mathbf{z},\mathbf{y};\boldsymbol{\theta})d\mathbf{z}d\mathbf{y} \\ &= \mathbf{K}_N\big((\mathbf{x}',t'),(\mathbf{x},t)\big) \end{aligned}$$

where $\Theta(\mathbf{y},\mathbf{z};\boldsymbol{\theta}) = \Theta(\mathbf{z},\mathbf{y};\boldsymbol{\theta})$ by the symmetry of $\Theta$. The symmetry of $\mathbf{K}_N$ makes $\mathcal{K}_N$ self-adjoint, i.e., $\langle\mathcal{K}_N[u],v\rangle = \langle v,\mathcal{K}_N[u]\rangle$ for any $v,u\in L^2(\Omega\times T)$.

**(Semi-Positive Definiteness).** To show semi-positive definiteness, we must verify that for any square-integrable function $f(\mathbf{x},t)$, the following double integral

$$\langle f, \mathcal{K}_N[f]\rangle = \int\int f(\mathbf{x},t)\mathbf{K}_N\big((\mathbf{x},t),(\mathbf{x}',t')\big)f(\mathbf{x}',t')d\mathbf{x}dt\mathbf{x}'dt' \geq 0. \tag{SM2.7}$$

Substituting the expression of WNTK in the equation 4.6, we have

$$\begin{aligned}
\langle f,\mathcal{K}_N[f]\rangle &= \iint f(\mathbf{x},t)\,\mathbf{K}_N\big((\mathbf{x},t),(\mathbf{x}',t')\big)f(\mathbf{x}',t')d\mathbf{x}dtd\mathbf{x}'dt' \\
&= \iiiint f(\mathbf{x},t)\Big[\int_\Omega\int_\Omega \frac{\partial u(\mathbf{x},t)}{\partial m(\mathbf{y})}\frac{\partial u(\mathbf{x}',t')}{\partial m(\mathbf{z})}\Theta(\mathbf{y},\mathbf{z};\boldsymbol{\theta})d\mathbf{y}d\mathbf{z}\Big]f(\mathbf{x}',t')d\mathbf{x}dtd\mathbf{x}'dt' \\
&= \int_{\Omega^2}\Theta(\mathbf{y},\mathbf{z};\boldsymbol{\theta})\left[\iint f(\mathbf{x},t)\frac{\partial u(\mathbf{x},t)}{\partial m(\mathbf{y})}d\mathbf{x}\,dt\right]\left[\iint f(\mathbf{x}',t')\frac{\partial u(\mathbf{x}',t')}{\partial m(\mathbf{z})}d\mathbf{x}'dt'\right]d\mathbf{y}d\mathbf{z} \\
&= \int_\Omega\int_\Omega \Theta(\mathbf{y},\mathbf{z};\boldsymbol{\theta})\,g(\mathbf{y})\,g(\mathbf{z})\,d\mathbf{y}\,d\mathbf{z} \geq 0,
\end{aligned} \tag{SM2.8}$$

where $g(\mathbf{y}) = \iint_{\Omega\times T} f(\mathbf{x},t)\frac{\partial u(\mathbf{x},t)}{\partial m(\mathbf{y})}d\mathbf{x}\,dt$ and $\Theta$ is semi-positive definite. Hence, we have $\langle f,\mathcal{K}_N[f]\rangle \geq 0$, proving that $\mathcal{K}_N$ is semi-positive definite.

**(Compactness)**. To show the compactness of the operator, we must verify that the WTK is square-integrable on the compact $\Omega\times T$. Under the Assumption 2.2 of neural network, the NTK $\Theta \in L^2(\Omega\times\Omega)$ when the activation function is smooth and continuous. Moreover, the sensitivity kernel is square-integrable on the compact $\Omega^2\times T$ under the assumption of wave equation in [SM3] holds. Then, we have

$$\begin{aligned}
I &= \int_{\Omega^2}\int_{T^2}\Big|\mathbf{K}_N\big((\mathbf{x},t),(\mathbf{x}',t')\big)\Big|^2 d\mathbf{x}dtd\mathbf{x}'dt \\
&= \int_{\Omega^2}\int_{T^2}\Big|\int_\Omega\int_\Omega \frac{\partial u(\mathbf{x},t)}{\partial m(\mathbf{y})}\frac{\partial u(\mathbf{x}',t')}{\partial m(\mathbf{z})}\Theta(\mathbf{y},\mathbf{z};\boldsymbol{\theta})\;d\mathbf{z}d\mathbf{y}\Big|^2 d\mathbf{x}dtd\mathbf{x}'dt \\
&\leq \int_{\Omega^2}\int_{T^2}\left(\int_{\Omega^2}\Big|\frac{\partial u(\mathbf{x},t)}{\partial m(\mathbf{y})}\frac{\partial u(\mathbf{x}',t')}{\partial m(\mathbf{z})}\Big|^2 d\mathbf{z}d\mathbf{y}\right)\left(\int_{\Omega^2}\Big|\Theta(\mathbf{y},\mathbf{z};\boldsymbol{\theta})\Big|^2 d\mathbf{z}d\mathbf{y}\right)d\mathbf{x}dtd\mathbf{x}'dt \\
&= ||\Theta||_{L^2(\Omega\times\Omega)}\int_{\Omega^2}\int_{T^2}\left(\int_{\Omega^2}\Big|\frac{\partial u(\mathbf{x},t)}{\partial m(\mathbf{y})}\Big|^2\Big|\frac{\partial u(\mathbf{x}',t')}{\partial m(\mathbf{z})}\Big|^2 d\mathbf{z}d\mathbf{y}\right)d\mathbf{x}dtd\mathbf{x}'dt \\
&= ||\Theta||_{L^2(\Omega\times\Omega)}\int_{\Omega^2}\left(\int_\Omega\int_T\Big|\frac{\partial u(\mathbf{x},t)}{\partial m(\mathbf{y})}\Big|^2 d\mathbf{x}dt\right)\left(\int_\Omega\int_T\Big|\frac{\partial u(\mathbf{x}',t')}{\partial m(\mathbf{z})}\Big|^2 d\mathbf{x}'dt'\right)d\mathbf{z}d\mathbf{y} \\
&= |\Theta||_{L^2(\Omega\times\Omega)}\left(\int_{\Omega^2}\int_T\Big|\frac{\partial u(\mathbf{x},t)}{\partial m(\mathbf{y})}\Big|^2 d\mathbf{x}d\mathbf{y}dt\right)^2 \leq \infty,
\end{aligned}$$

where the first inequality holds by the Cauchy-Schwarz inequality. Due to the compactness of $\Omega\times T$ and $\mathbf{K}_N$ is a continuous symmetric positive definite kernel, by Mercer's theorem, there exists a complete orthonormal system $\{\phi_k\}_{k=1}^\infty$ and non-negative eigenvalues $\{\lambda_k\}_{k=1}^\infty$ (for any square-integrable function $f$) such that

$$\mathcal{K}_N[f] = \sum_{k=1}^\infty \lambda_k\langle\phi_k, f\rangle\otimes\phi_k, \tag{SM2.9}$$

where the eigenvalue $\lambda_k \geq 0$ and eigenfunction $\phi_k$ is continuous for any $k\in\mathbb{R}$.

**SM2.4. Proof of Theorem 4.9.** Since $\mathcal{K}$ and $\mathcal{K}_N$ are self-adjoint, compact, and non-negative definite operators by Proposition 4.8 and Mercer's theorem, there exist eigenfunctions $\{\phi_k\}_{k=1}^\infty$ and non-negative eigenvalues $\{\lambda_k\}_{k=1}^\infty$ to decompose the WTK (see equation (SM2.9)). Then, the evolution of the data residual is given by

$$\frac{\partial u}{\partial\tau} = \frac{\partial\Delta D_m^\tau}{\partial\tau} = -\mathcal{K}_N\left[\Delta D_m^\tau\right] = -\sum_{k=1}^\infty\lambda_k\left\langle\Delta D_m^\tau,\phi_k\right\rangle\otimes\phi_k \tag{SM2.10}$$

Projecting both sides onto the eigenfunctions, i.e., taking the inner product:

$$\text{(SM2.11)}\qquad \begin{aligned} \Big\langle \frac{\partial}{\partial \tau}(\Delta D_m^\tau), \phi_m \Big\rangle &= \frac{\partial}{\partial \tau}\big\langle \Delta D_m^\tau, \phi_m \big\rangle = \Big\langle \sum_{k=1}^{\infty} \lambda_k \big\langle \Delta D_m^\tau, \phi_k \big\rangle \otimes \phi_k, \phi_m \Big\rangle \\ &= -\sum_{k=1}^{\infty} \lambda_k \Big\langle \Delta D_m^\tau, \phi_k \Big\rangle \Big\langle \phi_k, \phi_m \Big\rangle = -\lambda_m \Big\langle \Delta D_m^\tau, \phi_m \Big\rangle, \end{aligned}$$

where the last equality holds by the orthogonality of eigenfunctions $\langle \phi_k, \phi_m \rangle = \delta_{km}$. Solving the ordinary differential equation $\partial_\tau \big\langle \Delta D_m^\tau, \phi_m \big\rangle$ and reconstructing the wavefield data residual function, we have the following spectral data convergence

$$\text{(SM2.12)}\qquad \big\langle \Delta D_m^\tau, \phi_m \big\rangle = -\exp(-\lambda_m \tau)\big\langle \Delta D_m^0, \phi_m \big\rangle,\ \forall\ m \in [0, 1, \cdots],$$

where the eigenfunction $\phi$ controls the spectral space and its eigenvalue controls the reduction speed of the wavefield data residual in the spectral space.

**SM2.5. Proof of Theorem 4.10.** By Courant–Fischer–Weyl min-max principle, the eigenvalue of the compact self-adjoint operator $\mathcal{K}_i$ $(i = 1, 2)$ can be represented by the Rayleigh quotient. For $L^2(\Omega \times T)$ domain, the eigenvalue is given by:

$$\text{(SM2.13)}\qquad \lambda_j(\mathcal{K}_i) = \max_{\substack{S \subset L^2(\Omega \times T) \\ \dim(S) = j}} \min_{\substack{g \in S \\ g \neq 0}} \langle \mathcal{K}_i[g], g \rangle / |g|^2,$$

where $\lambda_j$ denotes the jth largest eigenvalue. Hence, comparing the magnitudes of the eigenvalues $\lambda_j(\mathcal{K}_1)$ and $\lambda_j(\mathcal{K}_2)$ is equivalent to comparing the magnitudes of the inner products $\langle \mathcal{K}_1[g], g \rangle$ and $\langle \mathcal{K}_2[g], g \rangle$. Let $\mathbf{U}(\mathbf{y}) = \int_{\Omega^2} \frac{\partial u(\mathbf{x},t)}{\partial m(\mathbf{y})} g(\mathbf{x}, t) d\mathbf{x} dt$ and the NTK satisfies $\Theta_1 \succeq \Theta_2$ (i.e., $\Theta_1 - \Theta_2$ is a positive semi-definite kernel). Then we have

$$\text{(SM2.14)}\quad \big\langle (\mathcal{K}_1 - \mathcal{K}_2)[g], g \big\rangle_{L^2(\Omega \times T)} = \int_\Omega \int_\Omega \mathbf{U}(\mathbf{y}) \big(\Theta_1 - \Theta_2\big)(\mathbf{y}, \mathbf{z}; \boldsymbol{\theta}) \mathbf{U}(\mathbf{z}) d\mathbf{y} d\mathbf{z} \geq 0,$$

where $g \in L^2(\Omega \times T)$. Then we have the following equalities hold:

$$\text{(SM2.15)}\qquad \begin{aligned} \mathcal{K}_1 \succeq \mathcal{K}_2 &\Leftrightarrow (\mathcal{K}_1 - \mathcal{K}_2) \text{ is positive semi-definite operator} \\ &\Leftrightarrow \big\langle (\mathcal{K}_1 - \mathcal{K}_2)[g], g \big\rangle_{L^2(\Omega \times T)} \geq 0 \quad \forall\ g \in L^2(\Omega \times T) \end{aligned}$$

where we obtain $\lambda_j(\mathcal{K}_1) \geq \lambda_j(\mathcal{K}_2), j = 1, 2$ by letting $g = \Delta D_m$ and the properties of positive semi-definite operator. This completes the proof.

**SM2.6. Proof of Theorem 5.1.** To simplify theoretical analysis, we drop the $n$ superscript to consider a single cell, with learnable weights $\theta_{11} = g(x_1, y_1), \theta_{12} = g(x_1, y_2), \theta_{21} = g(x_2, y_1), \theta_{22} = g(x_2, y_2)$. It follows that the gradient with respect to the weights inside this grid cell can be expressed as follows:

$$\text{(SM2.16)}\qquad \nabla_{\boldsymbol{\theta}} \tilde{g}(x, y) = \frac{1}{\Delta x \Delta y} \begin{bmatrix} \Delta x_2 \\ \Delta x_1 \end{bmatrix} \begin{bmatrix} \Delta y_2 & \Delta y_1 \end{bmatrix},$$

where the gradients of the grid parameters are independent as a result of concatenation. We therefore begin by computing the contribution from a single grid parameter, and the full kernel is subsequently obtained by summing the contributions from each grid cell. Let $\tilde{g}(\mathbf{x})$ be the output of the bilinear interpolation on a single grid cell, then the kernel element with the $i^{th}, j^{th}$ index of $\Theta_{mpe}$ is given by

$$
\begin{aligned}
\Theta_{mpe}^{(i,j)}(\mathbf{x}, \mathbf{x}') &= \mathbb{E}_{\boldsymbol{\theta}}\Big\langle \nabla_{\boldsymbol{\theta}}\sigma^*(\mathbf{x}, \boldsymbol{\theta}), \nabla_{\boldsymbol{\theta}}\sigma^*(\mathbf{x}', \boldsymbol{\theta})\Big\rangle \\
&= \mathbb{E}_{\boldsymbol{\theta}}\Big[\Big\langle \sigma\big(\mathbf{w}^\top \tilde{g}(\mathbf{x}) + b\big), \sigma\big(\mathbf{w}^\top \tilde{g}(\mathbf{x}') + b\big)\Big\rangle \\
&\quad + \Big\langle a\tilde{g}(\mathbf{x})\sigma'\big(\mathbf{w}^\top \tilde{g}(\mathbf{x}) + b\big), a\tilde{g}(\mathbf{x})\sigma'\big(\mathbf{w}^\top \tilde{g}(\mathbf{x}') + b\big)\Big\rangle \\
&\quad + \Big\langle a\sigma'\big(\mathbf{w}^\top \tilde{g}(\mathbf{x}) + b\big), a\sigma'\big(\mathbf{w}^\top \tilde{g}(\mathbf{x}') + b\big)\Big\rangle \\
&\quad + \Big\langle a\mathbf{w}\tilde{g}'(\mathbf{x})\sigma'\big(\mathbf{w}^\top \tilde{g}(\mathbf{x}) + b\big), a\mathbf{w}\tilde{g}(\mathbf{x}')\sigma'\big(\mathbf{w}^\top \tilde{g}(\mathbf{x}') + b\big)\Big\rangle\Big],
\end{aligned}
\tag{SM2.17}
$$

where the first three terms correspond to the original MLP with the input in the embedding space of the MPE instead of the original coordinate space. The last term is a new term induced by the learnable parameters in the grid. We denote $\Theta_{mlp}^{(i,j)}, \Theta_{grid}^{(i,j)}$ for the embedded kernel and the grid-based kernel, respectively. Moreover, the multigrid parametric encoding includes the original coordinate input in the Euclidean space, which induces the original NTK $\Theta_{inr}$ for neural representation without encoding. As layers along the $n_g$ dimension and trainable parameters along the $n_f$ dimension are independent, then we have the following kernel for the MPE-FWI method:

$$
\Theta_{mpe}(\mathbf{x}, \mathbf{x}') = \Theta_{inr}(\mathbf{x}, \mathbf{x}') + \sum_{i=1}^{n_g}\sum_{j=1}^{n_f} \Theta_{mlp}^{(i,j)}(\mathbf{x}, \mathbf{x}') + \sum_{i=1}^{n_g}\sum_{j=1}^{n_f} \Theta_{grid}^{(i,j)}(\mathbf{x}, \mathbf{x}'). \tag{SM2.18}
$$

Hence, we have $\Theta_{mpe} = \Theta_{inr} + \Theta^+$, where $\Theta^+$ is the positive semi-definite kernel produced by the sum over all learnable parameters in the grid. By the Weyl's inequality [SM12], we have $\lambda_i(\Theta_{inr}) \leq \lambda_i(\Theta_{inr}) + \lambda_n(\Theta^+) \leq \lambda_i(\Theta_{mpe})$. These inequalities demonstrate that the NTK eigenvalue decay rate of MPE-FWI is slower than the NTK eigenvalue decay rate of IFWI using the same activation functions [SM1], showing that MPE-FWI obtains more high-resolution inversion results.

**SM2.7. Proof of Theorem 5.3.** We compare the NTK using different neural networks. Let $\Theta^{HR}(\mathbf{x}, \mathbf{x}'), \Theta^{(1)}(\mathbf{x}, \mathbf{x}'), \Theta^{(2)}(\mathbf{x}, \mathbf{x}')$ be NTK for representation $\mathbf{F}_{\boldsymbol{\theta}}^1, \mathbf{F}_{\boldsymbol{\theta}}^2, \mathbf{F}_{\boldsymbol{\theta}}^{(HR)}$, respectively. These neural tangent kernels can be expressed as follows:

$$
\begin{aligned}
\Theta^{HR}(\mathbf{x}, \mathbf{x}') &= \mathbb{E}_{\boldsymbol{\theta}}\Big\langle \frac{\partial \mathbf{F}_{\boldsymbol{\theta}}(\mathbf{x})}{\partial \boldsymbol{\theta}}, \frac{\partial \mathbf{F}_{\boldsymbol{\theta}}(\mathbf{x}')}{\partial \boldsymbol{\theta}}\Big\rangle = \mathbb{E}_{\boldsymbol{\theta}}\Big[\Big\langle \frac{\partial \mathbf{F}_{\boldsymbol{\theta}}(\mathbf{x})}{\partial \boldsymbol{\theta}_0}, \frac{\partial \mathbf{F}_{\boldsymbol{\theta}}(\mathbf{x}')}{\partial \boldsymbol{\theta}_0}\Big\rangle \\
&\quad + a\Big\langle \frac{\partial \mathbf{F}_{\boldsymbol{\theta}}(\mathbf{x})}{\partial \boldsymbol{\theta}_1}, \frac{\partial \mathbf{F}_{\boldsymbol{\theta}}(\mathbf{x}')}{\partial \boldsymbol{\theta}_1}\Big\rangle + (1-a)\Big\langle \frac{\partial \mathbf{F}_{\boldsymbol{\theta}}(\mathbf{x})}{\partial \boldsymbol{\theta}_2}, \frac{\partial \mathbf{F}_{\boldsymbol{\theta}}(\mathbf{x}')}{\partial \boldsymbol{\theta}_2}\Big\rangle\Big], \\
\Theta^{(1)}(\mathbf{x}, \mathbf{x}') &= \mathbb{E}_{\boldsymbol{\theta}}\Big[\Big\langle \frac{\partial \mathbf{F}_{\boldsymbol{\theta}}(\mathbf{x})}{\partial \boldsymbol{\theta}_0}, \frac{\partial \mathbf{F}_{\boldsymbol{\theta}}(\mathbf{x}')}{\partial \boldsymbol{\theta}_0}\Big\rangle + \Big\langle \frac{\partial \mathbf{F}_{\boldsymbol{\theta}}(\mathbf{x})}{\partial \boldsymbol{\theta}_1}, \frac{\partial \mathbf{F}_{\boldsymbol{\theta}}(\mathbf{x}')}{\partial \boldsymbol{\theta}_1}\Big\rangle\Big], \\
\Theta^{(2)}(\mathbf{x}, \mathbf{x}') &= \mathbb{E}_{\boldsymbol{\theta}}\Big[\Big\langle \frac{\partial \mathbf{F}_{\boldsymbol{\theta}}(\mathbf{x})}{\partial \boldsymbol{\theta}_0}, \frac{\partial \mathbf{F}_{\boldsymbol{\theta}}(\mathbf{x}')}{\partial \boldsymbol{\theta}_0}\Big\rangle + \Big\langle \frac{\partial \mathbf{F}_{\boldsymbol{\theta}}(\mathbf{x})}{\partial \boldsymbol{\theta}_2}, \frac{\partial \mathbf{F}_{\boldsymbol{\theta}}(\mathbf{x}')}{\partial \boldsymbol{\theta}_2}\Big\rangle\Big].
\end{aligned}
\tag{SM2.19}
$$

Next, we calculate the difference of the above neural tangent kernel as follows:

$$
\begin{aligned}
(\Theta^{(1)} - \Theta^{HR})(\mathbf{x}, \mathbf{x}') &= (1-a)\mathbb{E}_{\boldsymbol{\theta}}\Big[\Big\langle \frac{\partial \mathbf{F}_{\boldsymbol{\theta}}(\mathbf{x})}{\partial \boldsymbol{\theta}_1}, \frac{\partial \mathbf{F}_{\boldsymbol{\theta}}(\mathbf{x}')}{\partial \boldsymbol{\theta}_1}\Big\rangle - \Big\langle \frac{\partial \mathbf{F}_{\boldsymbol{\theta}}(\mathbf{x})}{\partial \boldsymbol{\theta}_2}, \frac{\partial \mathbf{F}_{\boldsymbol{\theta}}(\mathbf{x}')}{\partial \boldsymbol{\theta}_2}\Big\rangle\Big], \\
(\Theta^{HR} - \Theta^{(2)})(\mathbf{x}, \mathbf{x}') &= a\mathbb{E}_{\boldsymbol{\theta}}\Big[\Big\langle \frac{\partial \mathbf{F}_{\boldsymbol{\theta}}(\mathbf{x})}{\partial \boldsymbol{\theta}_1}, \frac{\partial \mathbf{F}_{\boldsymbol{\theta}}(\mathbf{x}')}{\partial \boldsymbol{\theta}_1}\Big\rangle - \Big\langle \frac{\partial \mathbf{F}_{\boldsymbol{\theta}}(\mathbf{x})}{\partial \boldsymbol{\theta}_2}, \frac{\partial \mathbf{F}_{\boldsymbol{\theta}}(\mathbf{x}')}{\partial \boldsymbol{\theta}_2}\Big\rangle\Big].
\end{aligned}
$$

Let $\mathbf{K}_1 = \left\langle \frac{\partial \mathbf{F}_{\boldsymbol{\theta}}(\mathbf{x})}{\partial \boldsymbol{\theta}_1}, \frac{\partial \mathbf{F}_{\boldsymbol{\theta}}(\mathbf{x}')}{\partial \boldsymbol{\theta}_1} \right\rangle$ and $\mathbf{K}_2 = \left\langle \frac{\partial \mathbf{F}_{\boldsymbol{\theta}}(\mathbf{x})}{\partial \boldsymbol{\theta}_2}, \frac{\partial \mathbf{F}_{\boldsymbol{\theta}}(\mathbf{x}')}{\partial \boldsymbol{\theta}_2} \right\rangle$ be the NTK induced by parameters $\boldsymbol{\theta}_1, \boldsymbol{\theta}_2$, respectively. According to the eigenvalue decay rate of $\mathbf{F}^1_{\boldsymbol{\theta}}(\mathbf{x})$ and $\mathbf{F}^2_{\boldsymbol{\theta}}(\mathbf{x})$, i.e., $\lambda_j(\Theta^{(1)}) \geq \lambda_j(\Theta^{(2)})$ for any index $j$. We have $\mathbf{K}_1 \succeq \mathbf{K}_2$, i.e., all the above NTKs differences are semi-definite kernels. Hence, these NTKs are such that

$$\Theta^{(1)} \succeq \Theta^{HR} \succeq \Theta^{(2)}, \tag{SM2.20}$$

where we can obtain $\mathbf{K}^{(1)}_{wtk} \succeq \mathbf{K}^{(HR)}_{wtk} \succeq \mathbf{K}^{(2)}_{wtk}$ by the Theorem 4.10.

**SM3. Experimental Settings and Results.** This section provides detailed descriptions of the inversion (e.g., the 2D modified Overthrust model), along with the optimization settings and neural hyperparameter settings. Moreover, we present experimental results to validate the proposed selection criterion for the scaling factor.

**SM3.1. Neural Hyperparameters.** We carefully select widely used conventional FWI methods (i.e., ADFWI and MS-FWI) and existing state-of-the-art Neur-FWI methods (i.e., IFWI and LR-FWI) to compare with our proposed novel Neur-FWI methods (i.e., IG-FWI and LRG-FWI). Next, we provide further explanations and implementation details for each baseline and the proposed methods.

- (**Hyperparameter settings of IFWI**) For IFWI, we use the sine activation function with frequency hyperparameter $\omega_0 = 30$ (where $\omega_0$ controls the frequency of the sine function) for MLP, and the depth of the MLP is set to 4. The number of neurons of the MLP is set to 128, following the recommended settings in the paper [SM10]. For larger-scale FWI scenarios (i.e., BP and Foothills models), the number of MLP neurons is set to 256.
- (**Hyperparameter settings of LR-FWI**) For the LR-FWI method [SM2], we employ the same sine activation function as IFWI. The rank of the low-rank matrix factorization is set to half of the model dimension. The depth of the MLPs is set to 3, and the number of neurons in the MLPs is set to 128.
- (**Hyperparameter settings of MPE-FWI**) The MPE-FWI method utilizes a multi-resolution hash grid [SM8] encoding with 16 levels, a base resolution of 50, and a per-level scale factor of 1.05. This feature grid is combined with a compact MLP with two hidden layers of 64 neurons each.
- (**Hyperparameter settings of IG-FWI**) The proposed IG-FWI employs a hybrid architecture, combining an MPE (base resolution 50 with 16 levels and 1.05 per-level scale factor) with a sinusoidal INR (2 layers of 128 neurons with $\omega_0 = 30$). These extracted features are subsequently fused and processed by a compact MLP (2 layers of 64 neurons) to generate the velocity models.
- (**Hyperparameter settings of LRG-FWI**) The proposed LRG-FWI employs a hybrid architecture, combining an MPE (base resolution 50 with 16 levels and 1.05 per-level scale factor) with a low rank sinusoidal INR (2 layers of 128 neurons with $\omega_0 = 30$ and the output channel is 64). These extracted features are subsequently fused by a compact MLP (2 layers of 64 neurons).

The Adam optimizer is used to optimize the learnable parameters in all methods. The learning rate is set to 10 for conventional FWI methods and 0.0001 for Neur-FWI methods (0.001 for the hash encoding module). All numerical experiments are conducted using an RTX 3090 GPU, leveraging the PyTorch framework.

**SM3.2. Simple Stratified Model (Fig. 2).** The simple two-layer velocity model (2 km/s above 0.25 km depth and 3 km/s below 0.25 km depth) is discretized on a $100 \times 50$ grid (10 m spacing). The seismic wavefield generated by a 10 Hz Ricker wavelet source function at the center of the surface is recorded at the surface, and seismograms are recorded for 1 second with a 2 ms sampling interval. The initial model is a constant velocity model with 3km/s. The sensitivity kernels are calculated using automatic differentiation and Fourier transformation. The $L^2$ misfits are calculated using the seismic signal at 0.75 km with absorbing and free boundary conditions.

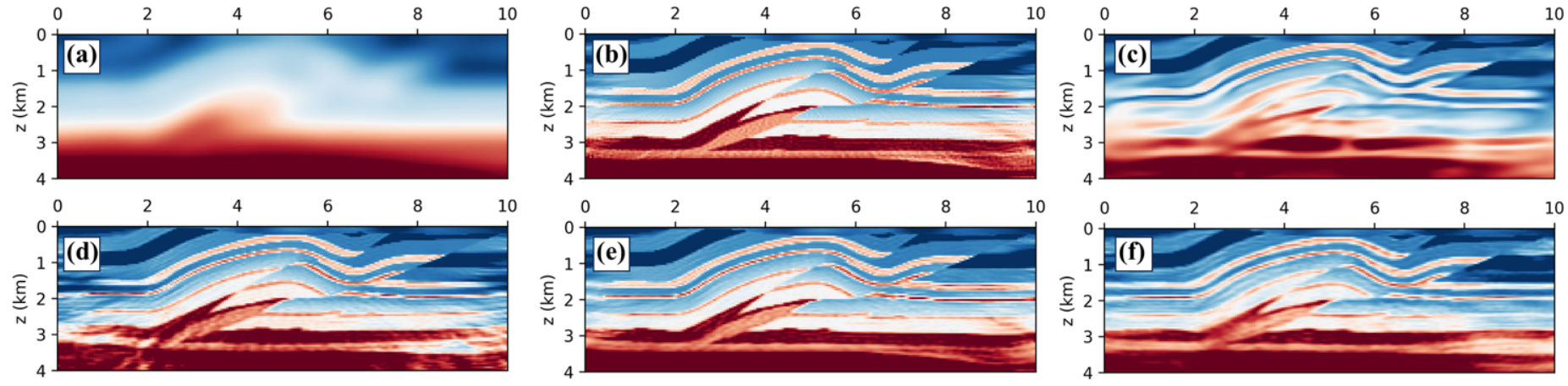

FIG. SM1. *Inversion results using different FWI methods on the Overthrust model. (a) Smooth initial model. (b) Conventional FWI. (c) IFWI. (d) MPE-FWI. (e) IG-FWI. (f) LRG-FWI.*

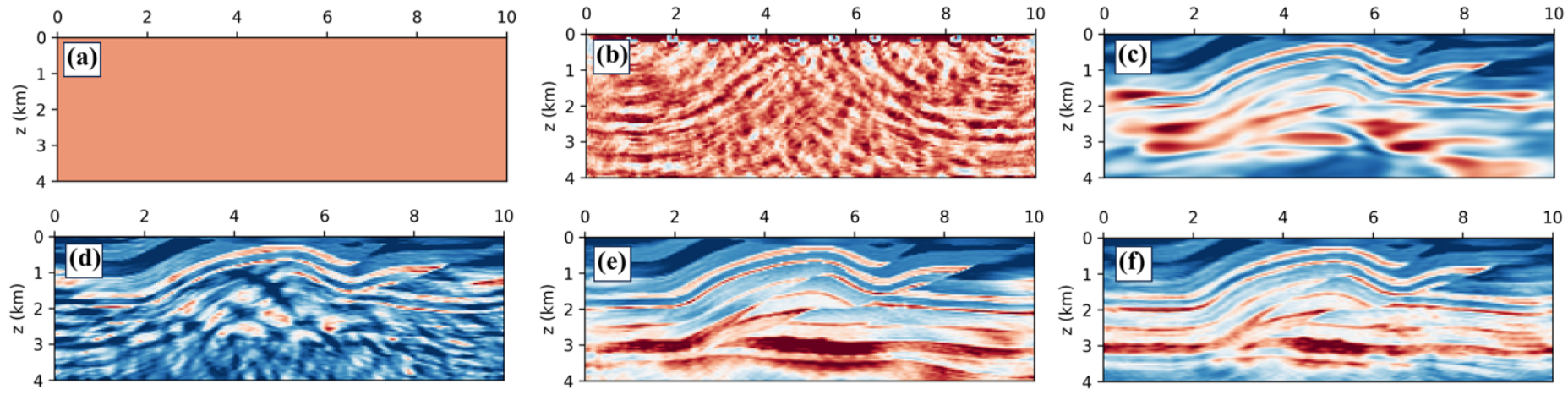

FIG. SM2. *Inversion results using different FWI methods on the Overthrust model. (a) Constant initial model. (b) Conventional FWI. (c) IFWI. (d) MPE-FWI. (e) IG-FWI. (f) LRG-FWI.*

**SM3.3. Overthrust Model (Fig. 3).** The 2D modified SEG/EAGE Overthrust velocity model with a grid size of $200 \times 80$ and 50-m cells. The 10-shot gathers are collected with a shot interval of 0.9 km at the surface, employing a Ricker wavelet source with a dominant frequency of 10 Hz. Receivers are placed on the surface, occupying each grid point at 50-m intervals. A free surface at the top and a perfectly matched layer absorbing condition are applied to the other boundaries. The time sample interval for wave propagation is 4 ms, and 1000 time samples are recorded. Moreover, we set the learning rate to 10 and 0.0001 for ADFWI and NeurFWI with the Adam optimizer, respectively. Here, other neural network hyperparameter of NeurFWI follows the standard settings in Section SM3.1. Here, we provide the inversion results using ADFWI, IFWI, MPE-FWI, IG-FWI, and LRG-FWI methods with smooth and constant initial velocity models, as shown in Figs. SM1 and SM2.

These inversion results demonstrate that ADFWI can achieve high-resolution inversion results using the smooth initial model, while falling into the local minimum using the constant initial model. In contrast, IFWI can stably invert the low-wavenumber background model using both smooth and constant initial models, at the cost of slow high-resolution convergence. The proposed MPE-FWI method can accelerate the high-wavenumber convergence due to its slower eigenvalue decay rate of WTK, but it still lacks inversion stability when using the constant initial model. Designing suitable and balanced eigenvalue decay rates using the proposed hybrid representation, including IG-FWI and LRG-FWI, can maintain the inversion stability while enhancing the high-wavenumber convergence speed.

**SM3.4. Hyperparameter Tests (Fig. 6).** To validate the Corollaries 3.10, 3.12, and 3.14 and guide the hyperparameter tuning of NeurFWI, we test the IFWI and WinFWI using different hyperparameters of activation functions and neural weights. For IFWI, we test different activation frequencies $\omega = 20, 30, 40$ and the variance of the weights $\sigma_{\mathbf{w}} = 0.5\sigma_0, 1.5\sigma_0, 2\sigma_0$ (where $\sigma_0$ denotes the standard variance), as shown in Fig. SM3. These inversion results demonstrate that the high-frequency components of the IFWI's inversion result increase as the frequency of the activation function or the variance of the weights increases. This phenomenon is consistent with our theoretical claims and results. Moreover, the high-frequency components of the WinFWI's inversion result also increase as frequency $\omega$, bandwidth $s_0$, and the output scaling. This phenomenon is consistent with our theoretical claims and results.

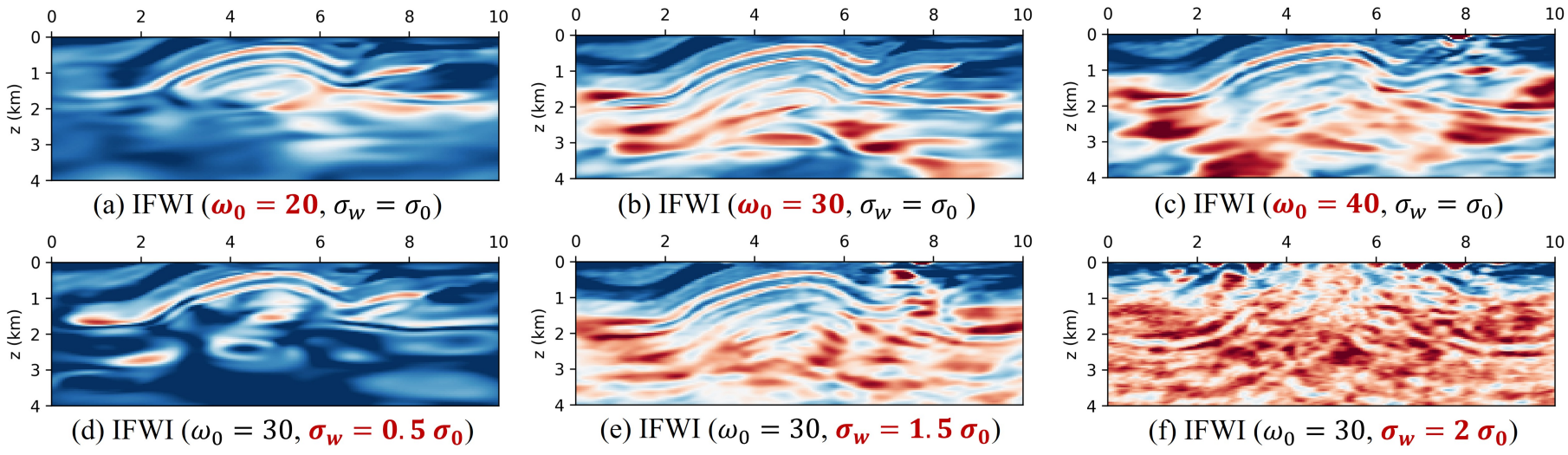

Fig. SM3. *The inversion results using IFWI with different frequency hyperparameters, starting from a constant initial model. (a)-(c) Inversion results using $\sigma_{\mathbf{w}} = \sigma_0$ with $\omega_0 = 20, 30, 40$, respectively. (d)-(e) Inversion results using $\omega = 30$ with $\sigma_{\mathbf{w}} = 0.5\sigma_0, 1.5\sigma_0, 2\sigma_0$, respectively.*

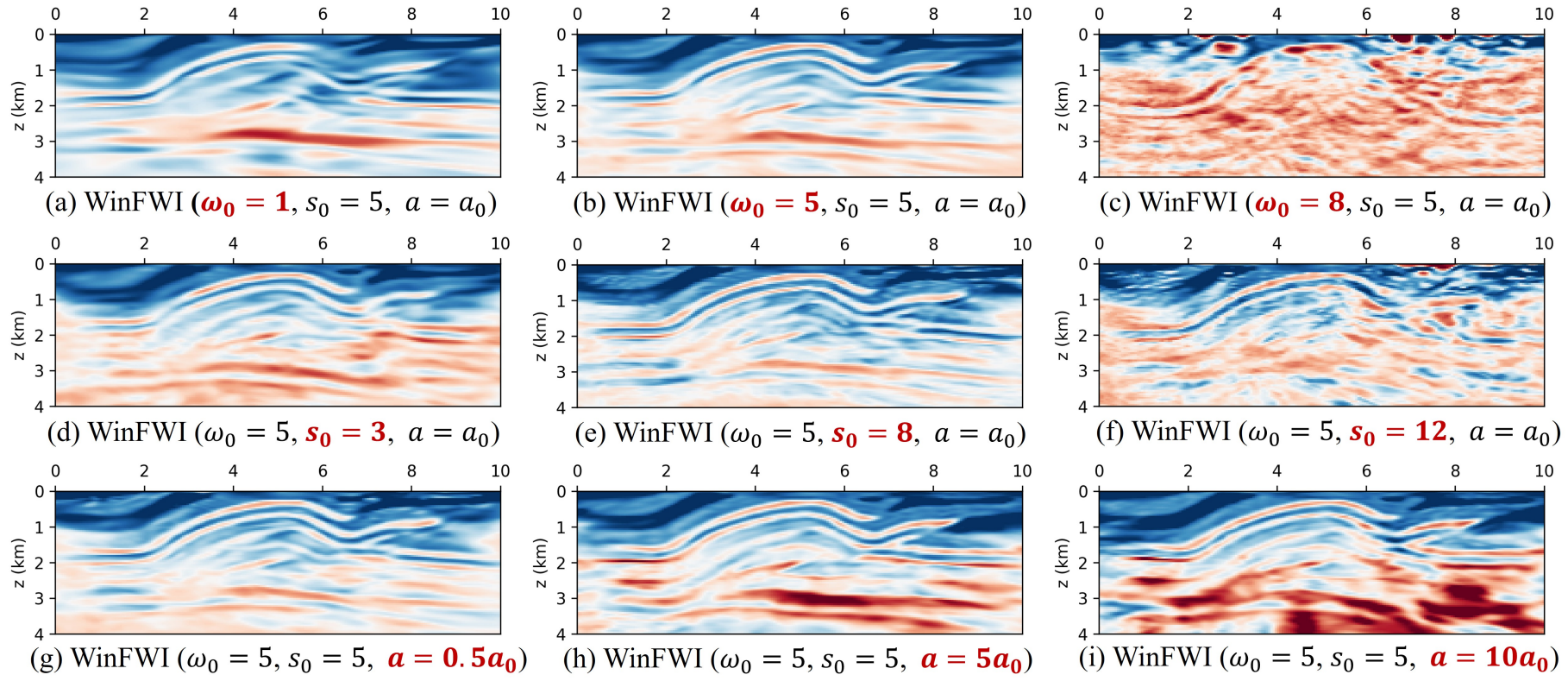

Fig. SM4. *The inversion results using WinFWI with different hyperparameters, starting from a constant initial model. (a)-(c) $s_0 = 5, a = a_0$ with $\omega = 1, 5, 8$, respectively. (d)-(f) $\omega_0 = 5, a = a_0$ with $s_0 = 3, 8, 12$, respectively. (g)-(i) $w_0 = 5, s_0 = 5$ with $a = 0.5a_0, 5a_0, 10a_0$, respectively.*

**SM3.5. Chevron Results.** Additional inversion results on Chevron model using ADFWI, LR-FWI, and the proposed LRG-FWI methods, as shown in Fig. SM5. ADFWI obtains the inversion result with abundant wrong high-frequency components due to inaccurate initial model and seismic noise pollution. Both our previously proposed LR-FWI and the proposed LRG-FWI methods can recover the low-wavenumber background velocity model, while achieving the high-resolution inversion results at the shallow and deep velocity structures, which can be attributed to their suitable eigenvalue decay rate of WTK. Moreover, the well log of Chevron model in Fig. SM5 further validate the performance and reliability of out proposed NeurFWI methods.

Since the ground truth was unavailable, we evaluated the results by comparing common-source shot gathers (Fig. SM6) using the initial model and IG-FWI inver-sion result. The results show that the initial value waves under the initial velocity model deviate from the observed seismic data, while the proposed FG-FWI method can accurately fit the first arrival waves between the simulated seismic data and the observed seismic data. Moreover, the reflected wave of synthetic seismic data using FG-FWI can also match that of observed data. These improvements achieved by FG-FWI can be attributed to its more suitable eigenvalue decay rate of WTK, which enhances the stability and accuracy of the inversion.

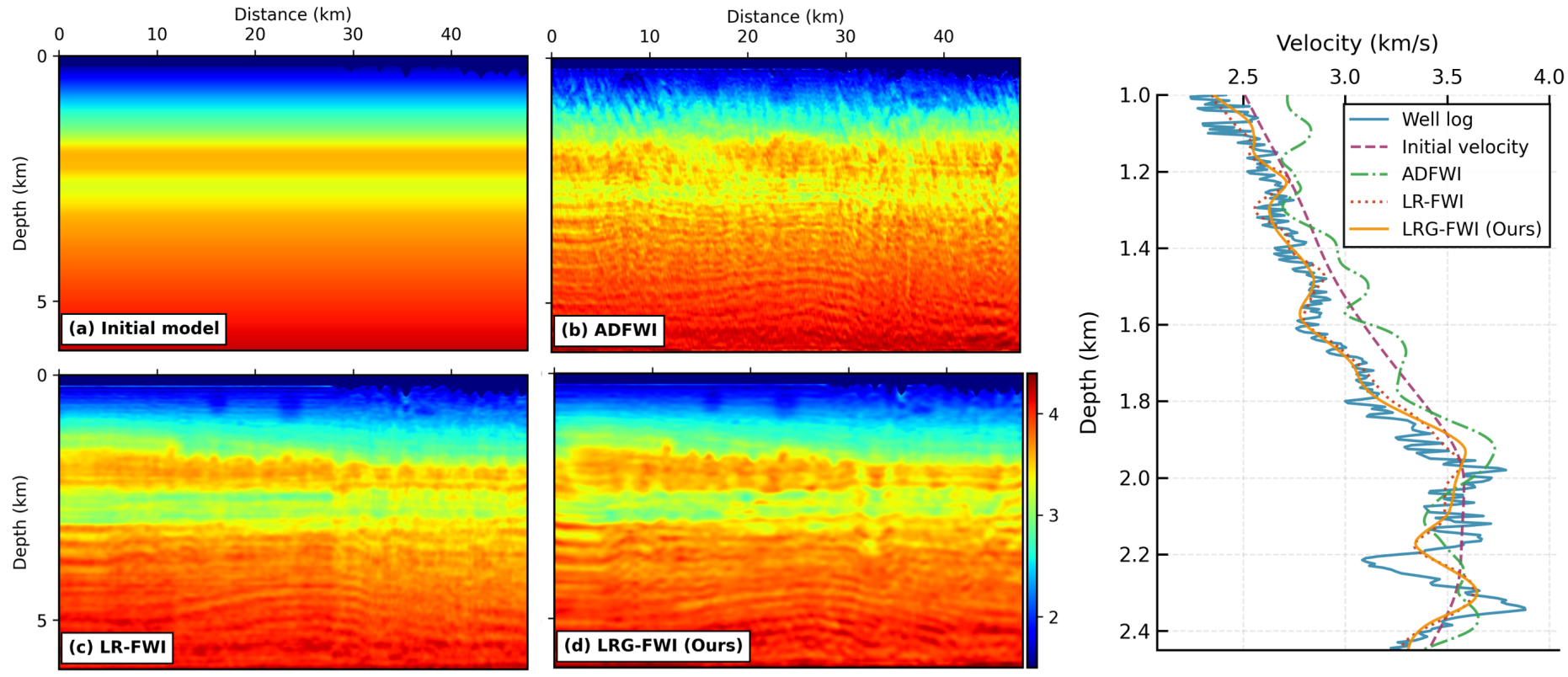

FIG. SM5. *Inversion results on the Chevron using different methods. (a) 1-D initial model. (b) ADFWI. (c) LR-FWI. (d) LRG-FWI. The right image is the well log of the Chevron model.*

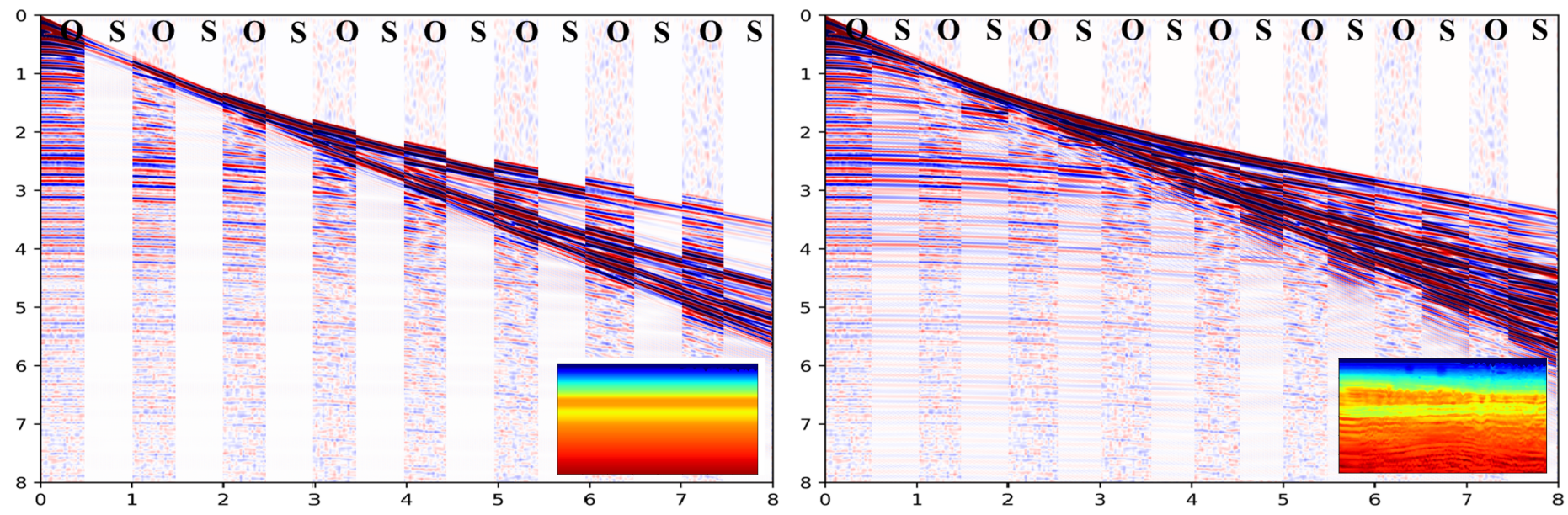

FIG. SM6. *Comparison of a common-source shot gather between the observed wavefield data and modeled records. Here S denotes simulated data and O denotes the observed data.*

**SM3.6. Scaling Selection Criterion.** We conducted additional experiments to validate the performance of the proposed scaling value selection criterion using the Overthrust model with smooth and constant initial velocity models, as shown in Fig. SM7. The results indicate that a higher scaling value is suitable for the smooth initial velocity model to achieve high-resolution inversion, while a smaller scaling value is suitable for the constant initial velocity model to ensure inversion stability. In both cases, the proposed criterion consistently yields relatively optimal results.

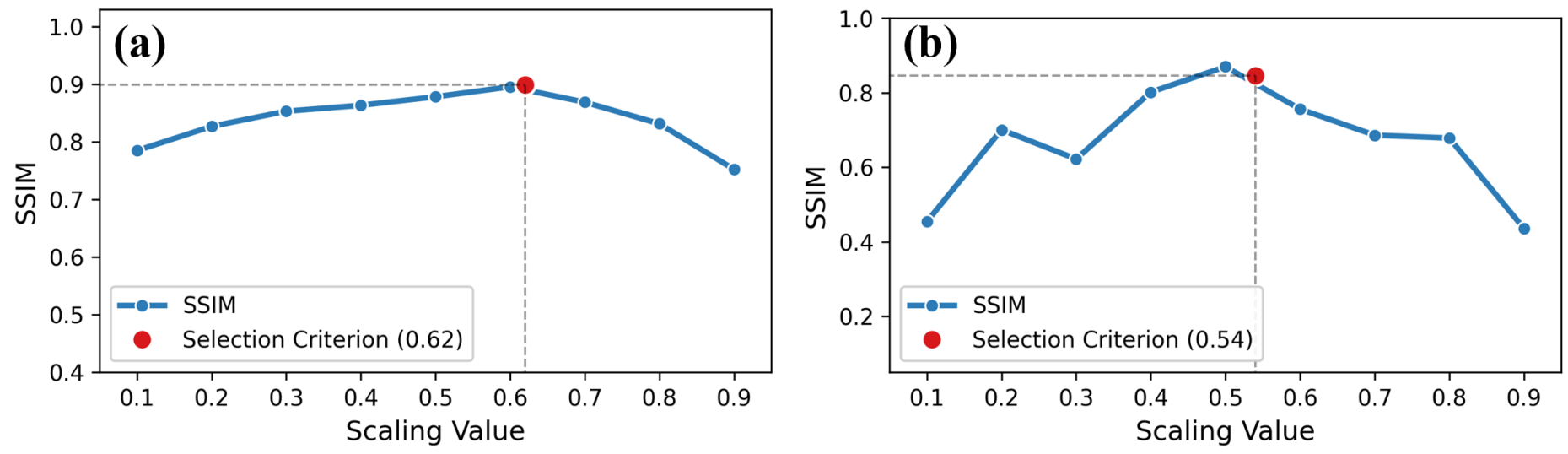

Fig. SM7. *Inversion results on the Chevron using different methods. (a) 1-D initial model. (b) ADFWI. (c) LR-FWI. (d) LRG-FWI. The right image is the well log of the Chevron model.*